\documentclass[pra,twocolumn,nofootinbib,superscriptaddress,showpacs,floatfix, longbibliography]{revtex4-2}
\usepackage{amsmath,amssymb,amsfonts}
\usepackage{mathcomp}
\usepackage{comment}
\usepackage{graphicx,subfigure}
\usepackage{txfonts}
\usepackage{epstopdf}
\usepackage[title]{appendix}
\usepackage{float}
\usepackage{adjustbox}
\makeatletter
\newsavebox{\@brx}
\newcommand{\llangle}[1][]{\savebox{\@brx}{\(\m@th{#1\langle}\)}%
  \mathopen{\copy\@brx\kern-0.5\wd\@brx\usebox{\@brx}}}
\newcommand{\rrangle}[1][]{\savebox{\@brx}{\(\m@th{#1\rangle}\)}%
  \mathclose{\copy\@brx\kern-0.5\wd\@brx\usebox{\@brx}}}
\makeatother

\usepackage[usenames, dvipsnames]{color}
\usepackage{soul}
\usepackage[normalem]{ulem}
\usepackage{physics}
\usepackage{bbold}
\usepackage{footnote}
\usepackage[colorlinks=true,citecolor=blue,linkcolor=blue,urlcolor=blue]{hyperref}

\setcitestyle{numbers,square}

\usepackage{xr-hyper}

\newcommand*\colvec[3][]{
\begin{pmatrix}\ifx\relax#1\relax\else#1\\\fi#2\\#3\end{pmatrix}
}

\begin{document}

\title{Perturbation-theory approach for predicting vibronic selectivity by entangled-photon-pair absorption}

\author{C. D. Rodr\'iguez-Camargo}
 \email{christian.rodriguez-camargo.21@ucl.ac.uk}
\affiliation{%
Department of Physics and Astronomy, University College London, London WC1E 6BT, United Kingdom
}%

\author{H. \'O. Gestsson}
\affiliation{%
Department of Physics and Astronomy, University College London, London WC1E 6BT, United Kingdom
}%

\author{C. Nation}
\affiliation{%
Department of Physics and Astronomy, University College London, London WC1E 6BT, United Kingdom
}%

\author{A. R. Jones}
\affiliation{%
Biometrology, Department of Chemical and Biological Sciences, National Physical Laboratory, Teddington, Hampton Road, Middlesex TW11 0LW, United Kingdom
}%

\author{A. Olaya-Castro}
\email{a.olaya@ucl.ac.uk}
\affiliation{%
Department of Physics and Astronomy, University College London, London WC1E 6BT, United Kingdom
}%

\begin{abstract}
Using second-order perturbation theory in the light-matter interaction, we derive an analytical approximation for the vibronic populations of a diatomic system excited by ultrabroadband frequency entangled photons and evaluate the population dynamics for different  degrees of entanglement between photon pairs. Our analytical approach makes the same predictions as previously derived via numerical solutions of the complete Schrödinger equation [H. Oka, Physical Review A 97, 063859 (2018)], with the added advantage of providing clear physical insights into the vibronic selectivity as a function of the degree of photon correlations while requiring significantly reduced computational effort. Specifically, our analytical expression for the probability of vibronic excitation includes a factor which predicts the enhancement of vibrational selectivity as a function of the degree correlation between the entangled photon pairs, the targeted vibrational energy level, and the vibrational molecular structure encoded in the Franck-Condon factors. Our results illustrate the importance of going beyond the usual approximations in second-order perturbation theory to capture the relevance of the vibrational structure of the molecular system of interest in order to gain a deeper understanding of the possible quantum-enhancement provided by the interaction between quantum light and matter.
\end{abstract}

\maketitle

\section{Introduction}
Quantum entanglement is currently understood as a physical resource that can provide significant quantum advantages for a variety of processes and technologies, such as quantum teleportation~\cite{pirandola2015advances, hu2023progress}, quantum communication~\cite{ursin2007entanglement, zou2021quantum, azuma2023quantum}, cryptography~\cite{pirandola2020advances}, metrology~\cite{RevModPhys.90.035005, polino2020photonic, TAYLOR20161}, imaging~\cite{lemos2014quantum, PhysRevA.92.013832, PhysRevResearch.4.033252, barreto2022quantum}, and sensing~\cite{RevModPhys.89.035002, pirandola2018advances}. In particular, entangled photon states have emerged as powerful and sensitive tools for probing complex molecular ~\cite{schlawin2013two, PhysRevA.82.013820, fujihashi2023probing} and photo-sensitive biological systems \cite{zheltikov2020photon, zong2021miniature, guzman2010spatial, upton2013optically, varnavski2022quantum}.

Entangled two-photon absorption (ETPA) processes in molecular systems have been intensively investigated over the last decade ~\cite{RevModPhys.88.045008, varnavski2017entangled, kang2020efficient, burdick2018predicting}.  In this process, the system transitions from the ground state to an excited state through the simultaneous absorption of a pair of entangled photons~\cite{RevModPhys.88.045008, eshun2022entangled, szoke2020entangled}. It has been proposed that quantum correlations embedded in the photon pair can provide an advantage in accessing  molecular information which remains inaccessible to classical light sources~\cite{schlawin2018entangled, Mukamel_2020}. Indeed, a variety of quantum states of light can offer new avenues for selective molecular excitation as well as for controlling relaxation and radiative processes ~\cite{burdick2018predicting, gu2021photoisomerization, PhysRevResearch.3.033154}.

A prominent advantage of employing two-photon non-classical light in spectroscopy is the occurrence of a two-photon absorption process with a resonant intermediate state at a rate linearly proportional to the light flux~\cite{PhysRevLett.78.1679, PhysRevB.69.165317,varnavski2020two}. For perfectly number-correlated light beams and low intensities, the flux of entangled photon pairs scales linearly with intensity, indicating that the biphoton effectively acts as a single unit ~\cite{PhysRevA.41.5088}. For larger intensities, theory predicts a crossover between linear and quadratic dependence on the photon flux, the latter being characteristic of classical or uncorrelated two-photon absorption (CTPA)~\cite{peticolas1967multiphoton}.

Several experimental efforts have been made to measure the ETPA cross sections of a variety of organic molecules~\cite{villabona2017entangled, eshun2018investigations, PhysRevLett.129.183601, PhysRevA.103.033701, villabona2018two, 10.1063/5.0193311}, with discrepancies reported for the results from different laboratories performing measurements on the same molecular system ~\cite{PhysRevLett.129.183601, PhysRevA.110.033708, doi:10.1021/acs.jpclett.2c00865, PhysRevApplied.15.044012} and other experiments reporting no ETPA detection for the same set-up and system, where previous reports indicated positive results for ETPA absorption cross sections \cite{PhysRevLett.129.183601, PhysRevA.110.033708}. For a comprehensive discussion of the bounds on ETPA absorption cross sections in common fluorophores and a detailed comparison between different experimental setups, we refer the reader to Ref.~\cite{PhysRevApplied.15.044012}. Hence, an open problem in the field is to understand the sources of experimental discrepancies and the inconsistencies between theory and experiment~\cite{PhysRevA.110.033708}. 

Most theoretical studies consider simplified models that do not necessarily capture the complexity of the systems of interest; for instance, they neglect the vibrational structure relevant to the systems~\cite{PhysRevLett.78.1679, PhysRevLett.80.3483, PhysRevA.57.3972, PhysRevA.57.3991, Leon-Montiel_2013, Schlawin_2017}. They also rely on perturbation approaches in limits that may not be compatible with the experimental conditions, for instance, finite-time pulses versus extended pulses~\cite{10.21468/SciPostPhysCore.4.4.028, PhysRevLett.78.1679, PhysRevB.69.165317, mollow1968two, RevModPhys.88.045008, 10.1063/5.0082500}. Indeed, it has been discussed that a more complete framework is necessary for a thorough understanding on the quantum advantage of ETPA~\cite{RevModPhys.88.045008, 10.1063/5.0049338}.  

In particularly, considerations of the molecular electronic and vibrational structures of the systems of interest appear crucial for accurate predictions~\cite{oka2011, oka2011control, PhysRevA.100.053844, PhysRevA.89.013830}. For instance, in Ref.~\cite{varnavski2023colors} by studying zinc tetraphenylporphyrin (ZnTPP) molecule, it was shown that considering the details of the molecular electronic structure, it is possible to explain the spectral shifting between ETPA and CTPA measured in the experiments. The experimental and theoretical results of this study discern from previous theoretical descriptions which neglect the electronic or vibronic structure of the matter~\cite{Leon-Montiel_2013, Landes21, 10.1063/5.0049338} and predicted similar shapes for ETPA and CTPA spectra. 

A few numerical studies have considered the interaction between entangled photons pairs and molecular systems modeled with their relevant vibrational structures~\cite{oka2011, oka2011control, oka2020enhanced, PhysRevA.97.063859}. In Ref.~\cite{PhysRevA.97.063859}, the author considered a diatomic molecule with a vibrational structure in the intermediate and excited electronic states to investigate vibrational selectivity in the excited state via two-step excitation (TSE) by ultrabroadband frequency-entangled photons. The numerical approach employed relies on solving the complete Schrödinger equation, which requires large computational resources owing to the need for precise discretisation of the frequency domain of the entangled photon states and the large number of vibrational levels that need to be considered. 

In this paper, we revise the findings of Ref.~\cite{PhysRevA.97.063859} and show that a carefully developed second-order perturbation theory that avoids common approximations, provides an efficient and insightful theoretical framework to investigate vibrational selectivity via ultra-broadband entangled photon pairs, as presented in~\cite{PhysRevA.97.063859}. Our approach significantly reduces the computational complexity of studying the interaction between the vibrational molecular structure and entangled photons, while providing clear physical insight into the factors dominating vibrational selectivity. Specifically, we derive a semi-analytical expression that predicts the population dynamics of the vibrational states in the excited electronic state for different degrees of correlation of the biphoton state. Our perturbative treatment assumes a finite-time light-matter interaction, without resorting to any in-resonance or far-resonance approximations. Using the same molecular parameters as those used in Ref.~\cite{PhysRevA.97.063859}, we show that the dynamics predicted by our approach fully compare with the numerical solutions presented in~\cite{PhysRevA.97.063859} up to a small relative error, which decreases as the degree of entanglement increases. 

Our analytical framework provides a clear understanding of how the excitation efficiency of vibronic excited states depends on the degree of photon correlation, the targeted vibrational energy level, and the molecular structure.  We are therefore able to present a detailed analysis of the probability of transition from the ground to an excited vibronic state for intermediate degrees of photon correlations, not previously considered, and show that vibrational selectivity can be achieved for non-perfectly entangled pairs. 

We also show how CTPA can be seen as a multiplicative single-photon process of transitions from ground to intermediate states, and from intermediate to excited states, while ETPA becomes a weighted average of two differentiated one-photon transitions, modulated by a Gaussian envelope that captures the resonance or off-resonance conditions and the degree of two-photon correlation. These results then allow us to understand the key physical differences in vibronic excitation with uncorrelated and quantum-correlated photons. 

Taking advantage of the low computational effort required by our approach, we further explore different resonance scenarios to understand the selectivity properties of different vibrational levels in the excited electronic state, the number of entangled modes in the biphoton state, and the molecular structure.  We discuss how the Franck-Condon structure is fundamental to understanding the differences that we observed, thereby highlighting the relevance of accounting for the vibronic structure of matter in predicting quantum enhancements given by entangled photon pairs, in line with recent remarks ~\cite{varnavski2023colors}.

The paper is organized as follows: In Sec.~\ref{preliminaries} we present the characterisation of the photon field and the molecular system. In Sec.~\ref{interaction} we present the interaction Hamiltonian, the perturbation theory approach and our analytical expression to evaluate the transition probabilities fo interest, and establish the set of Schrödinger equations to be solved numerically to benchmark with ~\cite{PhysRevA.97.063859} . In Sec.~\ref{results} we present the comparison between the numerical approach and our analytical framework.  Finally, in Sec.~\ref{conclusions} we summarise our results and present the outlook and perspectives. 

\section{Theoretical model}
\label{preliminaries}
In this section, we review the model and parameters presented in~\cite{PhysRevA.97.063859}. Hereafter, we use units such that $\hbar = c = 1$. Because we are developing an analytic structure to be compared with the numerical solutions of the complete Schrödinger equation, we do implement the same molecular parameters and simulation constants given in the aforementioned paper. First, we present the formalism involved in the configuration of frequency-entangled photons with energy anticorrelation. We present the one and two-photon states and their corresponding spectral amplitudes. We also present the molecular electronic-vibration configurations given by the Morse potentials and their vibrational eigenfunctions. Finally, we set the interaction Hamiltonian between the entangled photons and molecule given by the Franck-Condon approximation for transitions.  

\subsection{Frequency-entangled photons with energy anticorrelation}

We consider the interaction between two (propagating) pulses of a quantized radiation field (the “field” degrees of freedom) and three electronic energy levels of a molecule which has inner vibrational levels for each electronic level (the “molecular” degrees of freedom). In this section we describe the field Hamiltonian and its associated one-photon and two-photon states (uncorrelated and entangled). 

The incoming light pulses impinge on the three-level target located at the origin of the reference frame (See Fig.~\ref{figsys11}). Each field is quantized within a (cylindrical) quantization volume, along the same propagation direction, giving rise to modes labeled by a one-dimensional continuous variable, either the wave vector $k$ or frequency $\omega$~\cite{loudon2000}. Therefore, because we consider a one-dimensional input-output photon field interacting with a molecular system, we obtain annihilation $\hat{a}(k)$, and creation $\hat{a}^{\dagger}(k)$ operators satisfying the commutation relation $[\hat{a}(k), \hat{a}^{\dagger}(k')] = \delta (k-k')$. In this framework, the total Hamiltonian for both fields is:
\begin{equation}
\hat{H}_{F}=\int dk \, k\hat{a}^{\dagger}(k)\hat{a}(k) .
\label{fieldhamiltonian1}
\end{equation}

Since we are dealing with ultra-broadband photon fields, we avoid labels that differentiate signal and idler photons~\cite{PhysRevA.76.043813}.

Within the field Hamiltonian~(\ref{fieldhamiltonian1}) we define the one-photon state as~\cite{PhysRevA.42.4102, santos1997}
\begin{equation}
    |\psi ^{(1p)}\rangle = \int dk\,  \psi ^{(1p)} (k)\hat{a}^{\dagger}(k)|0\rangle ,
\end{equation}
where $\psi ^{(1p)} (k)$ is the momentum representation, or wave function, of the one-photon state $|\psi ^{(1p)}\rangle$ given by $\psi ^{(1p)}(k)= \langle k|\psi ^{(1p)}\rangle$, where $|0\rangle$ is the vacuum state defined as is usual, i.e., $\hat{a}(k) |0\rangle = 0$, and the continuous-mode single-photon state $|k\rangle$ is defined by $|k\rangle = |1_{k}\rangle = \hat{a}^{\dagger}(k)|0\rangle$~\cite{loudon2000, PhysRevA.76.043813, PhysRevA.42.4102, santos1997, migdall2013single}.

In an analogous way, we define the two-photon state as follows:
\begin{equation}
     |\psi ^{(2p)}\rangle = \int dk\, \int dk' \,  \psi ^{(2p)} (k, k')\hat{a}^{\dagger}(k) \hat{a}^{\dagger}(k')|0\rangle ,
\end{equation}
being $\psi ^{(2p)} (k, k') = \langle k, k' |\psi ^{(2p)}\rangle$ the two-photon joint spectral amplitude (JSA)~\cite{PhysRevA.50.5122, Sergienko:95}.

The description of the two-photon JSA is provided by the definition of the spatiotemporal one-photon wave packet $\varphi (r)$. For simplicity, this wave packet can be described by a Gaussian form as follows~\cite{oka2011}

\begin{equation}
\varphi (r) = \frac{1}{\sigma _{r}}\exp \left[-\frac{(r-r_{0})^{2}}{2\sigma _{r}^{2}}+ik_{0}(r-r_{0})\right],
\label{varphir1}
\end{equation}
where $r_{0}$ is the spatial center position of the wave packet at $t_{0}$, $\sigma _{r}$ is the coherent length of the wave packet, and $k_{0}$ is the central energy of the photon pulse.

The one-photon wave packet in the $k$ representation can be obtained using the Fourier transform of (\ref{varphir1}), which yields

\begin{equation}
\varphi (k)\exp \left(-ikr_{0}\right)=\frac{1}{\sqrt{2\pi}}\int _{-\infty}^{\infty}dr\, \varphi (r)\exp \left(-ikr\right) .
\end{equation}
Thus,

\begin{equation}
\varphi (k) = \frac{1}{\sqrt{2 \sigma ^{2}}} \exp \left[-\frac{(k-k_{0})^{2}}{4\sigma ^{2}}\right]
\end{equation}
where $\sigma = \sigma _{r}^{-1}$ denotes the spectral width of the wave packet.

Using the one-photon wave packet, we can now describe the two-photon JSA $\psi ^{(2p)}(k, k')$. 

Since an uncorrelated photon pair has no correlation between the two photons, the two-photon joint amplitude can be expressed as the product of one-photon wave packets, given by~\cite{Leon-Montiel_2013}:

\begin{equation}
\psi ^{(2p)}_{unc}(k, k')=\frac{1}{\sqrt{2\pi \sigma ^{2}}}\varphi (k)\varphi (k')\exp \left[-i(k+k')r_{0}\right] .
\label{psi2unco}
\end{equation}

On the other hand, the entangled photon pair with energy anticorrelation can be described as

\begin{equation}
\psi ^{(2p)}(k, k')=\varphi (k)\delta (k+k'-2k_{0})\exp \left[-i(k+k')r_{0}\right] .
\label{psientangled11}
\end{equation}

The Dirac delta function $\delta (k+k'-2k_{0})$ ensures the quantum-mechanical energy anticorrelation of the two photons: one photon with energy $k=k_{0}-\Delta$ is accompanied by another photon with energy $k'=k_{0}+\Delta$, conserving the total energy of $2k_{0}$. Using Fourier transform in the time domain, this property implies that the photon pair inherently has a time coincidence. Experimentally, this photon state can typically be obtained by spontaneous parametric down-conversion~\cite{PhysRevLett.59.2044, PhysRevA.50.23, gu2023photoelectron}.

In practice, the energy anticorrelation of experimentally generated entangled photons is not given by a perfect Dirac delta function, as shown in Eq. (\ref{psientangled11}), but instead exhibits a finite linewidth primarily due to the SPDC process bandwidth and the properties of the pump. This linewidth determines the degree of correlations between photon pairs. We therefore use the Gaussian definition of the Dirac delta function such that Eq. (\ref{psientangled11}) is given by

\begin{equation}
\psi ^{(2p)}(k, k')=\frac{1}{\sqrt{2\pi \sigma _{s}\sigma}}\varphi (k)\phi(k+k'-2k_{0})\exp \left[-i(k+k')r_{0}\right] .
\label{psientangled112}
\end{equation}
where $\phi(k)$ is defined as follows:
\begin{equation}
\phi (k) = \frac{1}{(\pi\sigma _{s}^{2})^{1/4}} \exp \left(-\frac{k^{2}}{4\sigma _{s}^{2}}\right) .
\end{equation}

 In the limit of $\sigma _{s}\rightarrow 0$, $\phi (k)$ becomes a perfect Dirac delta function and therefore corresponds to photon pairs with the maximum achievable correlation.  A finite linewidth, characterized by a finite $\sigma _{s}$, then implies variation of such correlations. As $\sigma_s$ becomes comparable to the spectral width $\sigma$, we are dealing with photons approaching the uncorrelated limit. Therefore, in what follows, we set $\sigma_s \equiv \sigma _{s}(\sigma) = \kappa \sigma$, with $0<\kappa \leq 1$, to investigate the population dynamics as function of the strength of photon correlations.

Note that expression (\ref{psientangled112}) is not symmetric; thus, we shall work with a symmetrized version of this given by~\cite{10.1063/5.0049338}
\begin{equation}
    \psi ^{(2p)}_{sym}(k, k')=\frac{\psi ^{(2p)}(k, k') + \psi ^{(2p)}(k', k)}{2} .
    \label{symmetrizedform1}
\end{equation}

Through the manuscript we consider the same parameters given in~\cite{PhysRevA.97.063859}.  For the photon field we set $\sigma = 10$ THz, and $\sigma _{s} = 500$ GHz as the upper bound in the correlation degree of entangled photons. We shall work with values of $\sigma _{s} = \sigma$, $\sigma _{s} = 0.5\sigma$, $\sigma _{s} = 0.25\sigma$, $\sigma _{s} = 0.1\sigma$, and $\sigma _{s} = 0.05\sigma$.  

To characterize the entanglement of correlated profiles, we use the Schmidt decomposition~\cite{10.1119/1.17904, PhysRevLett.85.1560, Sperling_2011}. This involves writing the JSA function $\psi ^{(2p)}(k, k')$ as a tensor product of two orthonormal sets $\{ |\phi \rangle _{j} \}$ and $\{ |\varphi \rangle _{j} \}$, such that,
\begin{equation}
|\psi ^{(2p)}(k, k')\rangle =\sum _{j}\lambda _{j}|\phi \rangle _{j} \otimes |\varphi \rangle _{j},
\end{equation}
which can similarly be written as,
\begin{equation}
\psi ^{(2p)}(k, k')=\sum _{j}\lambda _{j} \phi_{j}(k)\varphi _{j}(k'),
\end{equation}
where the coefficients $\lambda _{j}$ are the so called Schmidt coefficients and $\phi_{j}(k)\varphi _{j}(k')$ are the Schmidt modes~\cite{PhysRevLett.84.5304}.  

The entanglement encoded in $\psi ^{(2p)}(k, k')$ can be quantified by the Schmidt number~\cite{Carrington_2010, doi:10.1080/00107514.2013.878554}
\begin{equation}
K = \frac{\left( \sum _{j} \lambda _{j} \right)^{2}}{\sum _{j}\lambda _{j}^{2}}   .
\label{ksch}
\end{equation}
or entanglement entropy given by
\begin{equation}
S=-\sum _{j}\lambda _{j}^{2}\log _{2}\lambda _{j}^{2},
\label{entropysch}
\end{equation}

In Fig.~\ref{figsys11}, in the part $a)$, we depict a general form of a JSA (which is the fundamental object to characterize the propagating photon field) that departs from $r = r_{0}=-t_{0}$, and interacts with the molecular system at $r=0$. In Fig.~\ref{figsys11} $b)$ we show the behavior of the Schmidt number (\ref{ksch}) and the entanglement entropy (\ref{entropysch}) (inset) as a functions of the photons correlation degree given by $\sigma _{s} = \kappa \sigma$. Note that the Schmidt number (\ref{ksch}) increases significantly from $\sigma _{s}\leq 0.1\sigma$. This fact will help us to understand the selectivity dynamics derived from PT~\cite{varnavski2023colors}. In Fig.~\ref{figsys11} $c)$ We also depict the first three contributions of the Schmidt decomposition for a JSA with $\sigma _{s} = \sigma$.

\subsection{Molecular system}
We consider a molecular system (placed at $r=0$) which consists of three sets of vibronic states with inner vibrational modes: the ground state $|g\rangle$, the intermediate states $\{ |m_{\nu}\rangle \} _{\nu}$, and the excited states $\{ |e_{\alpha}\rangle \} _{\alpha}$. For the ground state, only the lowest vibrational mode was considered. The eigenenergies of these vibronic states are denoted as $\omega _{g}, \omega _{m_{\nu}}, \omega _{e _{\alpha}}$. 

\begin{figure*}
  \includegraphics[width=\textwidth,height=12.5cm]{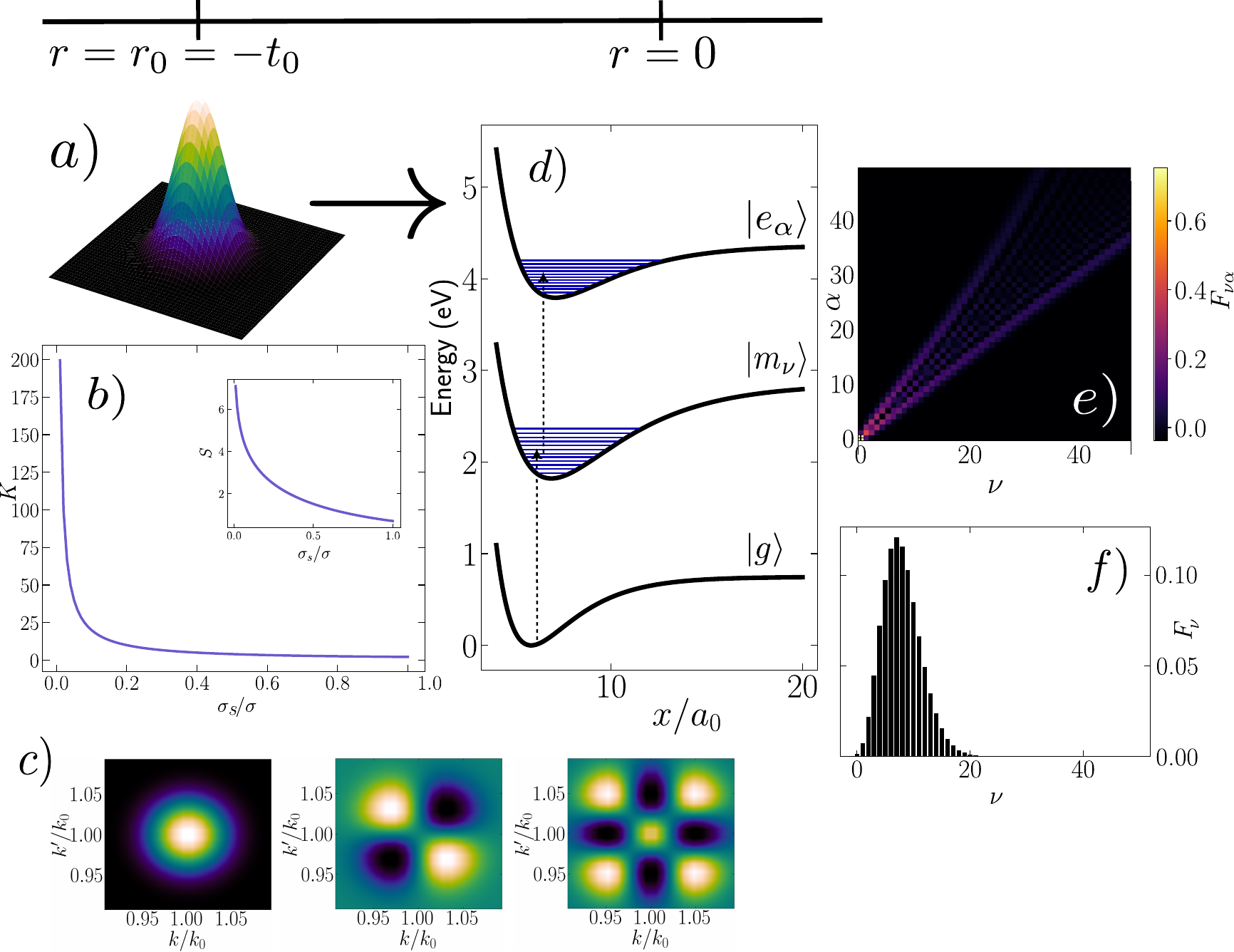}
  \caption{Schematics of the physical system. $a)$ A general form of a JSA which shall interact with the molecular system. The propagating photon field is departing from $r=r_{0}=-t_{0}$. $b)$ Behavior of the Schmidt number (\ref{ksch}) as a function of the photons correlation degree given by $\sigma _{s} = \kappa \sigma$. The inset is describing the behavior of the entanglement entropy (\ref{entropysch}). $c)$ The three firsts contributions of the Schmidt decomposition of a JSA with $\sigma _{s} =\sigma$. These Schmidt modes are used to characterize the JSA and its Schmidt number and entropy. We shall have more contributions as $\sigma _{s}\rightarrow 0$. $d)$ The three electronic levels of the molecular system, modeled by Morse potentials. The molecular system is placed at $r=0$. The intermediate vibrational sublevels are labeled with $\nu$, while the excited vibrational sublevels are labeled by $\alpha$. We are considering transitions between the ground state $|g\rangle$ and the intermediate sublevels $\{ |m_{\nu}\rangle \} _{\nu}$, and between the intermediate sublevels $\{ |m_{\nu}\rangle \} _{\nu}$ and excited sublevels $\{ |e_{\alpha}\rangle \} _{\alpha}$. These transitions are characterized by the Franck-Condon factors (\ref{fffactorspre}). $e)$ The Franck-Condon factors $F_{\nu \alpha}$ describing the transition from the intermediate sublevels and the excited sublevels. $f)$ The Franck-Condon factors $F_{0 \nu}\equiv F_{\nu}$ describing the transition from the ground state and the intermediate sublevels.}
  \label{figsys11}
\end{figure*}

The molecular system is a diatomic molecule, for which we approximate the adiabatic potential curve using the Morse potential. Thus, it is possible to obtain the vibrational eigenfunctions analytically. See Fig.~\ref{figsys11} $d)$. The Morse potential for the vibronic states of $\ell$ is defined as~\cite{PhysRev.34.57, PhysRev.42.210}:
\begin{equation}
V_{\ell}(x)=D_{\ell}\left[ \left( 1 - \exp \left(-\frac{x - x_{0}^{\ell}}{a_{\ell}}\right) \right) ^{2} - 1 \right] ,
\end{equation}
where $\ell$ is to label the levels $g$, $m$, and $e$; $x$ is the displacement of internuclear separation from the equilibrium position $x_{0}$. $D$ and $a$ are the depth and range of the potential, respectively. Therefore, the aforementioned eigenenergies of the vibronic states denoted by $\omega _{g}, \omega _{m_{\nu}}, \omega _{e _{\alpha}}$ are given by:
\begin{equation}
\omega _{\ell _{\beta}}=\epsilon _{\ell }+\omega _{\ell}\left( \beta +\frac{1}{2} \right) -\omega _{\ell}\chi _{\ell}\left( \beta +\frac{1}{2}\right) ^{2},
\end{equation}
where $\epsilon _{\ell}$ is the minimum of the potential energy, and
\begin{equation}
\omega _{\ell} = \sqrt{ \frac{2D_{\ell}}{a^{2}_{\ell}\mu} }
\end{equation}
with $D_{\ell}$ and $a_{\ell}$ are the depth and range of the potential, respectively. The expression for the anharmonicity of the Morse potential $\chi _{\ell}$ is
\begin{equation}
\chi _{\ell} = \frac{1}{\sqrt{8 a_{\ell}^{2}D_{\ell}\mu}} .
\end{equation}
Therefore, we shall adopt the following form for the molecular Hamiltonian
\begin{equation}
\hat{H}_{mol}=\sum _{\alpha}\omega _{e_{\alpha}}|e_{\alpha}\rangle \langle e_{\alpha}| + \sum _{\nu}\omega _{m _{\nu}} |m_{\nu }\rangle \langle m_{\nu }| .
\label{molecularhamiltonian1}
\end{equation}

Now, the corresponding vibrational eigenfunctions for each potential are given by
\begin{equation}
\xi _{\beta}^{\ell}(x)=N_{j_{\ell},\beta}\exp \left(-y_{\ell}(x)/2\right)\left( y_{\ell}(x) \right) ^{j_{\ell}/2 - \beta} L_{\beta}^{j_{\ell}-2\beta}(y_{\ell}(x))
\end{equation}
where $L$ represents the generalized Laguerre polynomials, and
\begin{equation}
y_{\ell}(x)=(j_{\ell}+1)\exp \left[ \frac{-(x-x_{0}^{(\ell)})}{a_{\ell}} \right],
\end{equation}
being
\begin{equation}
j_{\ell}=2a_{\ell}\sqrt{2\mu D_{\ell}}-1 .
\end{equation}
The normalization constant is defined as
\begin{equation}
N_{j_{\ell},\beta}=\left[  \frac{\beta ! (j_{\ell}-2\beta)}{a _{\ell}\Gamma (j_{\ell}-\beta+1)} \right] ^{1/2},
\end{equation}
where $\Gamma (z)$ denotes the gamma function.

For the optical transition, we adopt the Franck-Condon approximation. When the timescale of the light-molecule interaction is shorter than that of the molecular vibration, the optical transition rate between the two vibronic states can be approximated by the product of the electric-dipole transition rate and the Franck-Condon factor. The Franck-Condon factor is given by~\cite{10.1063/1.1725748}

\begin{equation}
F_{\alpha \nu}^{\ell \ell '}= \left|  \int dx\, \xi _{\nu}^{\ell '}(x)\xi _{\alpha}^{\ell}(x) \right| ^{2} .
\label{fffactorspre}
\end{equation}

In this study, we use the $\text{Na}_{2}$ molecule as an example of a diatomic system. The Morse parameters are listed in the Table~\ref{tableffactors}. The Franck-Condon factors for this molecule are shown in the Fig.~\ref{figsys11} part $e)$ and $f)$.

\begin{table}
    \centering
    \begin{tabular}{|c | c c c c|}  
 \hline
  & $\epsilon _{\ell}$ & $D_{\ell}$ & $a_{\ell}$ & $x_{0}^{\ell}$ \\ [0.5ex] 
 \hline\hline
 $|g\rangle$\, $(1^{1}\Sigma _{g}^{+})$ & 0 eV & 0.7466 eV & $2.2951 a_{B}$ & $5.82 a_{B}$ \\ 
 \hline
 $|m\rangle$\, $(1^{1}\Sigma _{u}^{+})$ & 1.8201 eV & 1.0303 eV & $3.6591 a_{B}$ & $6.87 a_{B}$ \\
 \hline
 $|e\rangle$ $(2^{1}\Pi _{g})$ & 3.7918 eV & 0.5718 eV & $3.1226 a_{B}$ & $7.08 a_{B}$ \\[1ex] 
 \hline
    \end{tabular}
 \caption{Morse parameters for the $\text{Na}_{2}$ molecule~\cite{PhysRevLett.67.3753, 10.1063/1.464755}. Here $a_{B}$ is the Bohr radius. $\mu = 19800$ is used in the units of electron mass.}
\label{tableffactors}
\end{table}

\subsection{Interaction Hamiltonian}

We consider only the transitions between $|g\rangle$ and $\{ |m_{\nu}\rangle \} _{\nu}$, and between $\{ |m_{\nu}\rangle \} _{\nu}$ and $\{ |e_{\alpha}\rangle \} _{\alpha}$, as illustrated by the vertical dashed arrows in Fig. \ref{figsys11}(d). The direct transition from $|g\rangle$ to $\{ |e_{\alpha}\rangle \} _{\alpha}$ by one photon, and the vibrational relaxations within $\{ |m_{\nu}\rangle \} _{\nu}$ and $\{ |e_{\alpha}\rangle \} _{\alpha}$ are ignored on the assumption of an ultracold molecule, for simplicity. 

Thus, the Hamiltonian of the whole system is expressed as
\begin{eqnarray}
\hat{H}=&&\hat{H}_{F}+\hat{H}_{mol} 
\nonumber \\
&+& \sum _{\nu} \int dk\, \sqrt{\frac{\gamma _{m}F_{0\nu}^{gm}}{\pi}}(| m _{\nu} \rangle \langle g _{0} |\hat{a}(k) + \hat{a}^{\dagger}(k) | g _{0} \rangle \langle m _{\nu} |) 
\nonumber \\
&+& \sum _{\nu ,\alpha} \int dk\, \sqrt{\frac{\gamma _{e}F_{\nu \alpha}^{me}}{\pi}}(| e _{\alpha} \rangle \langle m _{\nu} |\hat{a}(k) + \hat{a}^{\dagger}(k) | m _{\nu} \rangle \langle e _{\alpha} |) ,
\nonumber \\
\label{hamiltonian1}
\end{eqnarray}
where $\gamma _{m _{\nu}}$ is the relaxation rate between $|g\rangle$ and $|m _{\nu}\rangle$ and $\gamma _{e_{\nu \alpha}}$ is the relaxation rate between $| m_{\nu}\rangle$ and $|e _{\alpha}\rangle$. In this study, we set $\gamma = \gamma _{m _{\nu}} = \gamma _{e_{\nu \alpha}}$. Under this assumption, the Hamiltonian~(\ref{hamiltonian1}) can be written as
\begin{equation}
\hat{H}=\hat{H}_{F}+\hat{H}_{mol} +\hat{H}_{\text{int}} ,
\end{equation} 
where 
\begin{eqnarray}
\hat{H}_{\text{int}}=\gamma _{s} \left\lbrace \sum _{\nu}\int dk\, F_{\nu}\left[ |m_{\nu }\rangle \langle g | \hat{a}(k) +\hat{a}^{\dagger}(k)|g\rangle \langle m_{\nu}| \right] \right.
\nonumber\\
+ \left.  \sum _{\nu ,\alpha}\int dk\, F_{\nu \alpha}\left[ |e_{\alpha}\rangle \langle m_{\nu} | \hat{a}(k) +\hat{a}^{\dagger}(k)|m_{\nu}\rangle \langle e_{\alpha}| \right]\right\rbrace ,
\nonumber\\
\end{eqnarray}
with $\gamma _{s}=\sqrt{\gamma}$, $F_{\nu}=\sqrt{F_{0\nu}^{gm} / \pi}$ and $F_{\nu \alpha} = \sqrt{F_{\nu \alpha}^{me} / \pi}$.

\section{Perturbation theory and Quantum Dynamics}
\label{interaction}
In this section, we translate the Hamiltonian to the interaction picture, and we present the main result of this paper: the analytical expression for the vibrational populations derived from second order perturbation theory. Furthermore, we present the set of Schrödinger equations to be solved numerically, in order to compare the accuracy of our expression.

If we work within the interaction picture, we have
\begin{equation}
\hat{H}_{\text{int}}(t)=\text{e}^{i\hat{H}_{0}(t-t_{0})}(\hat{H}_{\text{int}})_{S}\text{e}^{-i\hat{H}_{0}(t-t_{0})}, \qquad  \hat{H}_{0}=\hat{H}_{F}+\hat{H}_{mol},
\end{equation}
which yields
\begin{eqnarray}
\hat{H}_{\text{int}}(t)=\gamma _{s} \left\lbrace \sum _{\nu} F_{\nu} \hat{\tau}_{\nu}(t)   \right. \left. + \sum _{\nu ,\nu '} F_{\nu \nu'}\hat{\tau}_{\nu \nu '}(t)\right\rbrace .
\end{eqnarray}
If the light-matter interaction is switched on at $t=-t_{0}$, the operators $\hat{\tau}_{\nu}(t)$ and $\hat{\tau}_{\nu \nu '}(t)$ have the following form
\begin{equation}
\hat{\tau}_{\nu}(t) = \text{e}^{i\omega _{m_{\nu }}(t+t_{0})}|m_{\nu }\rangle \langle g | \hat{a}(t) +\text{e}^{-i\omega _{m_{\nu }}(t+t_{0})}\hat{a}^{\dagger}(t)|g\rangle \langle m_{\nu}| ,
\end{equation}
and
\begin{eqnarray}
\hat{\tau}_{\nu \nu '}(t) &=&  \text{e}^{i(\omega _{e_{\nu '}}-\omega _{m_{\nu}})(t+t_{0})} |e_{\nu '}\rangle \langle m_{\nu} | \hat{a}(t) 
\nonumber \\
&+&\text{e}^{-i(\omega _{e_{\nu '}}-\omega _{m_{\nu}})(t+t_{0})}\hat{a}^{\dagger}(t)|m_{\nu}\rangle \langle e_{\nu '}| .
\end{eqnarray}

In order to apply these operators, we can define an initial state as
\begin{equation}
|\psi _{0}\rangle = |\varphi _{0}\rangle \otimes |g\rangle ,
\end{equation}
where $|\varphi _{0}\rangle$ is the initial state of the field and $|g\rangle$ ground state of the molecule.

Using a perturbative expansion, the evolution operator satisfies
\begin{eqnarray}
\hat{U}(t, t_{0}) = 1-i\int _{-t_{0}}^{t}dt' \, \hat{H}_{\text{int}}(t') 
\nonumber \\
+ (-i)^{2}\int _{-t_{0}}^{t}dt'\, \hat{H}_{\text{int}}(t') \int _{-t_{0}}^{t'}dt''\, \hat{H}_{\text{int}}(t'')  .
\end{eqnarray}
Therefore, the transition between state $|g, \psi _{2p}\rangle$, which is the molecular ground state and the entangled pair of photons, to state $|e_{\alpha}, 0\rangle$, which is the vibrational excited state $\alpha$ and the vacuum in the field, is given by
\begin{equation}
\langle e_{\alpha}, 0 | \hat{U}(t, t_{0})|g, \psi _{2p}\rangle = (-i\gamma _{s})^{2}\int _{-t_{0}}^{t}dt'\int _{-t_{0}}^{t'}dt'' \sum _{\nu} G_{\nu \alpha}(t',t'') ,
\label{perturbation11}
\end{equation}
where
\begin{eqnarray}
G_{\nu \alpha}(t',t) = F_{\nu} F_{\nu \alpha}\exp \left[i\Omega _{\alpha \nu}(t'+t_{0})+i\omega _{m_{\nu}}(t''+t_{0})\right] \phi _{q}(t, t')    ,
\nonumber \\
\end{eqnarray}
with $\Omega _{\alpha \nu}=\omega _{e_{\alpha}}-\omega _{m_{\nu}}$, and
\begin{equation}
    \phi _{q}(t, t') = \langle 0 |\hat{a}(t)\hat{a}(t')|\psi_{2p}\rangle
\end{equation}
is the so-called two-photon wavefunction~\cite{PhysRevA.41.5088}.

After some algebraic manipulations, Eq. (\ref{perturbation11}) reads
\begin{equation}
\langle e_{\alpha}, 0 | \hat{U}(t, t_{0})|g, \psi _{2p}\rangle =2(-i\gamma _{s})^{2}\sum _{\nu}F_{\nu}F_{\nu \alpha}T_{\nu \alpha}(t, t_{0}),
\label{prob1}
\end{equation}
where 
\begin{widetext}
\begin{eqnarray}
T_{\nu \alpha}(t, t_{0}) = \int _{-t_{0}}^{t}dt'\int _{-t_{0}}^{t'}dt''\int dk\, \int dk '\, \exp \left[-i(k ' - \Omega _{\alpha \nu})(t'+t_{0})-i(k - \omega _{m_{\nu}})(t''+t_{0})\right] \psi _{sym}(k ,k ') ,
\label{prob2}
\end{eqnarray}
being $\psi _{sym}(k ,k ')$ the symmetrized form of the JSA of the entangled photons.

Since we are in the regime in which the incoming photon pulses of bandwidth $\Delta k$ are much smaller than their central frequency $k_{0}$, we extend the range of all the above frequency integrals to $(-\infty , \infty)$, such that, Eq. (\ref{prob2}) reads

\begin{eqnarray}
T_{\nu \alpha}(t, t_{0}) = &-&\int _{-\infty}^{\infty}dk \int _{-\infty}^{\infty}dk ' \, \frac{\exp \left[-i(k + k ' - \omega _{e _{\alpha}})t\right]\psi _{sym}(k ,k ')}{(k - \omega _{m _{\nu}})(k + k ' - \omega _{e _{\alpha}})} - \int _{-\infty}^{\infty}dk\, \int _{-\infty}^{\infty}dk ' \, \frac{\exp\left[-i(k + k ' - \omega _{e _{\alpha}})t_{0}\right]\psi _{sym}(k ,k ')}{(k ' - \omega _{e _{\alpha}} + \omega _{m _{\nu}})((k + k ' - \omega _{e _{\alpha}}))} 
\nonumber \\
&+& \int _{-\infty}^{\infty}dk\, \int _{-\infty}^{\infty}dk ' \, \frac{\exp\left[-i(k ' - \omega _{e _{\alpha}} + \omega _{m _{\nu}})t + (k - \omega _{m _{\nu}})t_{0}\right]\psi _{sym}(k ,k ')}{(k - \omega _{m _{\nu}})(k ' - \omega _{e _{\alpha}} + \omega _{m _{\nu}})} .
\end{eqnarray}

In the case of uncorrelated photons, we use the expression~(\ref{psi2unco}), such that Eq. (\ref{prob1}) yields
\begin{equation}
\langle e_{\alpha}, 0 | \hat{U}(t, t_{0})|g, \psi _{2p}\rangle =\frac{\pi \sqrt{\pi}(-i\gamma _{s})^{2} }{\sqrt{2\sigma  ^{2}}}\exp \left(i\omega _{e_{\alpha}}t\right)\sum _{\nu}F_{\nu}F_{\nu \alpha}  \zeta _{\nu \alpha}^{(u)} \left\lbrace  \varrho _{\nu \alpha}^{(u)}(t, t_{0}) + I_{\nu \alpha}^{(u)}(t, t_{0})  \right\rbrace ,
\label{tranunco1}
\end{equation}
being
\begin{equation}
\zeta _{\nu \alpha}^{(u)} = \exp \left[-\frac{(k_{0}-\omega _{m_{\nu}})^{2}}{4\sigma ^2} -\frac{(k_{0}- \Omega _{\alpha \nu})^{2}}{4\sigma ^2}\right],    
\end{equation}
\begin{eqnarray}
\varrho _{\nu \alpha}^{(u)}(t, t_{0}) = \text{erfi}\left( i \sigma (t - t_{0}) + \frac{1}{2\sigma} (k _{0} - \Omega _{\alpha \nu}) \right) \left[ 2\, \text{erf}\left( \sigma (t - t_{0}) - \frac{i}{2\sigma} (k _{0} - \omega _{m_{\nu}}) \right) - \text{erf}\left( 2\sigma t - \frac{i}{2\sigma} (k _{0} - \omega _{m_{\nu}}) \right)  \right]
\nonumber \\
- \text{erfi}\left( 2 i \sigma t + \frac{1}{2\sigma} (k _{0} - \Omega _{\alpha \nu}) \right)  \text{erf}\left( 2\sigma t - \frac{i}{2\sigma} (k _{0} - \omega _{m_{\nu}}) \right)
\end{eqnarray}
and 
\begin{eqnarray}
I_{\nu \alpha}^{(u)}(t, t_{0}) = \frac{i}{\pi \sigma ^{2}}\sum _{n,l = 0}^{\infty} \frac{(-1)^{l}\sigma ^{-2(n+l)}}{2^{n+l}n!l!(2l+1)}\left[2t^{2}\sigma^{4}f^{(u,1)}_{n,l}(t,t_{0}) + \frac{2t\sigma ^{2}-i(k _{0} - \omega _{m_{\nu}})}{(2n+1)(\omega _{e _{\alpha}} - 2\omega _{m_{\nu}})} f^{(u,2)}_{n,l}(t,t_{0}) \right],
\label{uncoexp1}
\end{eqnarray}
where $\text{erf}(z)$ is the error function of the complex variable $z$ defined as,
\begin{equation}
    \text{erf}(z) = \frac{2}{\sqrt{\pi}}\int _{0}^{z}dt\, \exp (-t^{2}) ,
\end{equation}
$\text{erfi}(z)$ is the imaginary error function given by $\text{erfi}(z) = -i\, \text{erf}(iz)$, and the explicit forms of $f^{(u,1)}_{n,l}(t,t_{0})$ and $f^{(u,2)}_{n,l}(t,t_{0})$ are given in the Appendix.

The expression (\ref{tranunco1}) describes the product of two Gaussians, which envelop a function composed of the product of the error functions. The first Gaussian function is centered at the resonance of the intermediate state energies and half of the central energy, $k_{0}$. One may observe that this part accounts for one-photon processes from the ground state to intermediate states. The second term is centered where the difference in energies between excited and intermediate states, that is, $\Omega _{\alpha \nu}$ is in resonance with half of the central energy, $k_{0}$. Therefore, this structure is describing one-photon processes which occur between intermediate and excited states. 

To obtain an expression for the transition amplitude of correlated photons, we use Eqs. (\ref{psientangled112}) and (\ref{symmetrizedform1}) to define $\psi _{sym}(k, k ')$, such that the symmetrized form reads
\begin{equation}
\psi _{sym}(k, k ') = \frac{1}{2\sqrt{2\pi \sigma _{s}\sigma}N} \phi(k+k'-2k_{0})\exp \left[-i(k + k ')r_{0}\right] \left\lbrace \varphi (k) + \varphi (k ')   \right\rbrace ,
\end{equation}
where we have introduced a normalization factor $N$ given by
\begin{equation}
    N = \sqrt{\frac{1}{2}+\frac{\sigma}{\sqrt{4\sigma ^{2}+\sigma _{s}^{2}}}},
\end{equation}
in order to have $\int \int \, dk \, dk' \, |\psi _{sym}(k, k ') | ^{2} = 1$. After some manipulations, Eq. (\ref{prob1}) yields
\begin{equation}
\langle e_{\alpha}, 0 | \hat{U}(t, t_{0})|g, \psi _{2p}\rangle =\frac{\pi \sqrt{\pi}(-i\gamma)^{2} }{2\sqrt{2\sigma \sigma _{s}}N}\exp(i\omega _{e_{\alpha}}t)\zeta _{\alpha}\sum _{\nu}F_{\nu}F_{\nu \alpha}   \left\lbrace \exp \left[ -\frac{(k_{0}-\omega _{m_{\nu}})^{2}}{4\sigma ^2}\right] \rho _{\nu \alpha}^{(en,1)}(t, t_{0}) + \exp \left[-\frac{(k_{0}- \Omega _{\alpha \nu})^{2}}{4\sigma ^2}\right] \rho _{\nu \alpha}^{(en,2)}(t, t_{0})  \right\rbrace ,
\label{tranen1}
\end{equation}
where
\begin{equation}
\zeta _{\alpha} = \exp \left[-\frac{(2k_{0}-\omega _{e_{\alpha}})^{2}}{4\sigma _{s}^{2}}\right],
\label{envgaussiane}
\end{equation}
\begin{eqnarray}
\rho _{\nu \alpha}^{(en,1)}(t, t_{0}) = \text{erf}\left(\frac{i}{2\sigma}(k _{0} - \omega _{m_{\nu}}) \right) \left[  \text{erf}\left( 2\sigma _{s}t - \frac{i}{2\sigma _{s}}(2k _{0} - \omega _{e_{\alpha}}) \right) - \text{erf}\left( \sigma _{s}(t - t_{0}) - \frac{i}{2\sigma _{s}}(2k _{0} - \omega _{e_{\alpha}}) \right)  \right] + I_{\nu \alpha}^{(en,1)}(t, t_{0}) ,
\nonumber \\ 
\label{entexp1}
\end{eqnarray}
and
\begin{eqnarray}
\rho _{\nu \alpha}^{(en,2)}(t, t_{0}) =  \text{erfi}\left( \frac{-2i\sigma ^{2} \sigma _{s}^{2}(t-t_{0})-\sigma ^{2}(2k _{0} - \omega _{e_{\alpha}})-\sigma _{s}^{2}(k _{0} - \Omega _{\alpha \nu})}{2\sigma \sigma _{s} \sqrt{\sigma ^{2} + \sigma _{s}^{2}}} \right) \left[ (i+1)\, \text{erf}\left(  \frac{2\sigma _{s}^{2}(t-t_{0}) - i(k _{0} - \omega _{m_{\nu}})}{2\sqrt{\sigma ^{2} + \sigma _{s}^{2}}} \right) \right] + I_{\nu \alpha}^{(en,2)}(t, t_{0}) .
\label{entexp2}
\end{eqnarray}
\end{widetext}

\begin{figure}[t]
	\includegraphics[width=0.49\textwidth]{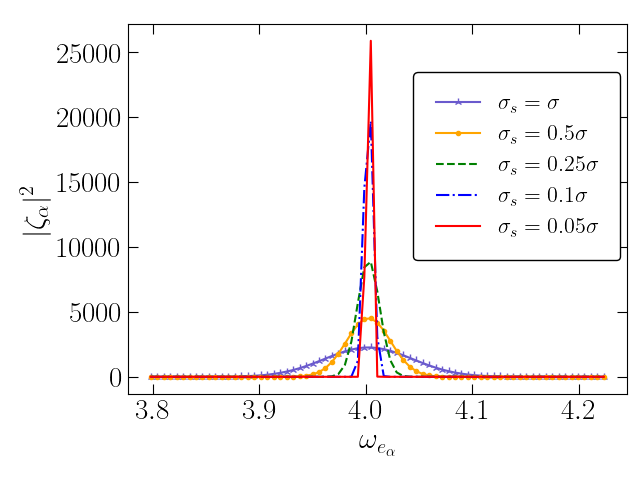}\\
	\caption{Envelope Gaussian function which establish selectivity.}
	\label{gausiane1}
 \end{figure}

In this case, we can observe a clear separation between the excitation processes from the ground states to the intermediate states, and the excitation processes from the intermediate states to the excited states. However, both are weighted by the product of the Franck-Condon factors from the ground to intermediate levels $\nu$, $F_{\nu}$, and the Franck-Condon factors from intermediate levels $\nu$, to excited levels $\alpha$, $F_{\nu \alpha}$, and a Gaussian function centered at the resonance between the target level eigenenergy and the correlated photons central energy $2k_{0}$, defined by Eq. (\ref{envgaussiane}). See Fig. \ref{gausiane1}. Within this fact, we may evidence that second-order PT predicts the strongly enhanced vibrational-mode-selective two-step excitation by ultrabroad-band frequency-entangled photons, and provides a function to control in a more precise way this vibrational selectivity at the experimental level with other kinds of molecular systems. Notice that by maintaining the integrals in Eq. (\ref{prob2}) at a finite temporal interval, gives us information about memory processes weighted by the spectral width of the wave packet, and the correlation parameter $\sigma _{s}$, given in the terms $\sigma(t + t_{0})$ and $\sigma_{s}(t-t_{0})$. Furthermore, the structures of the arguments of the error functions provide information on the relationships that arise between the photons correlations and vibronic correlations. The explicit forms of expressions (\ref{uncoexp1}), (\ref{entexp1}), and (\ref{entexp2}) can be found in the Appendix \ref{explicitfapp}.

\subsubsection{Quantum dynamics}
To compare the results from Perturbation Theory, we solve numerically the complete Schrödinger equation 

\begin{equation}
\frac{d}{dt}|\Psi (t)\rangle = -i\hat{H}|\Psi (t)\rangle,
\end{equation}
where $|\Psi (t)\rangle$ is a superposition state given by

\begin{eqnarray}
|\Psi (t)\rangle &=& \frac{1}{\sqrt{2}}\int dk \int dk' \psi ^{(2p)}_{sym}(k, k', t) \hat{a}^{\dagger}(k)\hat{a}^{\dagger}(k')|0\rangle |g_{0}\rangle 
\nonumber \\
&+& \sum _{\nu}\int dk \, \psi ^{(1pm)}(k, \nu ,t) \hat{a}^{\dagger}(k) |0\rangle |m_{\nu}\rangle 
\nonumber \\
&+& \sum _{\nu '}\psi ^{(e)}(\nu ', t)|0\rangle |e_{\nu '}\rangle,
\label{statepsi11}
\end{eqnarray}
where $\psi ^{(2p)}_{sym}(k, k', t)$ is the two-photon joint amplitude of the incident pulse at $|g_{0}\rangle$, $\psi ^{(1pm)}(k, \nu ,t)$ is the one-photon state at $|m_{\nu}\rangle$, and $\psi ^{(e)}(\nu ', t)$ is the zero-photon state at $|e_{\nu '}\rangle$.

For simplicity, in the calculation of quantum dynamics, we omit the degrees of freedom of the polarization of light by assuming linearly polarized light.

Applying the Hamiltonian operator to state (\ref{statepsi11}), we obtain the following set of Schrödinger equations: 
\begin{eqnarray}
\frac{d}{dt}\psi ^{(2p)}_{sym}(k, k', t) = &-&i(k+k')\psi ^{(2p)}_{sym}(k, k', t)
\nonumber \\
&-&i\gamma _{s}\sum _{\nu}\frac{1}{\sqrt{2}}F _{\nu}[\psi ^{(1pm)}(k, \nu ,t) 
\nonumber \\
&+& \psi ^{(1pm)}(k', \nu ,t)],
\label{sch1}
\end{eqnarray}

\begin{eqnarray}
\frac{d}{dt}\psi ^{(1pm)}(k, \nu ,t) = &-&i(k+\omega _{m _{\nu}})\psi ^{(1pm)}(k, \nu ,t) 
\nonumber \\
&-& i \gamma _{s} \sqrt{2}F _{\nu} \int dk' \, \psi ^{(2p)}_{sym}(k, k', t)
\nonumber \\
&-& i \gamma _{s} \sum _{\alpha}F _{\nu \alpha}\psi ^{(e)}(\alpha , t),
\label{sch2}
\end{eqnarray}

\begin{eqnarray}
\frac{d}{dt}\psi ^{(e)}(\alpha , t) = &-&i \omega _{e_{\alpha}}\psi ^{(e)}(\alpha , t)
\nonumber \\
&-&i\gamma _{s}\sum _{\nu} F _{\nu \alpha}\int dk \, \psi ^{(1pm)}(k, \nu ,t) .
\label{sch3}
\end{eqnarray}
The initial state of the whole system is given by $\psi ^{(1pm)}(k, \nu ,0) = \psi ^{(e)}(\alpha, 0) = 0$.

In order to solve Eqs. (\ref{sch1}), (\ref{sch2}), and (\ref{sch3}), we discretize the photon fields by converting from $\int dk$ to $\sum _{k} \delta k$, and $\hat{a}(k)$ to $\hat{a}_{k}$, and set $\delta k = 100$ GHz. We refer the reader to Appendix \ref{app:disc_schro_eq} for a detailed description of the discretized Schrödinger equations.

\section{Population dynamics}
\label{results}
In this section, we analyze the temporal behavior of Eqs. (\ref{tranunco1}) and (\ref{tranen1}), and we compare them with the numerical solutions of Eqs. (\ref{sch1}), (\ref{sch2}), and (\ref{sch3}). Recall that the central energy of the entangled photons is set to $2k_{0}=\omega_{e_{\alpha= 18}}\approx 4.0024$ eV, and $\gamma = 6$ MHz. We define $\langle e_{\nu}\rangle = |\langle e_{\nu}, 0 | \hat{U}(t)|g, \psi _{2p}\rangle |^{2}$.

In Fig.~\ref{fig_results_1a}, we depict the results for the uncorrelated JSA (\ref{psi2unco}), and low entanglement degrees $\sigma _{s} = \sigma$, and $\sigma _{s} = 0.5\sigma$. In Fig.~\ref{fig_results_1b}, we plot for the JSA with entanglement degrees $\sigma _{s} =0.25\sigma$, $\sigma _{s}=0.1\sigma$, and the most entangled degree $\sigma _{s}=0.05\sigma$. In all the Figures, the first column corresponds to the visualization of the JSA. The second column presents the results obtained by numerically solving Eqs. (\ref{sch1}), (\ref{sch2}), and (\ref{sch3}). In the third column we depict Eqs. (\ref{tranunco1}) and (\ref{tranen1}).

For uncorrelated photons, our expression predicts the simultaneously exciting of many vibrational modes owing to short pulse excitation with a spectrally broad bandwidth. The same phenomena were obtained using the numerical method. In addition, our expression reproduces the order (in terms of the numerical value achieved in the steady state) of the depicted levels. We can see that the differences between PT and the numerical method, in general, are given by: i) The slope of the transition between the null population value and steady state. For all levels, is higher for our expression; ii) Our expression reproduces the formation of Gaussian peaks in the neighborhood of $r\sigma = 0$ (where the molecular system is placed), for all levels. However, there is a substantial difference between the numerical results given by the narrowing, and the maximum values. In addition, for the maximum population, achieved for $\alpha = 12$, we observed the occurrence of a small Gaussian peak that was not observed in the numerical results. 

In Ref.~\cite{PhysRevA.97.063859} it is observed that the maximum population is achieved for $\alpha = 12$ owing to the value of the Franck-Condon factors $F_{\nu , \alpha = 18}$, which are very small for $\nu < 12$, and $F_{\nu = 10, \alpha = 12}$ becomes larger. For $\sigma _{s} = \sigma$ we have that the maximum population is now achieved for $\alpha = 14$, followed by $\alpha = 16$, and $\alpha = 15$.

Since the correlation between the photons is low, we have simultaneous excitations, as in the classical situation. This can be explained by the structure of Eqs. (\ref{tranunco1}) and (\ref{tranen1}), respectively. The Gaussian structure of Eq. (\ref{tranen1}) is now dominated by Eq. (\ref{envgaussiane}). For higher $\sigma _{s}$, this factor has a broad form, which allows simultaneous excitations. However, the factor is pushing towards to the targeted level, supported by the Gaussians shaped by the spectral width of the wave packet $\sigma$ and the energy differences $(k_{0}-\omega _{m_{\nu}})$ and $(k_{0}-\Omega _{\alpha \nu})$, which could be encoded in an \emph{effective} entangled transition factor defined by
\begin{equation}
\zeta _{\nu \alpha}^{(e)} =  \zeta _{\alpha}\left\lbrace \exp \left[-\frac{(k_{0}-\omega _{m_{\nu}})^{2}}{4\sigma ^2}\right] + \exp \left[-\frac{(k_{0}- \Omega _{\alpha \nu})^{2}}{4\sigma ^2}\right] \right\rbrace  .
\label{entangledfactor}
\end{equation}

To illustrate this point, let us define the transition matrix elements $\Theta _{\mu \alpha}^{(u, e)}$ as
\begin{equation}
\Theta _{\nu \alpha}^{(u, e)} = F_{\nu}F_{\nu \alpha}\zeta _{\nu \alpha} ^{(u, e)}  .
\label{tran_matrix1}
\end{equation}

In Fig.~\ref{fig_fcsigmas_1}, we plot the matrix elements~(\ref{tran_matrix1}) for the uncorrelated, and correlated cases. Panel $a)$ shows the uncorrelated case. Panels $b)$, $c)$, $d)$, show $\sigma_{s} = \sigma$, $\sigma_{s} = 0.5\sigma$, and $\sigma_{s} = 0.25\sigma$, respectively. We can evidence the displacement of the maximum achieved in the CTPA to the maximum achieved to each $\sigma _{s}=\sigma$-ETPA predicted using Eq. (\ref{entangledfactor}). Furthermore, notice the maximum values in each panel; our factor predicts (and can quantify) the quantum enhancement of the ETPA.

For $\sigma _{s} = 0.5\sigma$, the levels near to the resonance, that is, $\alpha = 17$ and $\alpha = 16$, achieve the maximum population. This is due to the fact that the factor Eq. (\ref{envgaussiane}) becomes narrower. We can also observe that the three main differences mentioned above are still present. 

For $\sigma _{s} = 0.25\sigma$, the maximum population is achieved by the targeted level, followed by levels near the resonance $\alpha = 17$, and $\alpha = 19$. Other levels with a significant population were $\alpha = 16$, and $\alpha = 20$. This means that the factor (\ref{envgaussiane}) becomes dominant. In this framework, for $\sigma _{s} = 0.1\sigma$, the targeted level achieved a dominant contribution, and for $\sigma _{s} =0.05\sigma$, the targeted level was completely selected, with a significant increase in the numerical value with respect to the other correlation degrees.

In Fig.~\ref{fig_evolution_sigmas}, we depict the evolution of the targeted level population for different values of $\sigma _{s}$. First of all, note that the results obtained using our analytical expressions (Fig.~\ref{fig_evolution_sigmas} $b)$) closely match those obtained via the numerical method (Fig.~\ref{fig_evolution_sigmas} $a)$). As $\sigma_s$ decreases from $\sigma_s=\sigma$ to $\sigma_s \rightarrow 0$ so does the the steady-state population of the targeted level, $\langle e _{\nu} \rangle _{s}$. In Fig.~\ref{figsys11} $b)$, we have shown that the Schmidt number $K$, indicating the number of entangled modes, grows nonlinearly as $\sigma_{s}$ goes from $\sigma_s=\sigma$ to $\sigma_s \rightarrow 0$; similar behavior is observed for the entropy of entanglement $S$ (see Fig.~\ref{figsys11} $b)$ (inset)). Hence, to understand the functional relation between $\langle e _{\nu} \rangle _{s}$ and the degree of entanglement quantified by  $K$ and $S$, in Fig.~\ref{en_vs_pop}, we present parametric plots of such steady state population as a function of $K$ (Fig.~\ref{en_vs_pop} $a)$) and of $S$ (Fig.~\ref{en_vs_pop} $b)$), indicating for each curve the $\sigma_s$ values we have considered. Interestingly, we see a linear relationship between the target state steady population and the number of entangled Schmidt modes, while we observe a non-linear relationship between such population and the entropy of entanglement, suggesting that the number of Schmidt modes plays a fundamental role in the vibronic selectivity investigated here.

\onecolumngrid

\begin{figure}[H]
    \centering
      \begin{tabular}{@{}ccc@{}}
    \begin{adjustbox}{valign=c}
    \includegraphics[width=.32\textwidth]{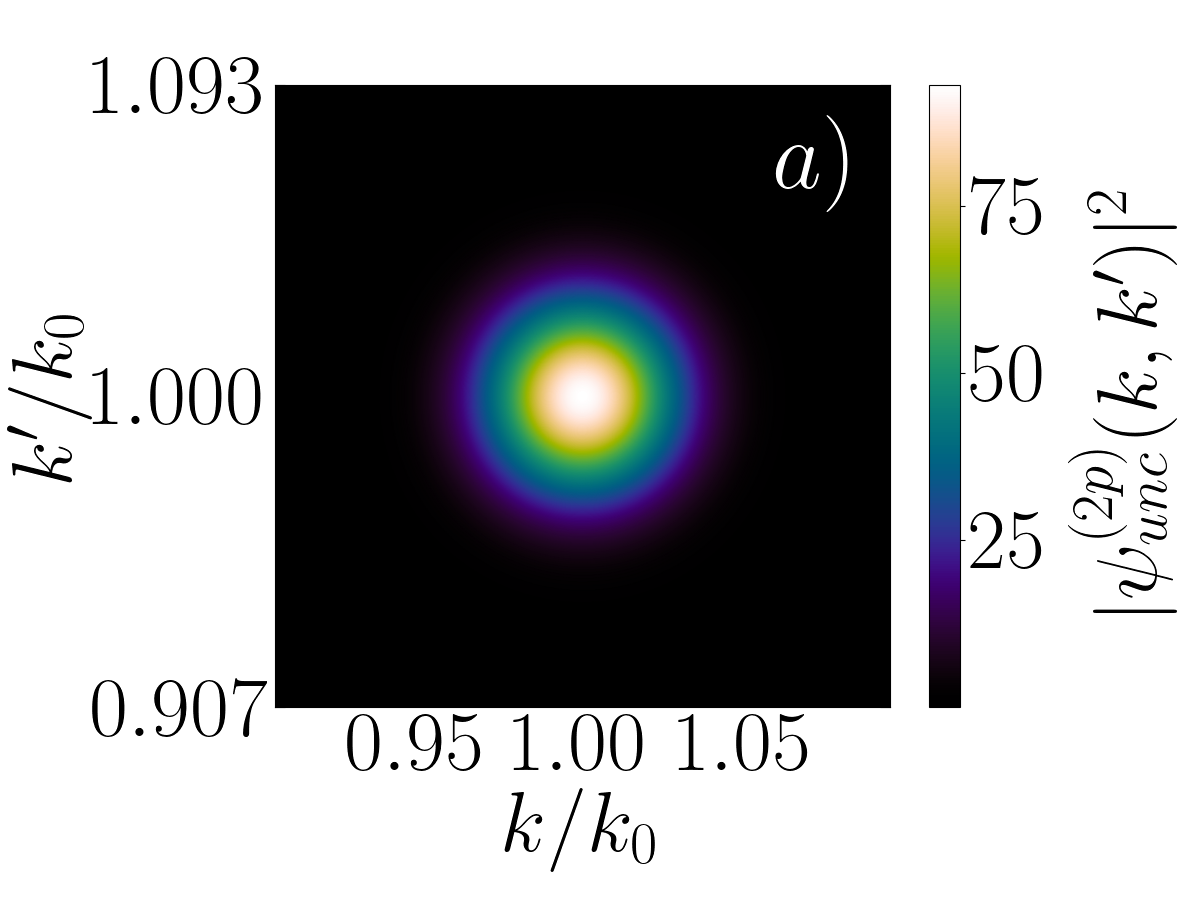}
    \end{adjustbox}&    
    \begin{adjustbox}{valign=c}
    \includegraphics[width=.27\textwidth]{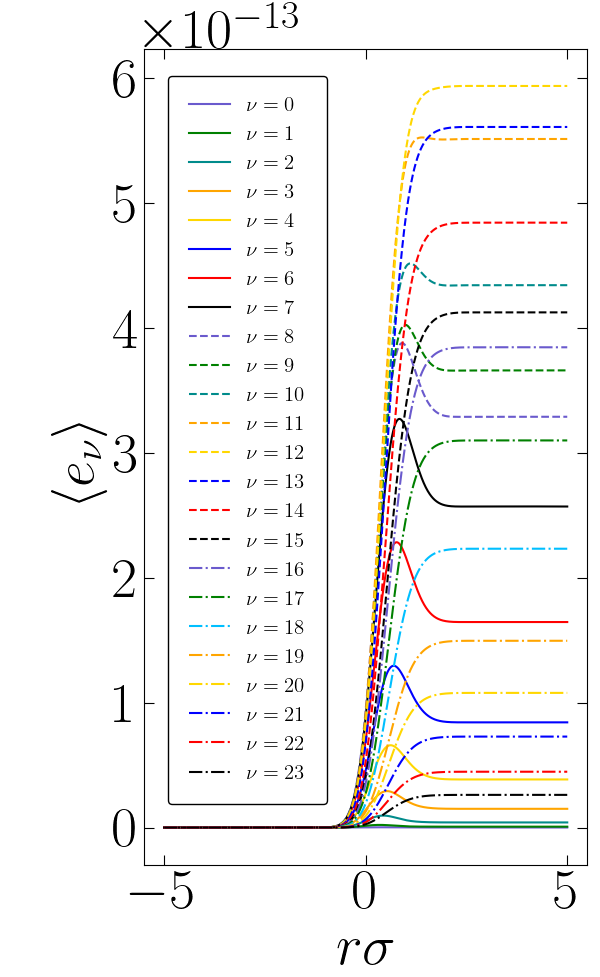}
    \end{adjustbox}& 
    \begin{adjustbox}{valign=c}
    \includegraphics[width=.27\textwidth]{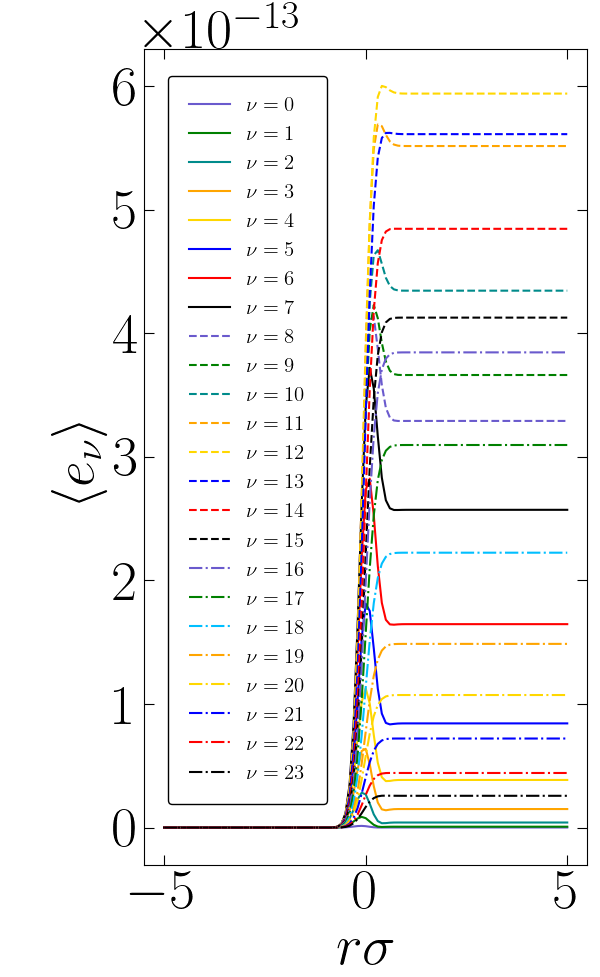}
    \end{adjustbox}\\    
    \begin{adjustbox}{valign=c}
    \includegraphics[width=.32\textwidth]{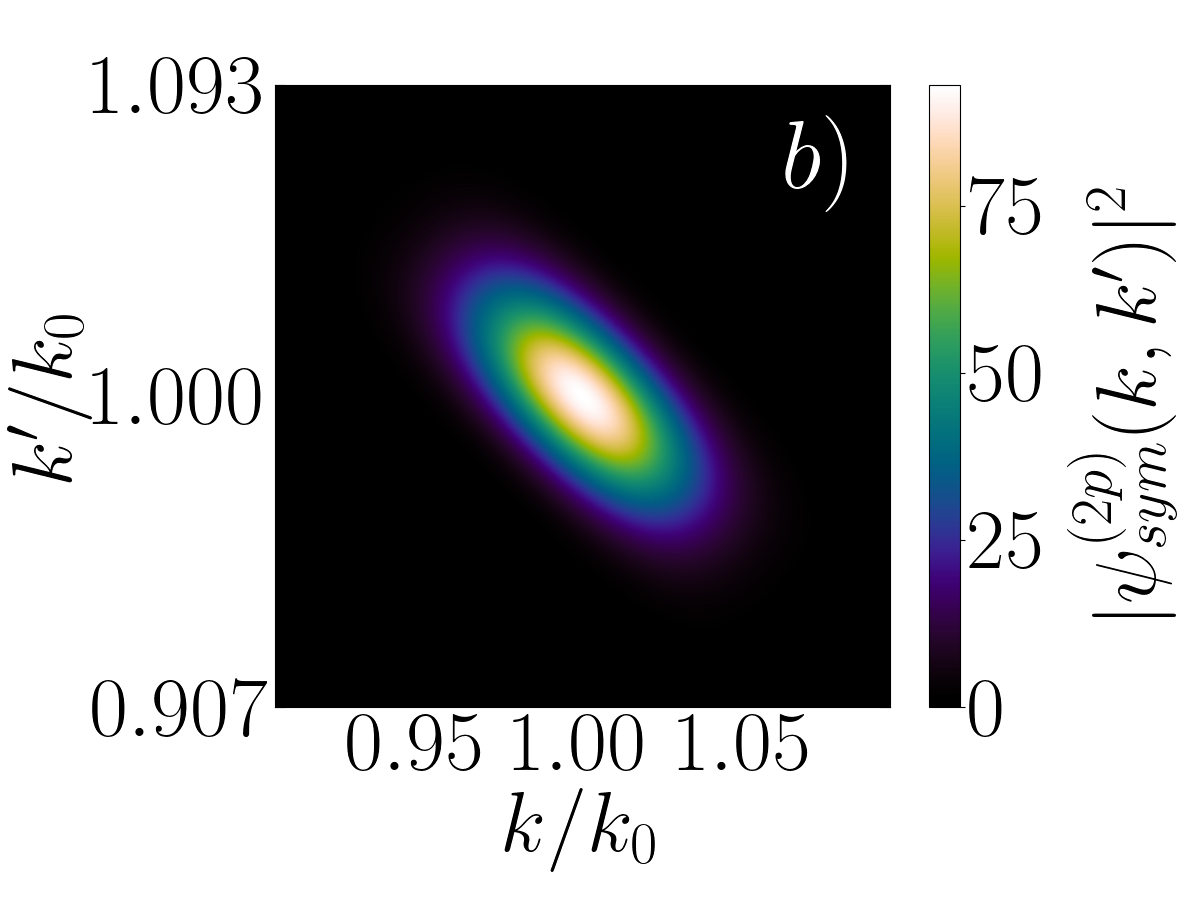}
    \end{adjustbox}& 
    \begin{adjustbox}{valign=c}
    \includegraphics[width=.27\textwidth]{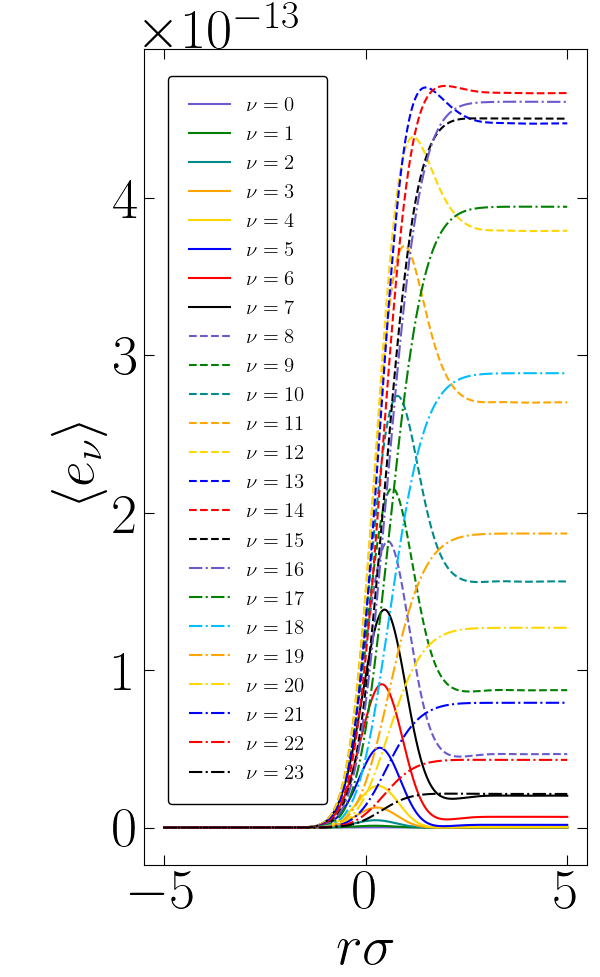}
    \end{adjustbox}&
    \begin{adjustbox}{valign=c}
    \includegraphics[width=.27\textwidth]{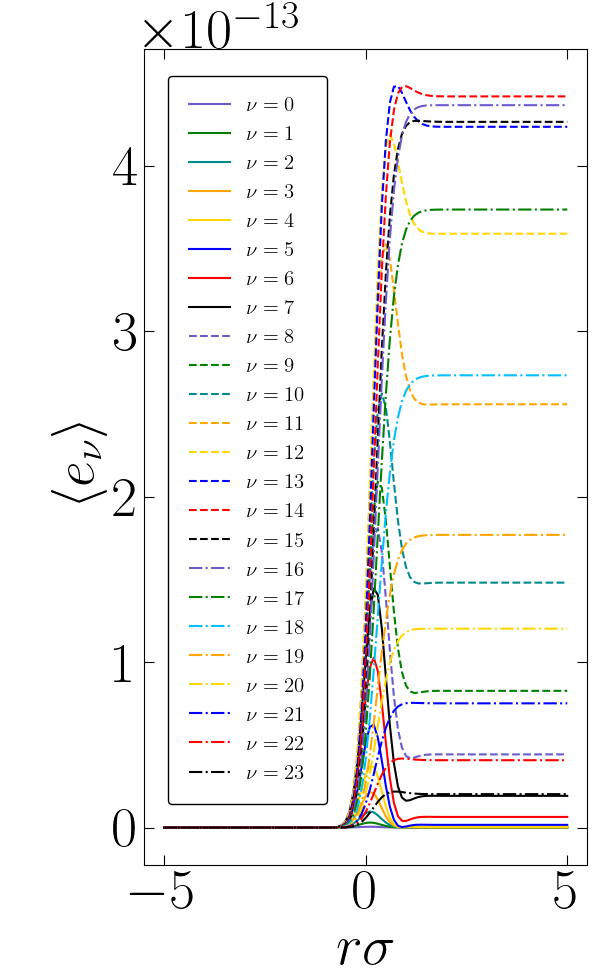}
    \end{adjustbox}\\
    \begin{adjustbox}{valign=c}
    \includegraphics[width=.32\textwidth]{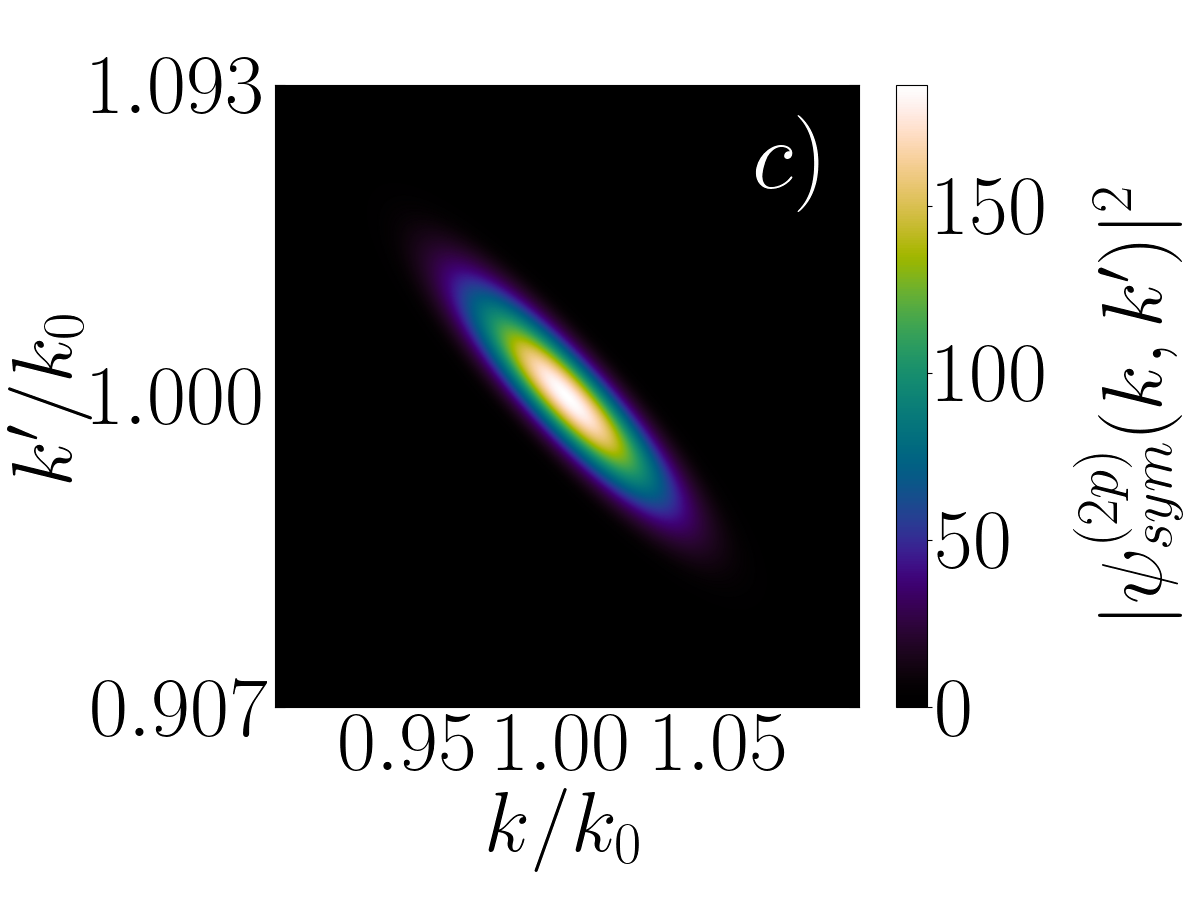}
    \end{adjustbox}&
    \begin{adjustbox}{valign=c}
    \includegraphics[width=.27\textwidth]{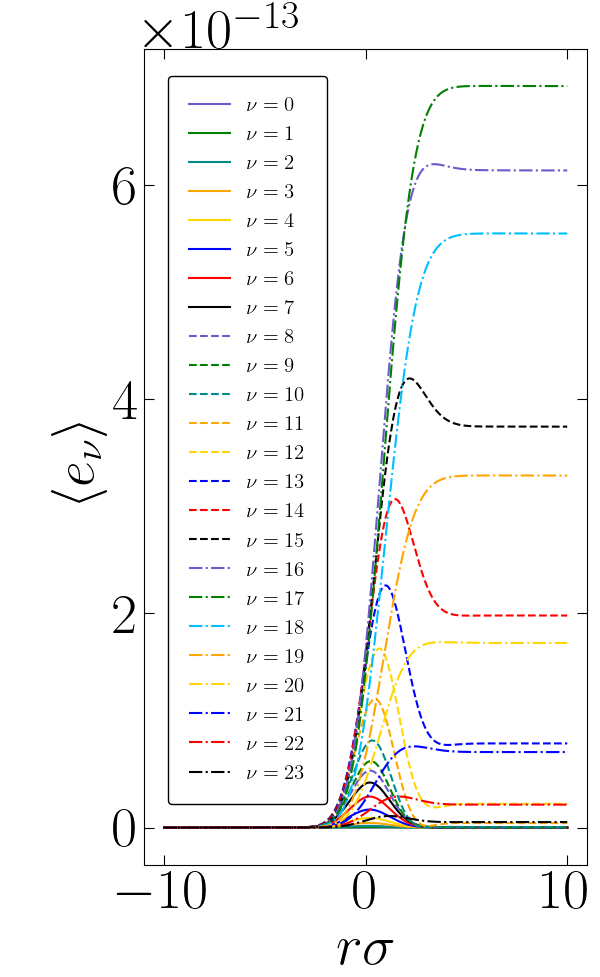}
    \end{adjustbox}& 
    \begin{adjustbox}{valign=c}
    \includegraphics[width=.27\textwidth]{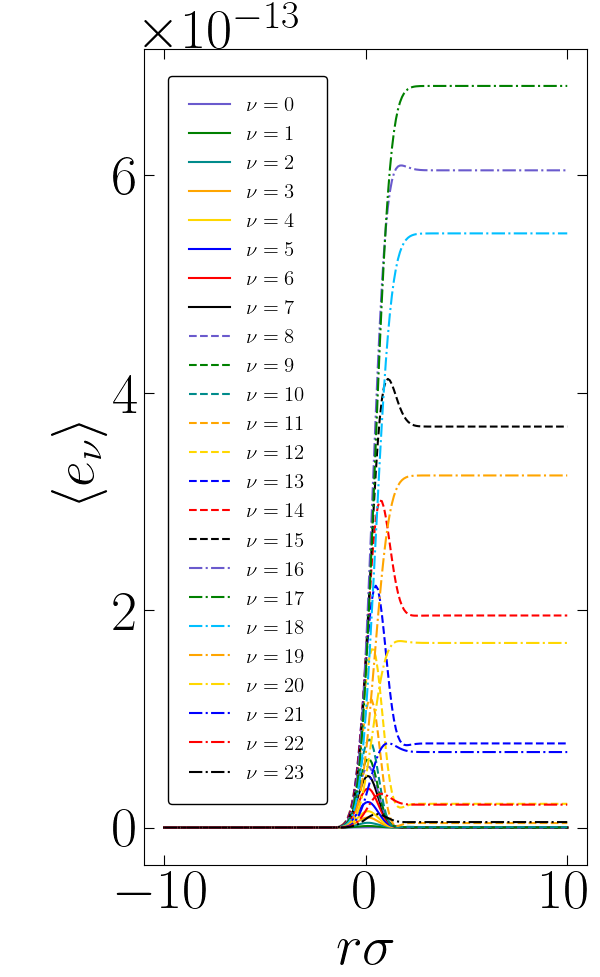}
    \end{adjustbox}\\      
    
  \end{tabular}
 
  \caption{Population probability of the first 23 excited levels for: a) Uncorrelated photons; b) $\sigma _{s} = \sigma$; and c) $\sigma _{s} = 0.5\sigma$. In the first column we depict the corresponding JSA. The second column corresponds to the populations calculated numerically .The third column corresponds to the populations calculated by PT.}
  \label{fig_results_1a}

\end{figure}   

\begin{figure}[H]
    \centering
      \begin{tabular}{@{}ccc@{}}
     \begin{adjustbox}{valign=c}
      \includegraphics[width=.32\textwidth]{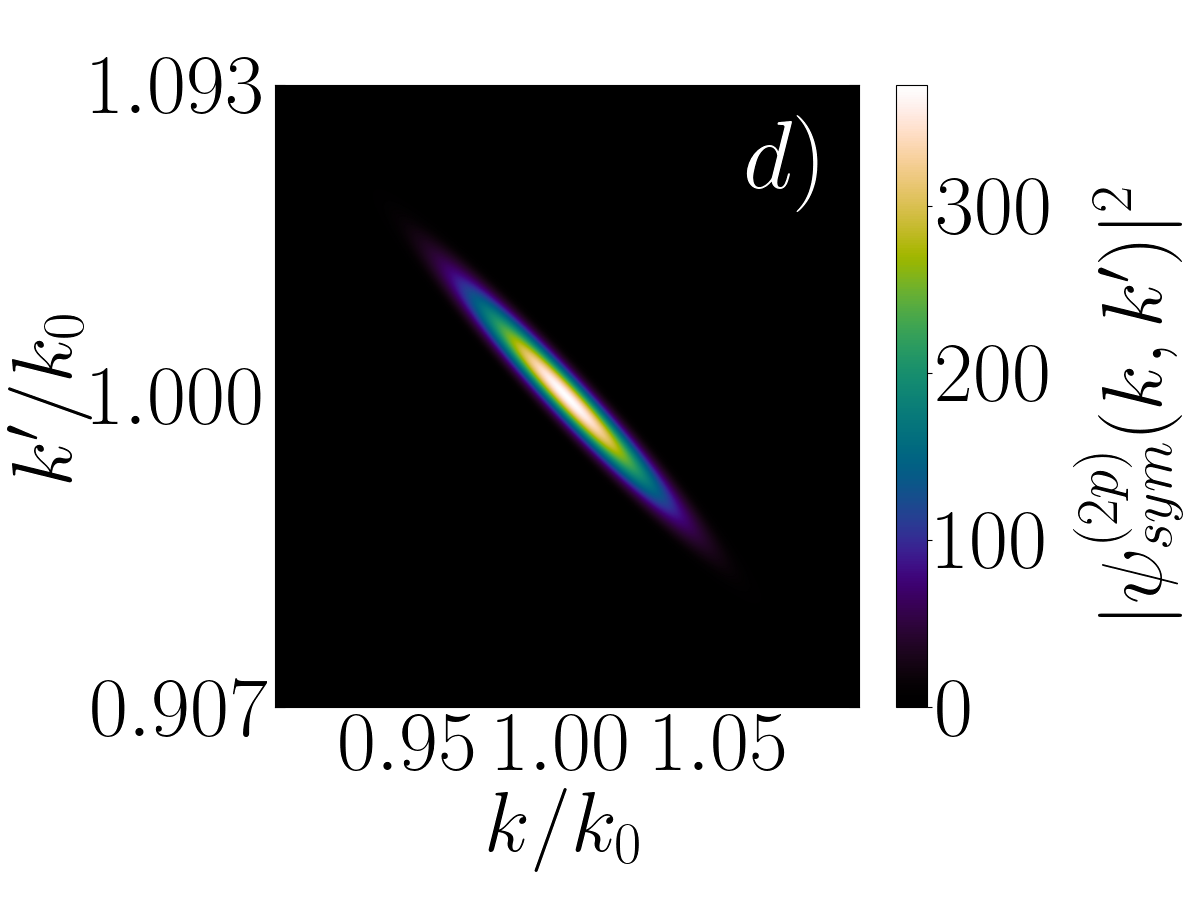}
    \end{adjustbox} &
    \begin{adjustbox}{valign=c}
      \includegraphics[width=.27\textwidth]{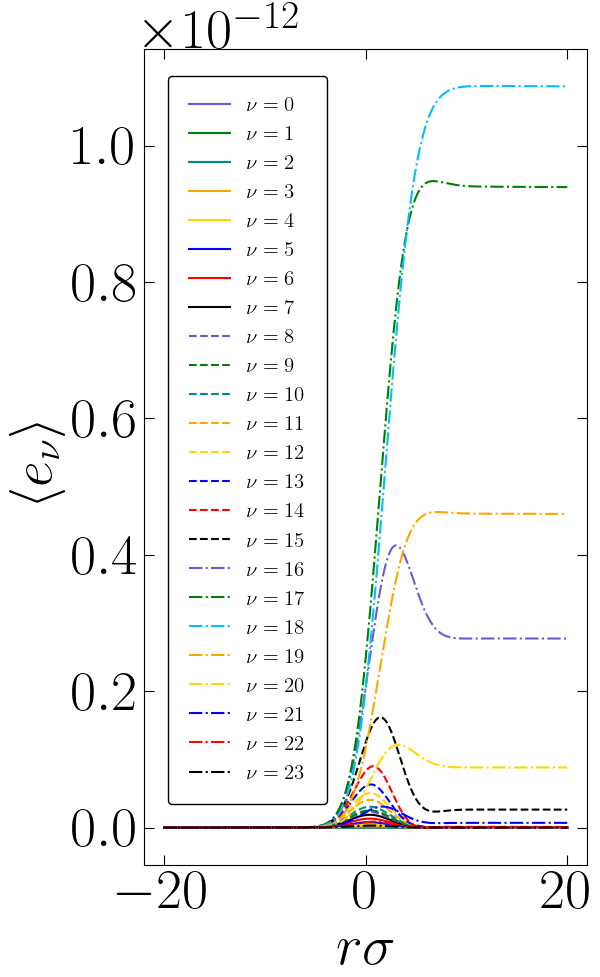}
    \end{adjustbox} &
    \begin{adjustbox}{valign=c}
      \includegraphics[width=.27\textwidth]{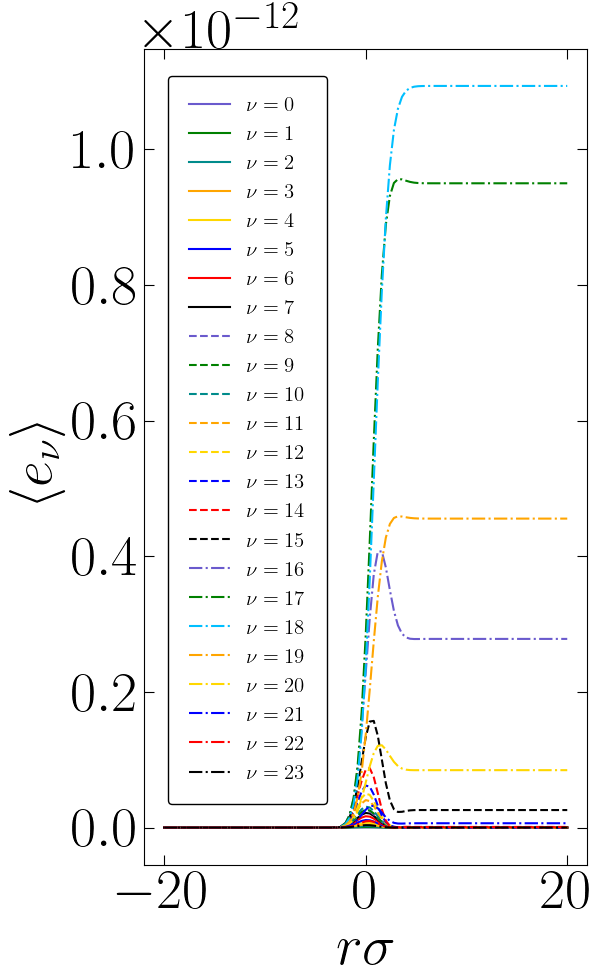}
    \end{adjustbox} \\
    \begin{adjustbox}{valign=c}
      \includegraphics[width=.32\textwidth]{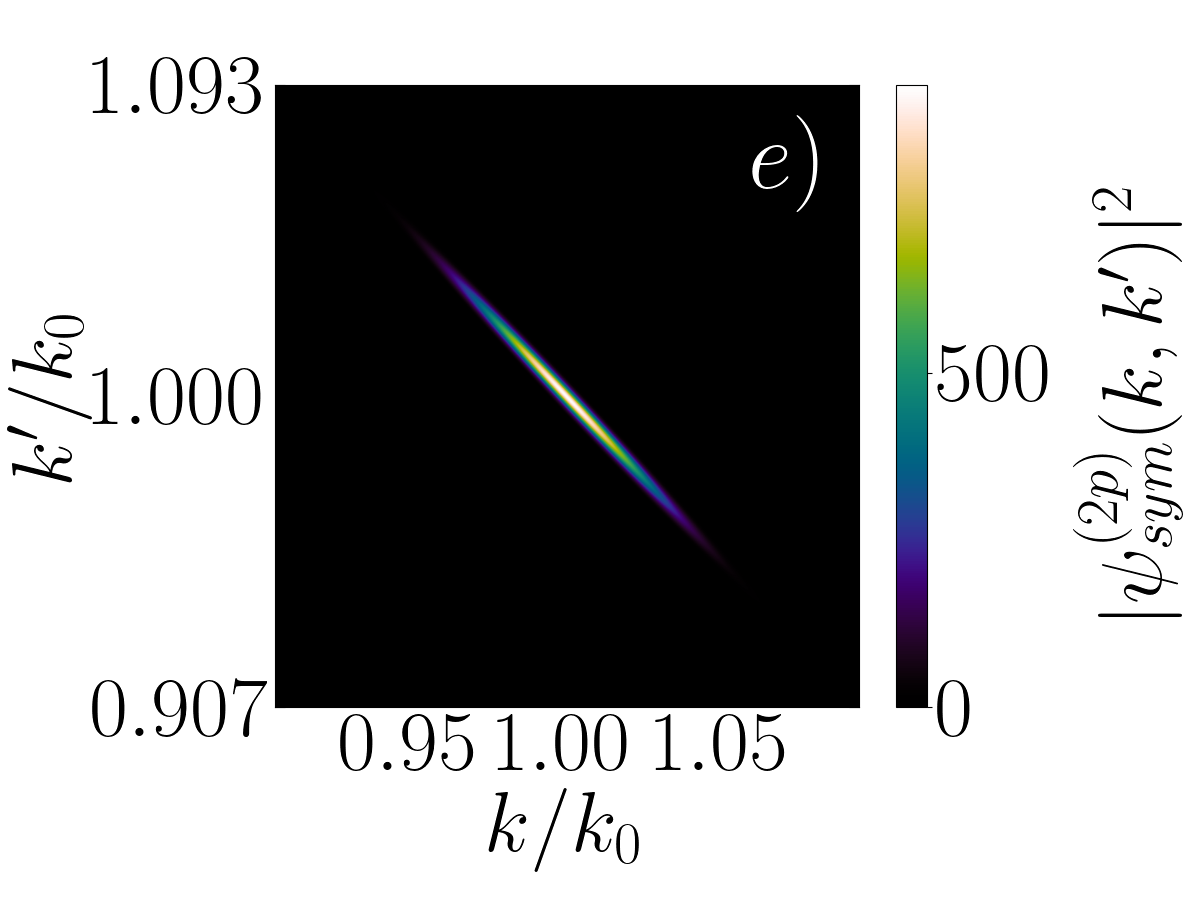}
    \end{adjustbox} &
    \begin{adjustbox}{valign=c}
      \includegraphics[width=.27\textwidth]{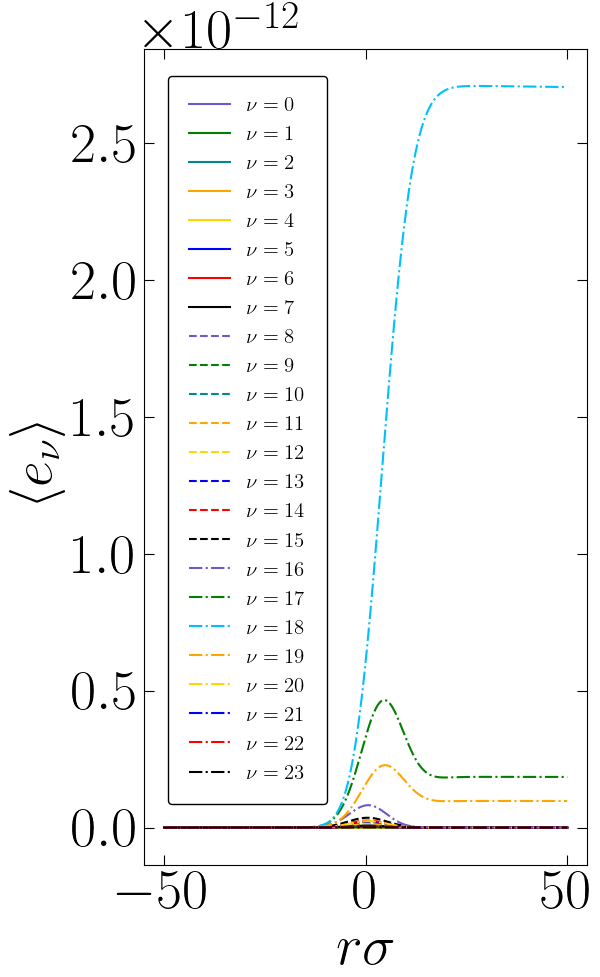}
    \end{adjustbox} &
    \begin{adjustbox}{valign=c}
      \includegraphics[width=.27\textwidth]{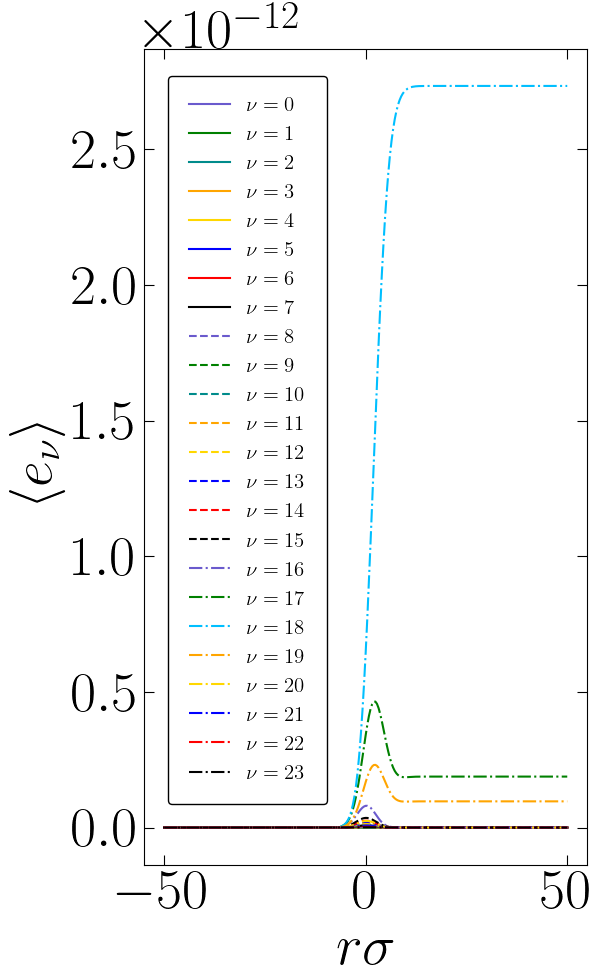}
    \end{adjustbox} \\
    \begin{adjustbox}{valign=c}
      \includegraphics[width=.32\textwidth]{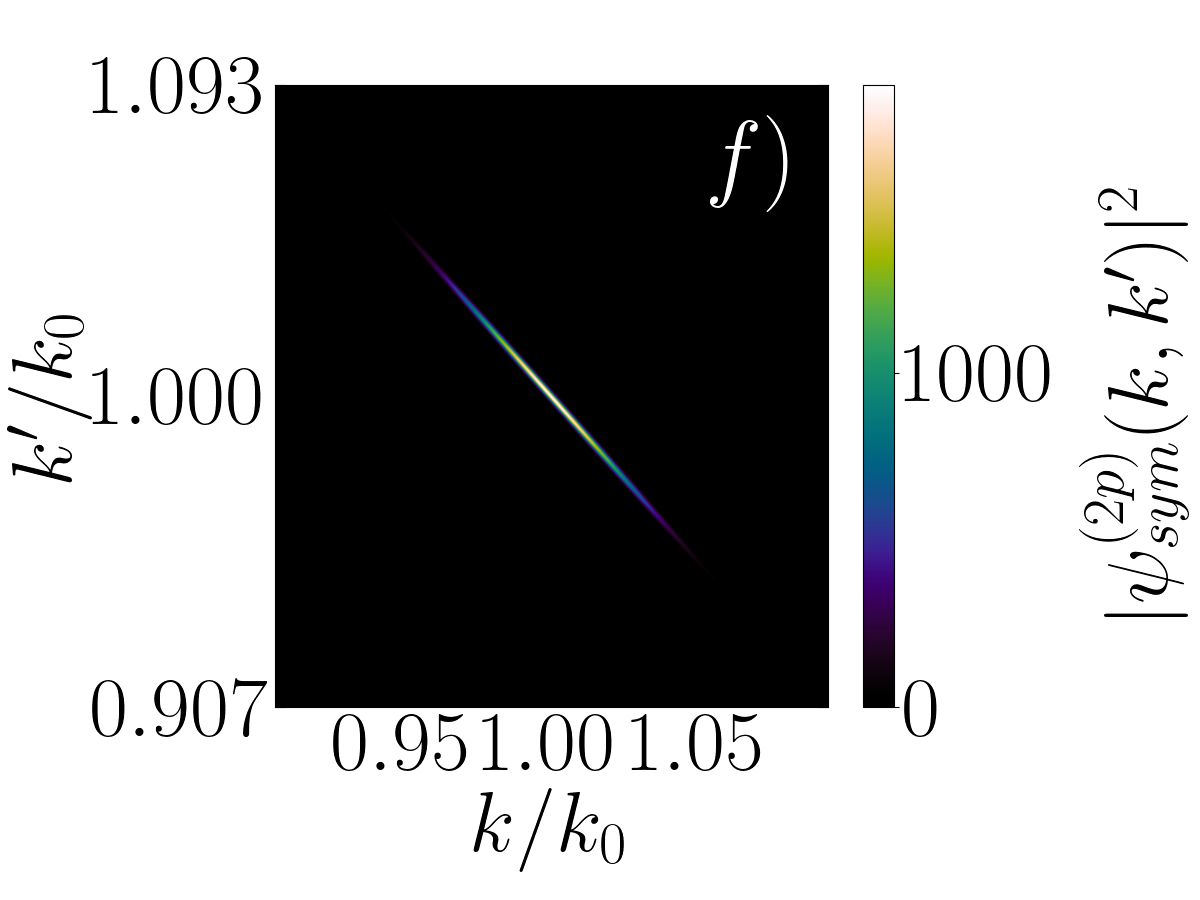}
    \end{adjustbox} &
    \begin{adjustbox}{valign=c}
      \includegraphics[width=.27\textwidth]{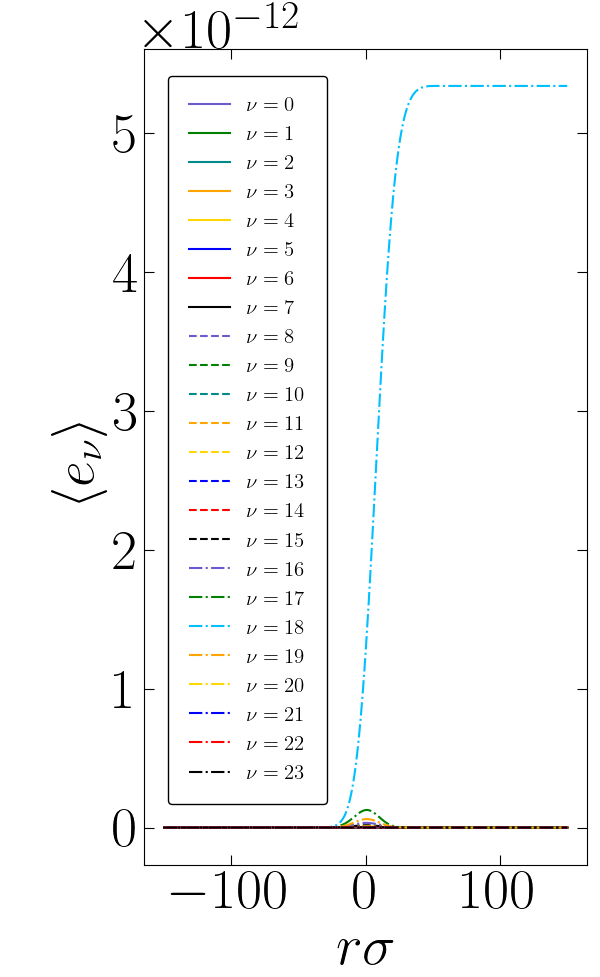}
    \end{adjustbox} &  
    \begin{adjustbox}{valign=c}
      \includegraphics[width=.27\textwidth]{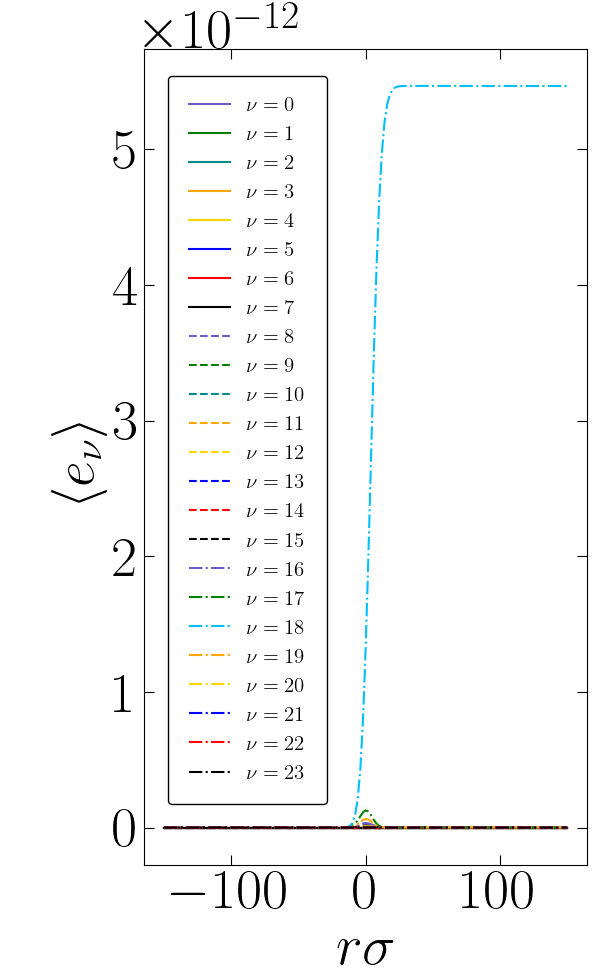}
    \end{adjustbox} \\

  \end{tabular}
 
  \caption{Continued from Fig.~\ref{fig_results_1a}. Population probability of the first 23 excited levels for: d) $\sigma _{s} = 0.25\sigma$; e) $\sigma _{s} = 0.1\sigma$; and f) $\sigma _{s} = 0.05\sigma$. In the first column we depict the corresponding JSA. The second column corresponds to the populations calculated numerically .The third column corresponds to the populations calculated by PT.}
  \label{fig_results_1b}

\end{figure}

\twocolumngrid

\onecolumngrid

\begin{figure}[t]
    \centering
      \begin{tabular}{@{}cc@{}}
    \begin{adjustbox}{valign=c}
    \includegraphics[width=.34\textwidth]{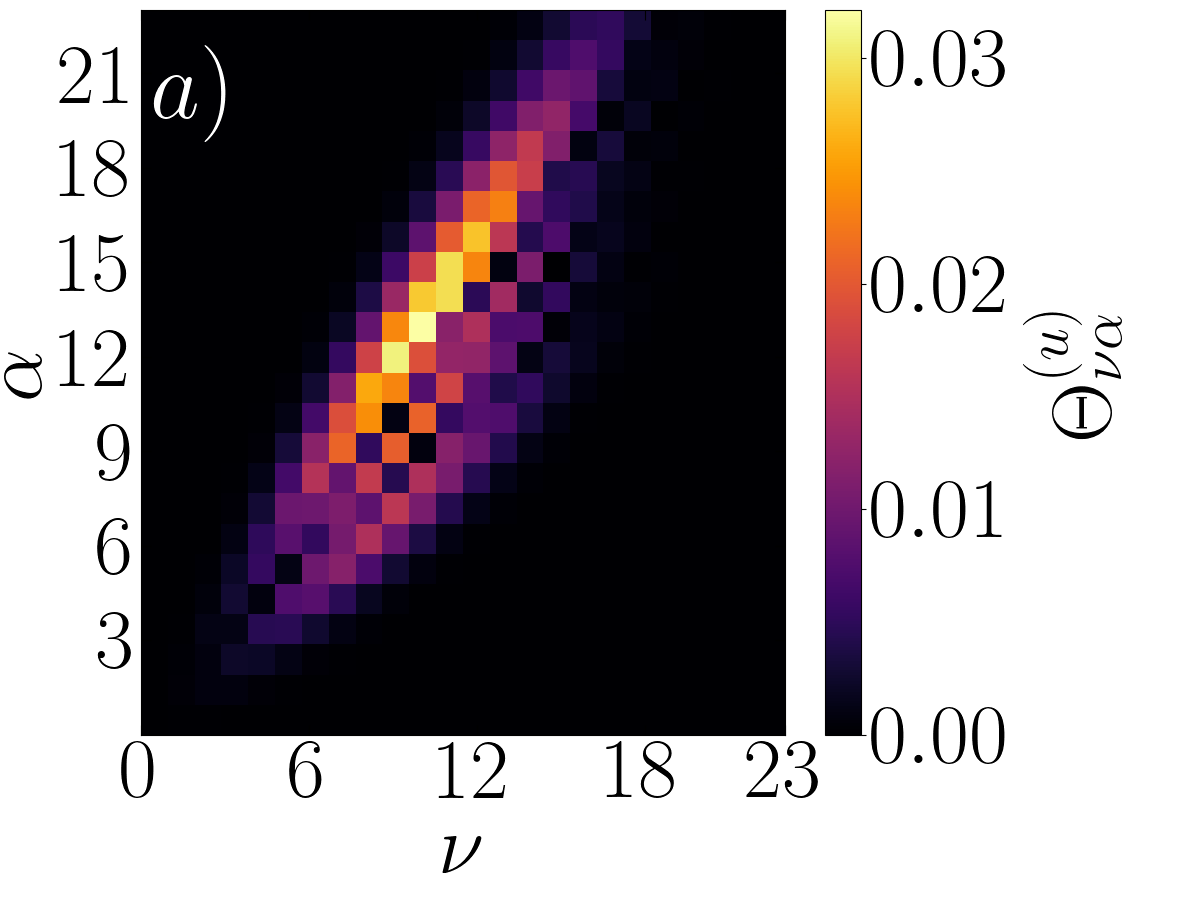}
    \end{adjustbox}&    
    \begin{adjustbox}{valign=c}
    \includegraphics[width=.34\textwidth]{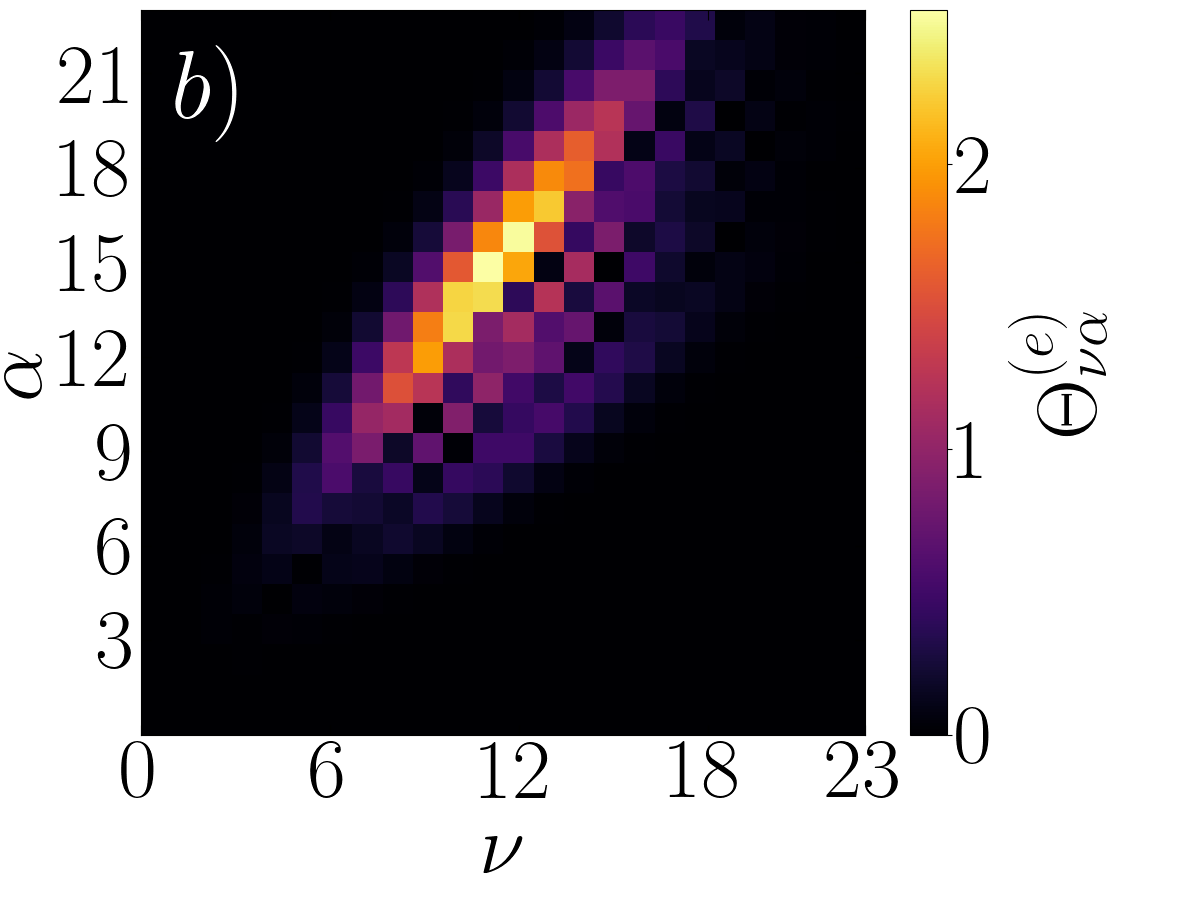}
    \end{adjustbox}\\      
    \begin{adjustbox}{valign=c}
    \includegraphics[width=.34\textwidth]{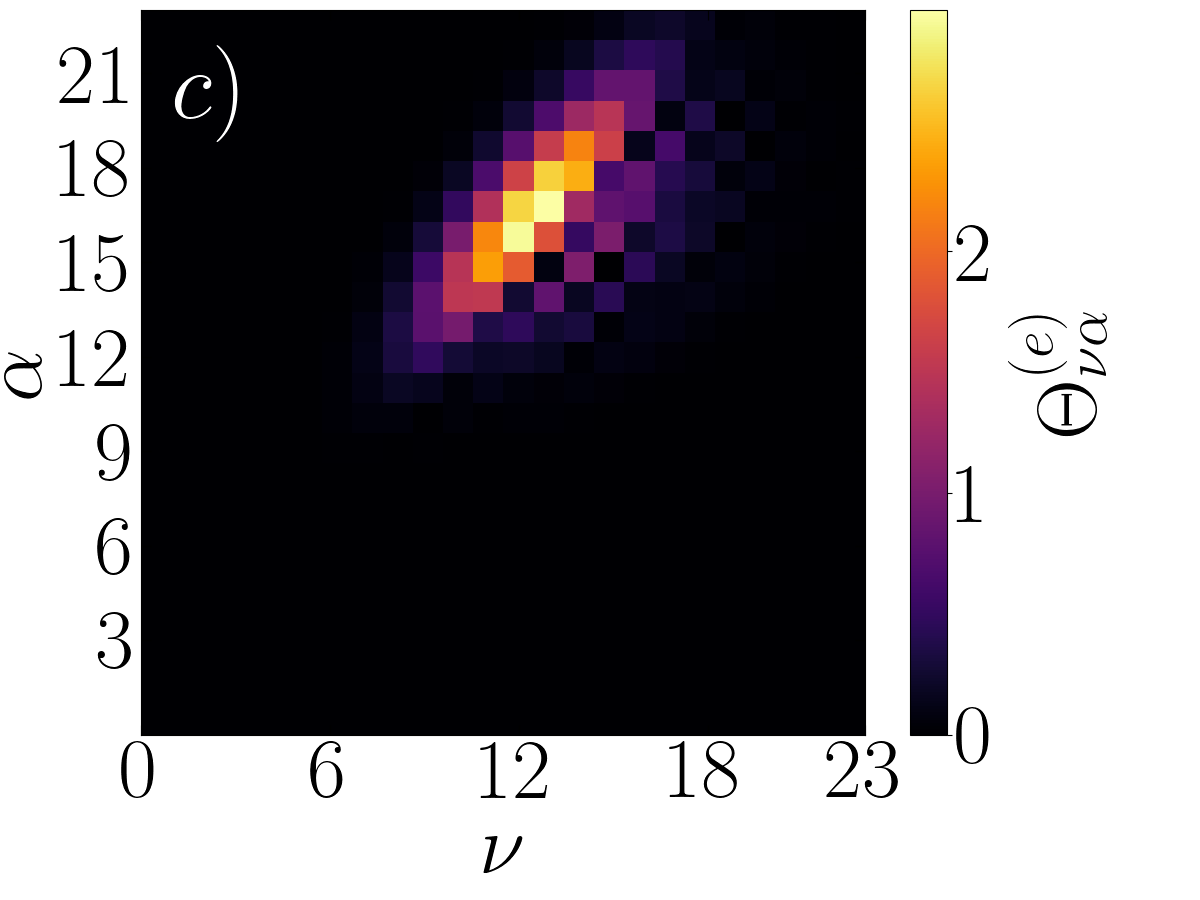}
    \end{adjustbox}&    
    \begin{adjustbox}{valign=c}
    \includegraphics[width=.34\textwidth]{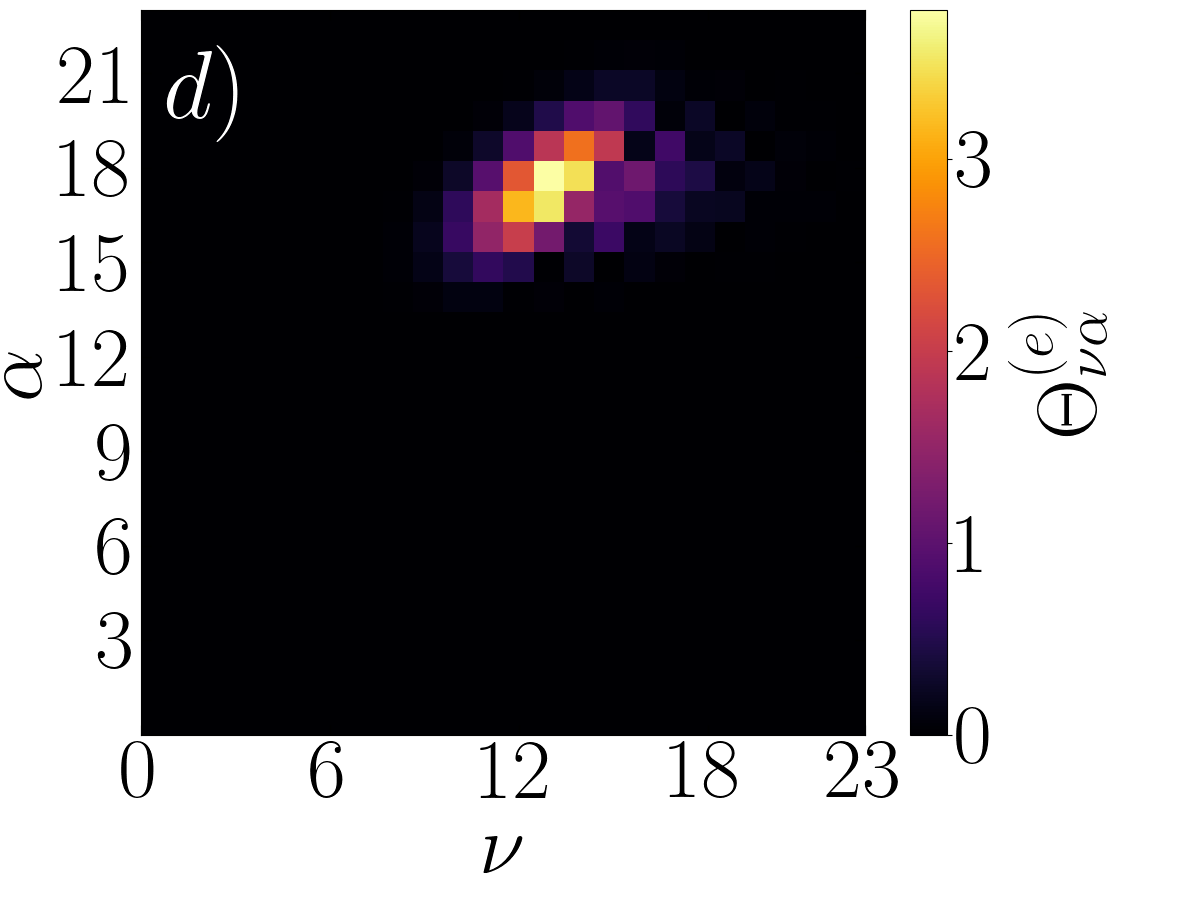}
    \end{adjustbox}\\ 
  \end{tabular}
 
  \caption{Behaviour the transition matrices $\Theta _{\mu \alpha}^{(u, e)}$ as a function of the correlation degree. $a)$ The transition matrix $\Theta _{\mu \alpha}^{(u)}$. The transition matrix $\Theta _{\mu \alpha}^{(e)}$ with: $b)$ $\sigma_{s} = \sigma$, $c)$ $\sigma_{s} = 0.5\sigma$, and $d)$ $\sigma_{s} = 0.25\sigma$. Compare with the maximum values achieved in Figs. \ref{fig_results_1a}  and \ref{fig_results_1b}.}
  \label{fig_fcsigmas_1}

\end{figure} 

\twocolumngrid

\begin{figure}[H]
\centering
  \begin{tabular}{@{}cc@{}}
    \includegraphics[width=.24\textwidth]{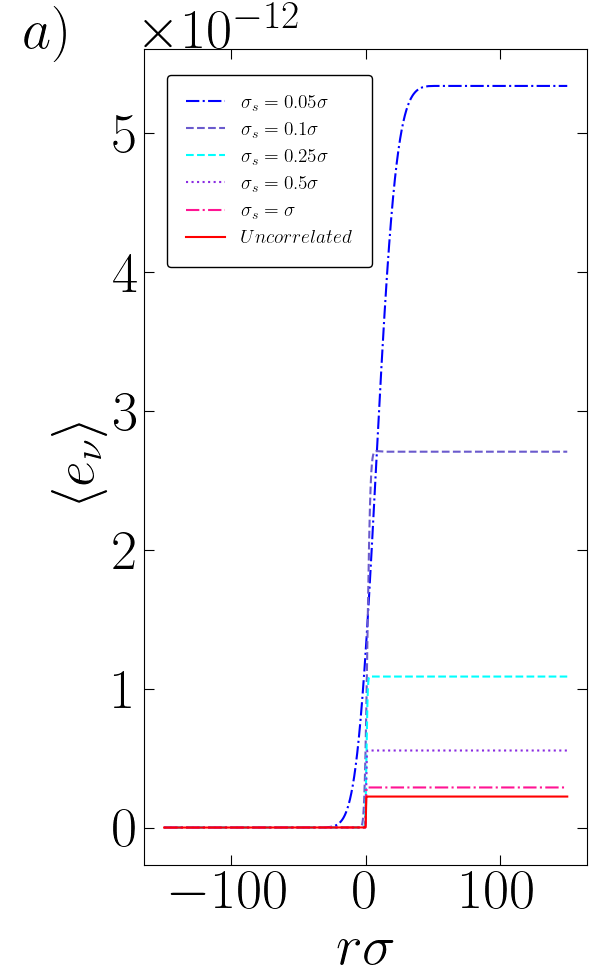} & 
    \includegraphics[width=.24\textwidth]{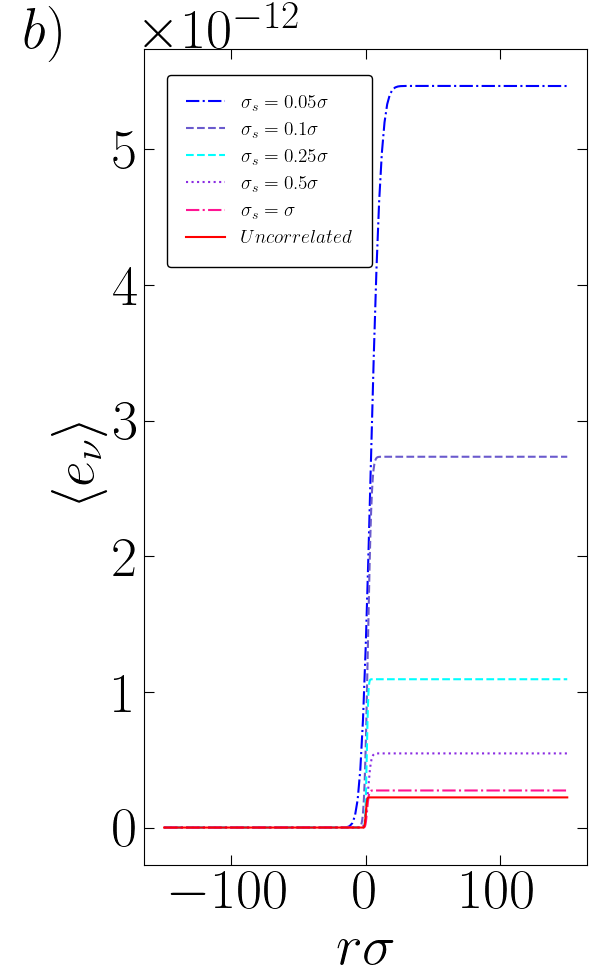} \\    
  \end{tabular}
  \caption{Evolution of the population of targeted excited sub-level with respect to the correlation degree. a) Populations calculated numerically; b) Populations calculated by PT.}
  \label{fig_evolution_sigmas}
\end{figure}

\begin{figure}[h]
	\includegraphics[width=0.47\textwidth]{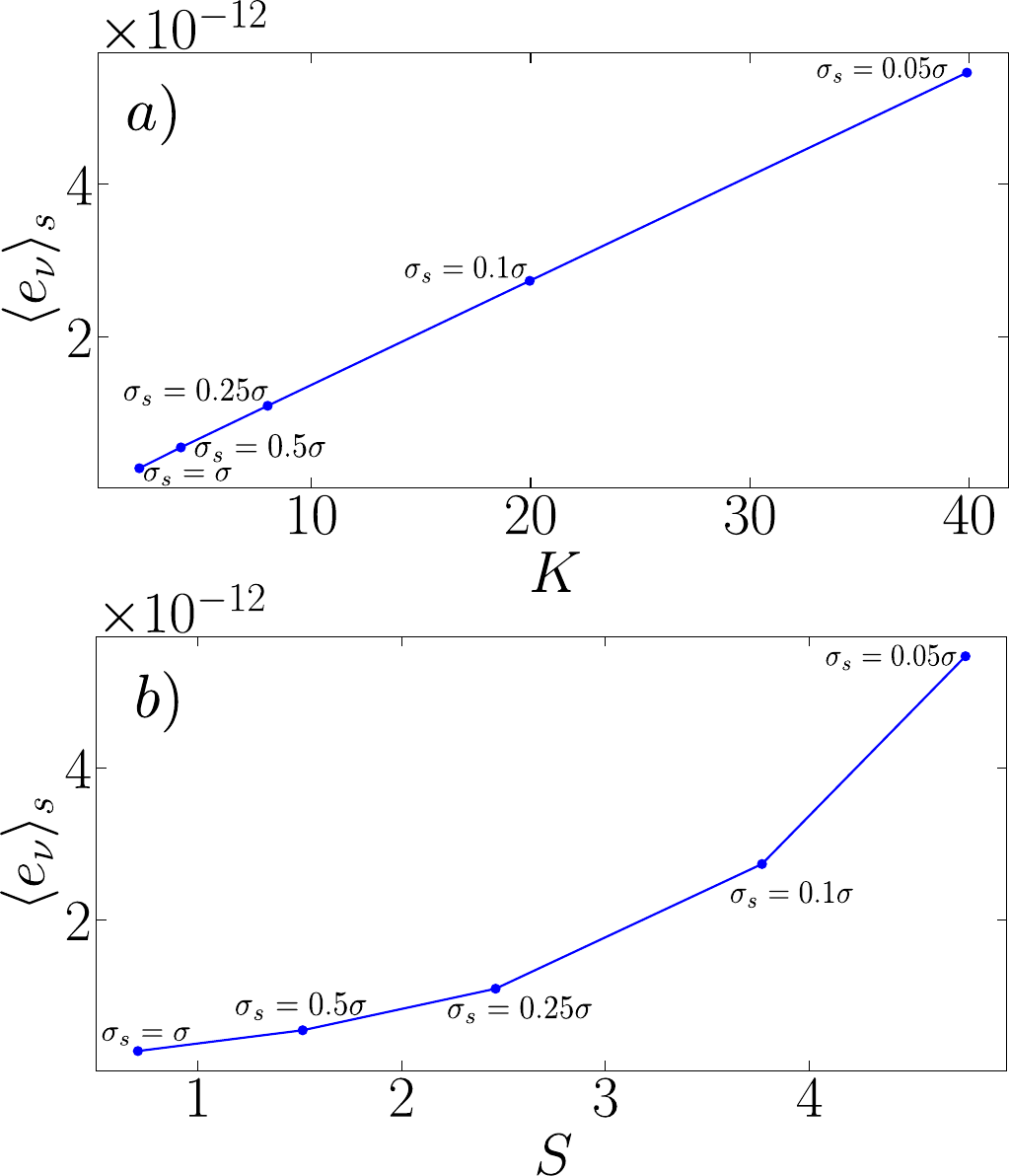}\\
	\caption{$a)$ Steady state value of the population of the targeted excited sub-level as function of the Schmidt number $K$ and  $b)$ as a function of the entanglement entropy $S$. For both plots each point represents the studied correlation degrees, i.e., $\sigma_{s} = \sigma$, $\sigma_{s} = 0.5\sigma$, $\sigma_{s} = 0.25\sigma$, $\sigma_{s} = 0.1\sigma$, and $\sigma_{s} = 0.05\sigma$.}
	\label{en_vs_pop}
 \end{figure}

We remark that the use of our analytical approach drastically reduces the computational effort required by numerical methods. Our analytic method avoids the problem of frequency discretization  that is required in the numerical exact method. In~\cite{PhysRevA.97.063859} the total number of photon modes used was 49014001 (7001 modes per photon) and a mode spacing of 100 GHz. In Appendix A we quantify the computational effort in each case, showing that even when such the frequency discretization considers an appropriate number of modes depending on $\sigma _{s}$, our analytical method implies a reduction in memory and time consumption of approximately three and two orders of magnitude, respectively (see Figs. 10 and 11 and associated discussion in Appendix A).

\begin{figure}[h]
	\includegraphics[width=0.47\textwidth]{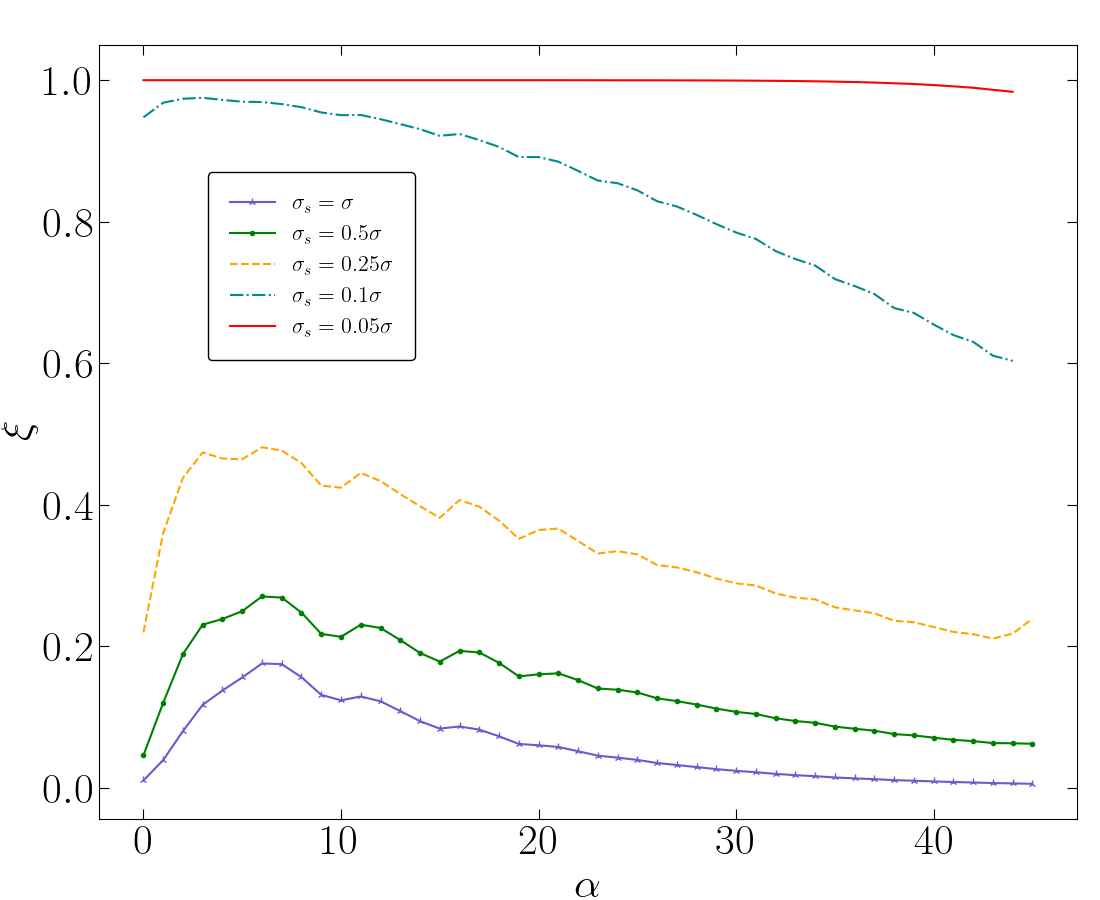}\\
	\caption{Selectivity for each energy level in resonance with the central frequency of the photon field.}
	\label{res_selec}
 \end{figure}

Taking advantage of the low computational effort of our analytical method, we now analyze the selectivity dynamics for different resonance scenarios. That is, we change the central frequency of the entangled photons to be resonant with each of the first 45 vibrational excited levels, and evaluate the increase in the targeted population to study the dependence of the selectivity on the energy of the chosen level. Specifically, in Fig.~\ref{res_selec} we evaluate the selectivity $\xi$ defined as
\begin{equation}
    \xi = \frac{\langle e_{\nu = \text{Targeted level}} \rangle}{\sum _{\nu}\langle e_{\nu}\rangle} .
\end{equation}

We observe behavior resembling a Poisson distribution at low degrees of correlation i.e. $\sigma_{s}>0.1 \sigma$, exhibiting a maximum for the excited vibrational level $\alpha = 7$. This shape is similar to the  distribution of the Franck-Condon factors $F_{\nu}$ (Fig.~\ref{figsys11} $f)$). We attribute the fluctuations observed in each curve in Fig.~\ref{res_selec} to numerical errors in computing the populations for different $k_{0}$ and $\sigma _{s}$. For a higher degree of correlation i.e. $\sigma _{s}\leq 0.1\sigma$, we see a high degree of selectivity for all $\alpha$ such that for $\sigma _{s} = 0.05\sigma$ all vibrational levels have equal selectivity regardless of the strength of the Franck-Condon transitions. Thus, for this molecular system, if we want to obtain a rapid enhancement of the population as the the degree of correlation is increased, we must concentrate the efforts in obtaining entangled photons with central frequency near to the energy of the lowest excited vibrational levels i.e. $5<\alpha <10$. In other words, if we have experimental access only to an intermediate degree of photon entanglement, we can still take advantage of the vibrational structure of the molecule to achieve high selectivity of specific excited states.

\begin{figure*}
\centering
  \begin{tabular}{@{}ccc@{}}
    \includegraphics[width=.27\textwidth]{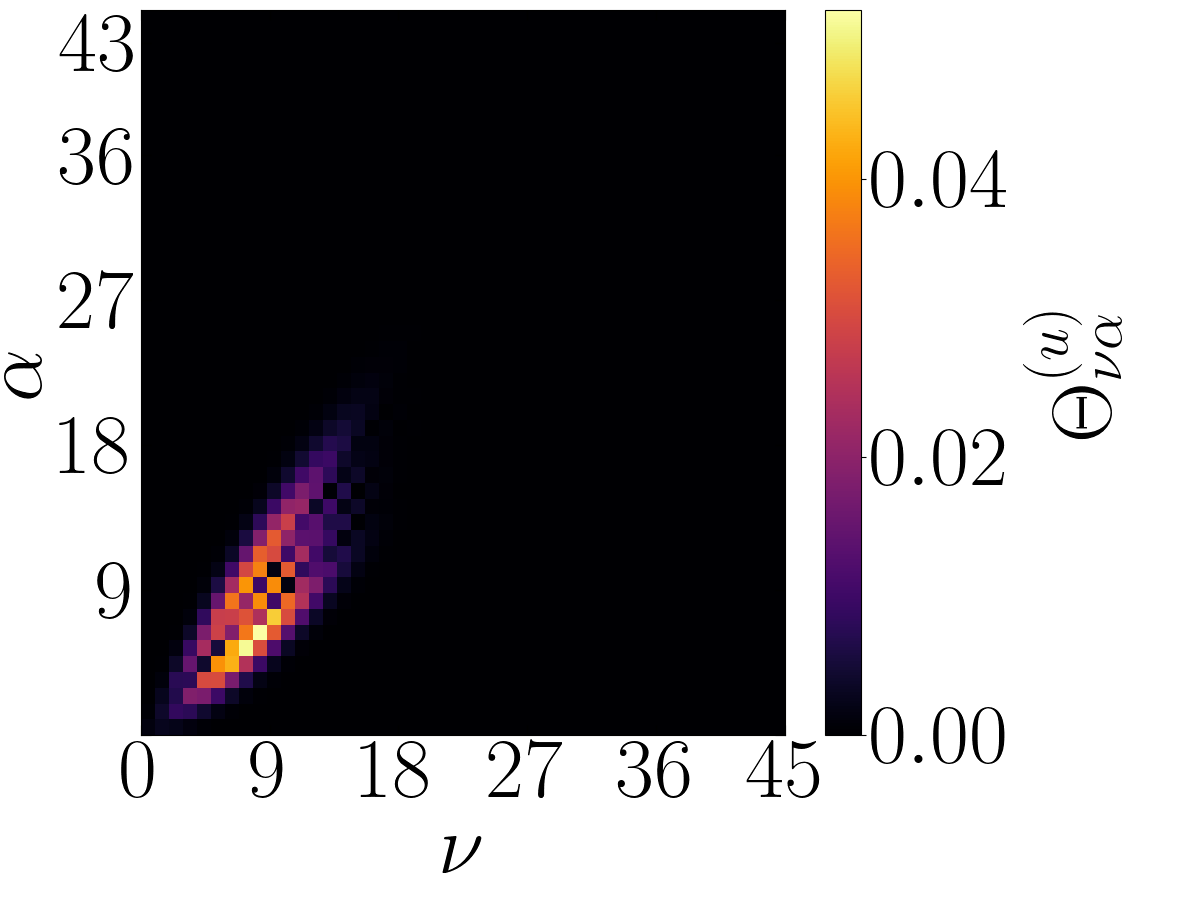} & 
    \includegraphics[width=.27\textwidth]{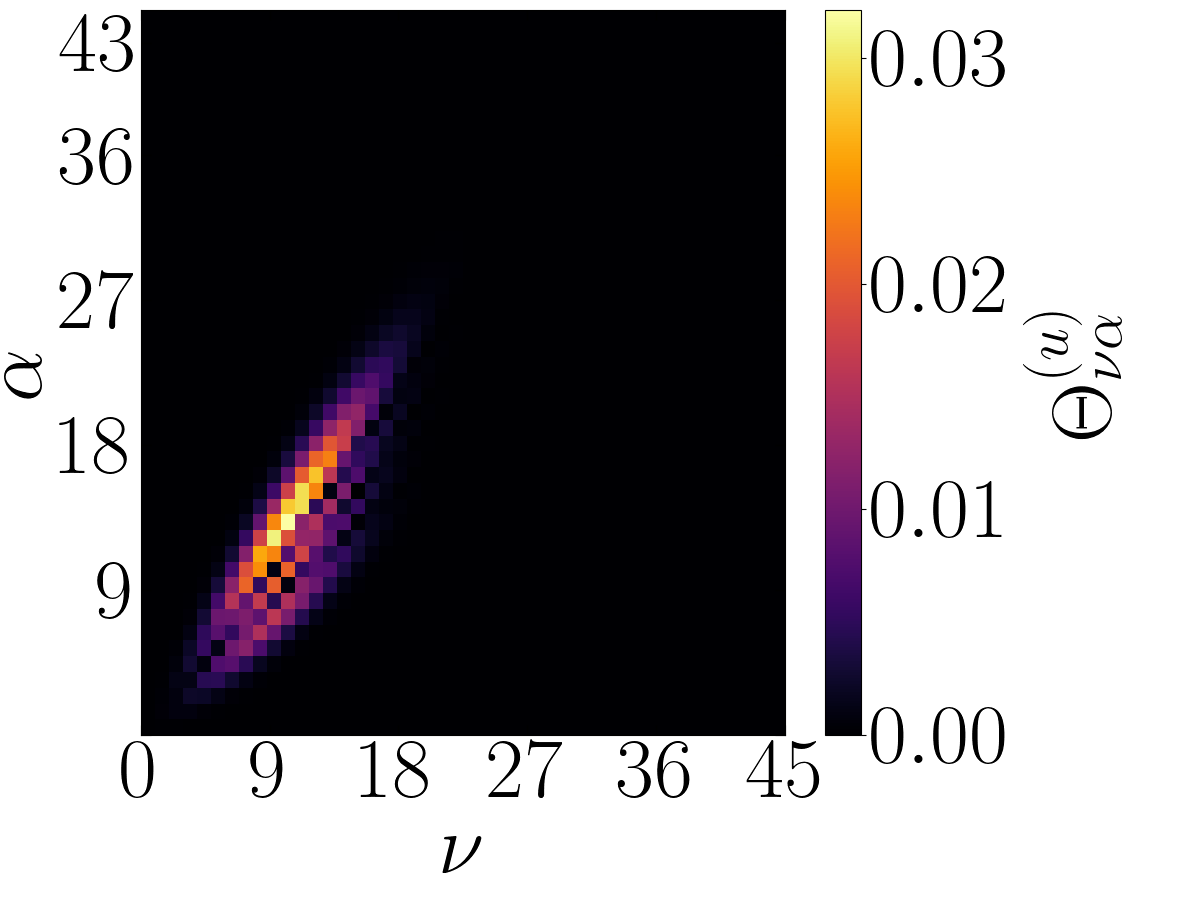} & 
    \includegraphics[width=.27\textwidth]{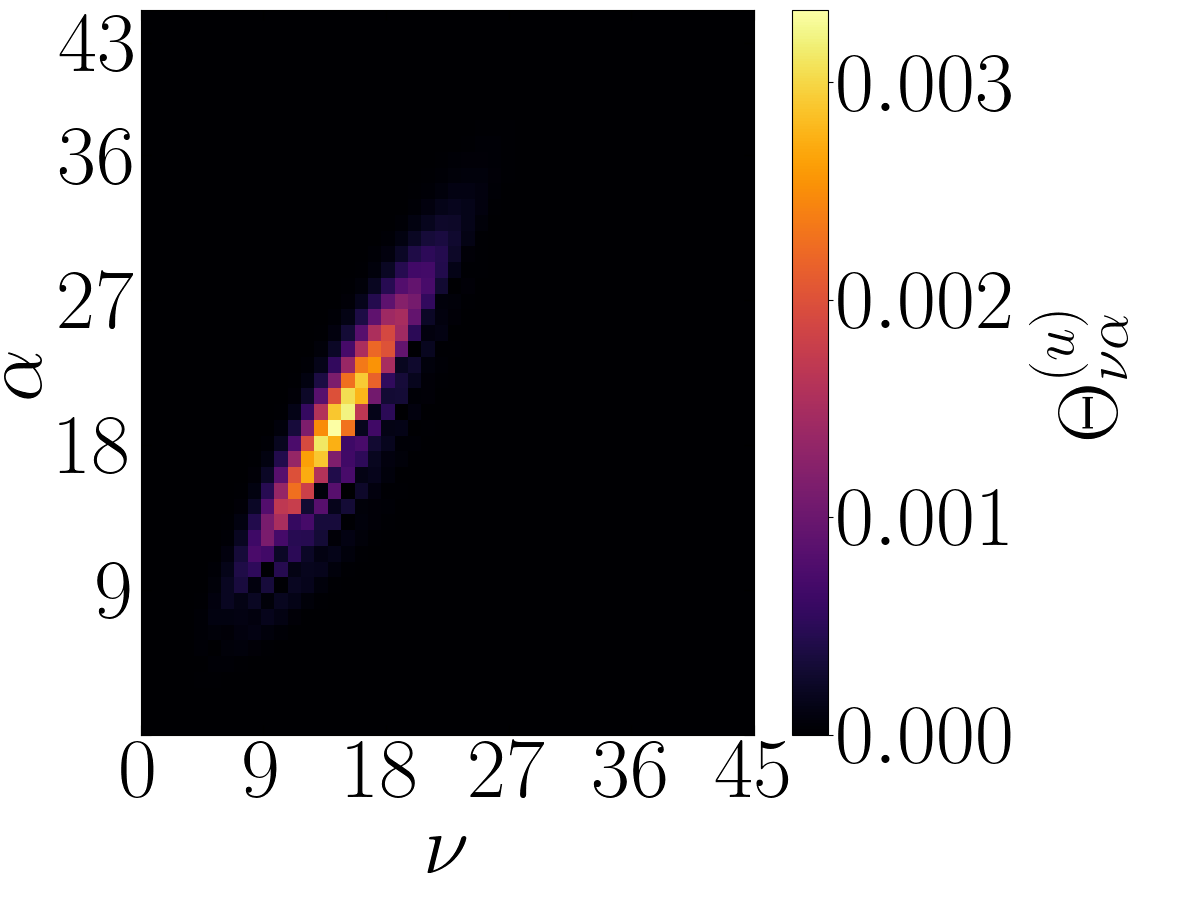}\\
    \includegraphics[width=.27\textwidth]{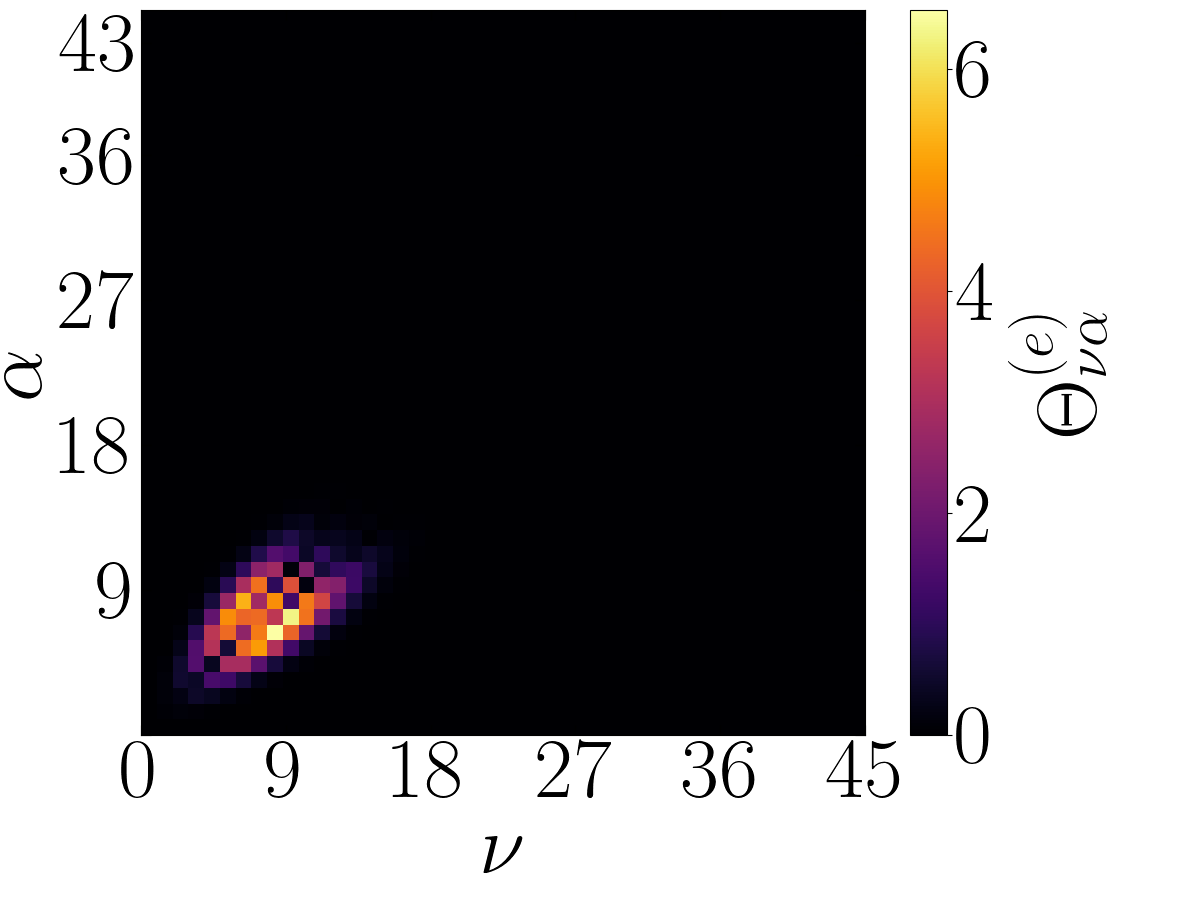} & 
    \includegraphics[width=.27\textwidth]{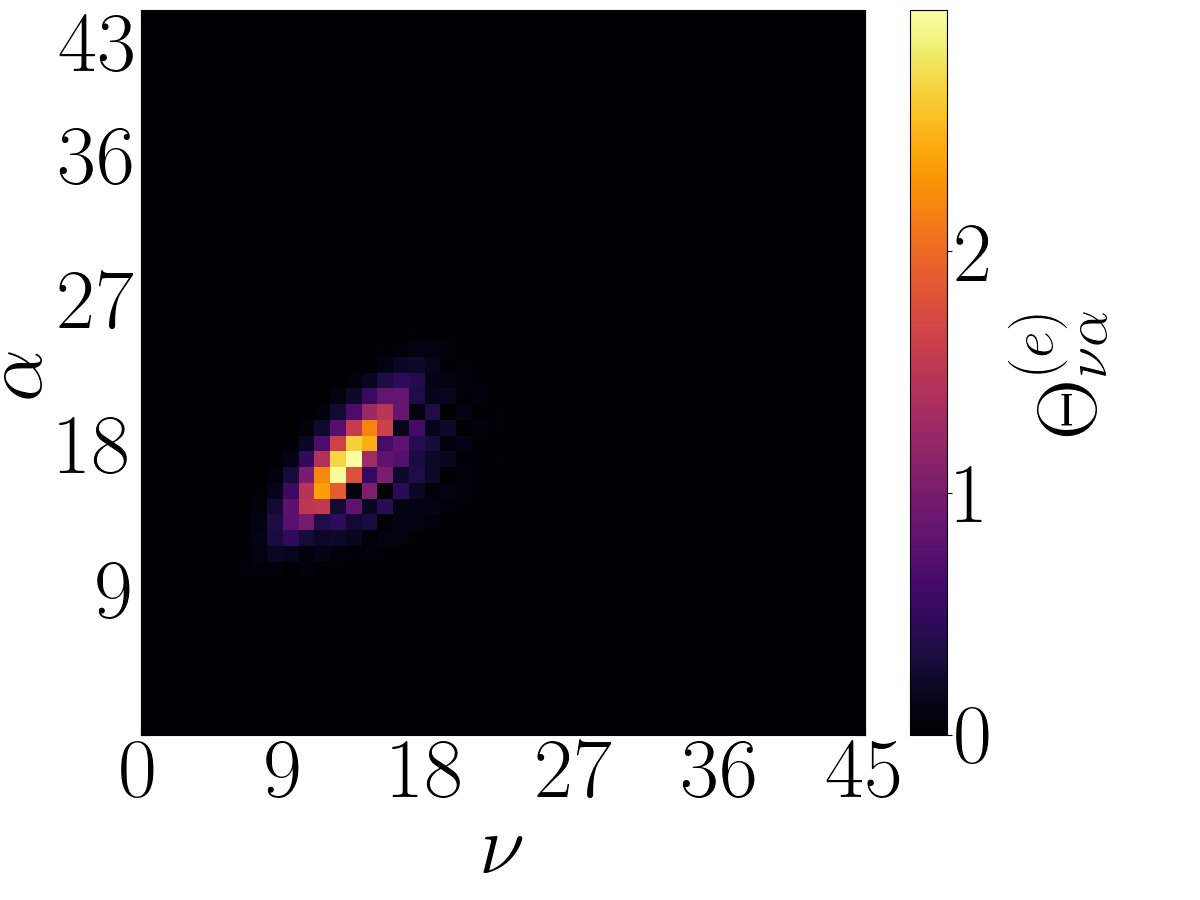} & 
    \includegraphics[width=.27\textwidth]{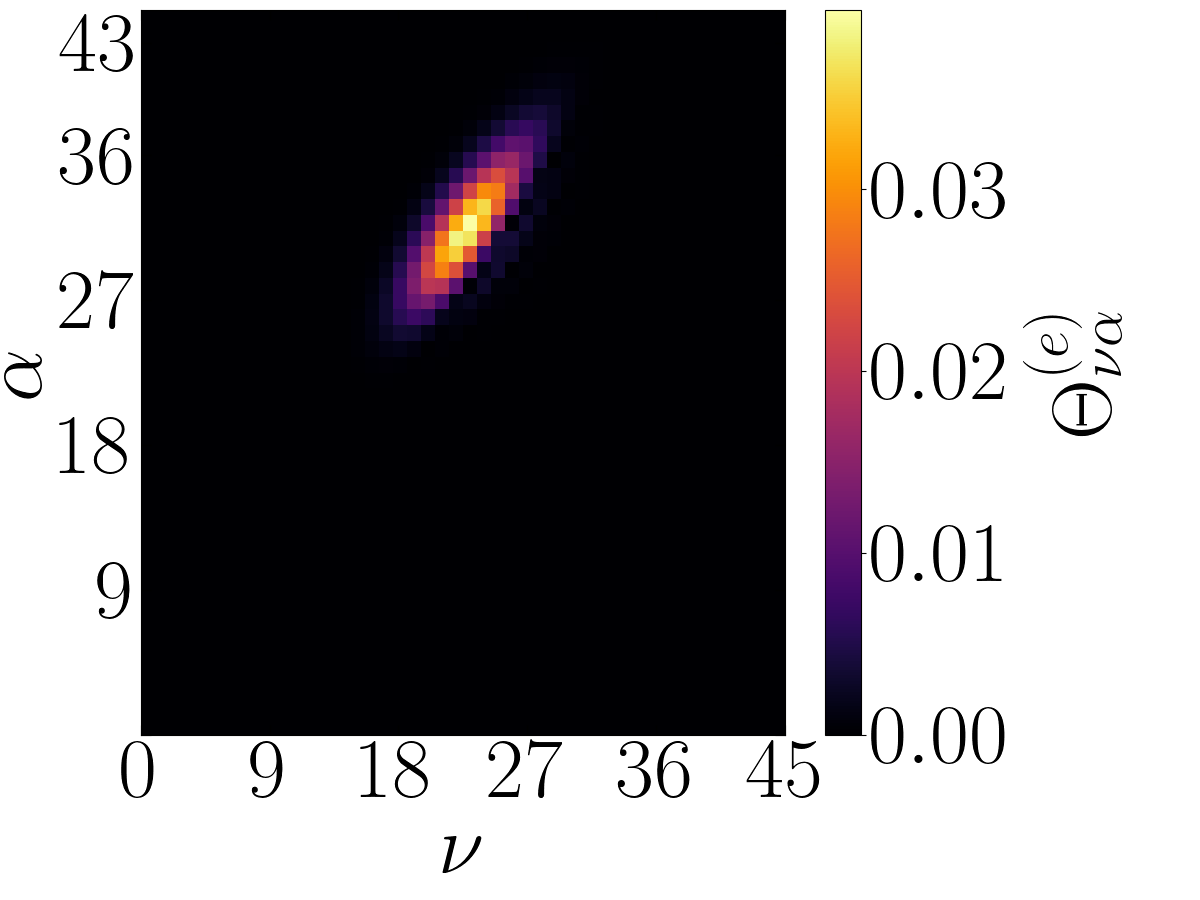}\\
    \includegraphics[width=.27\textwidth]{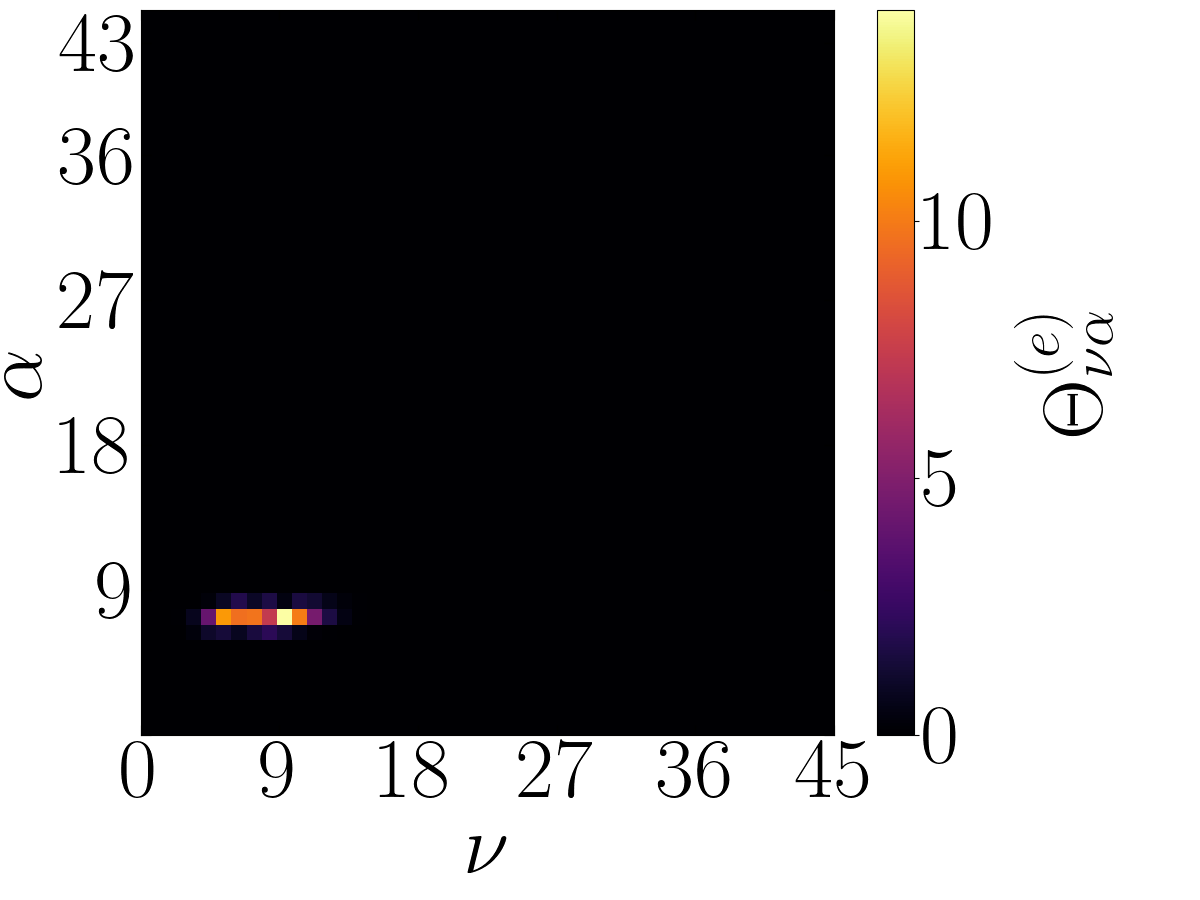} & 
    \includegraphics[width=.27\textwidth]{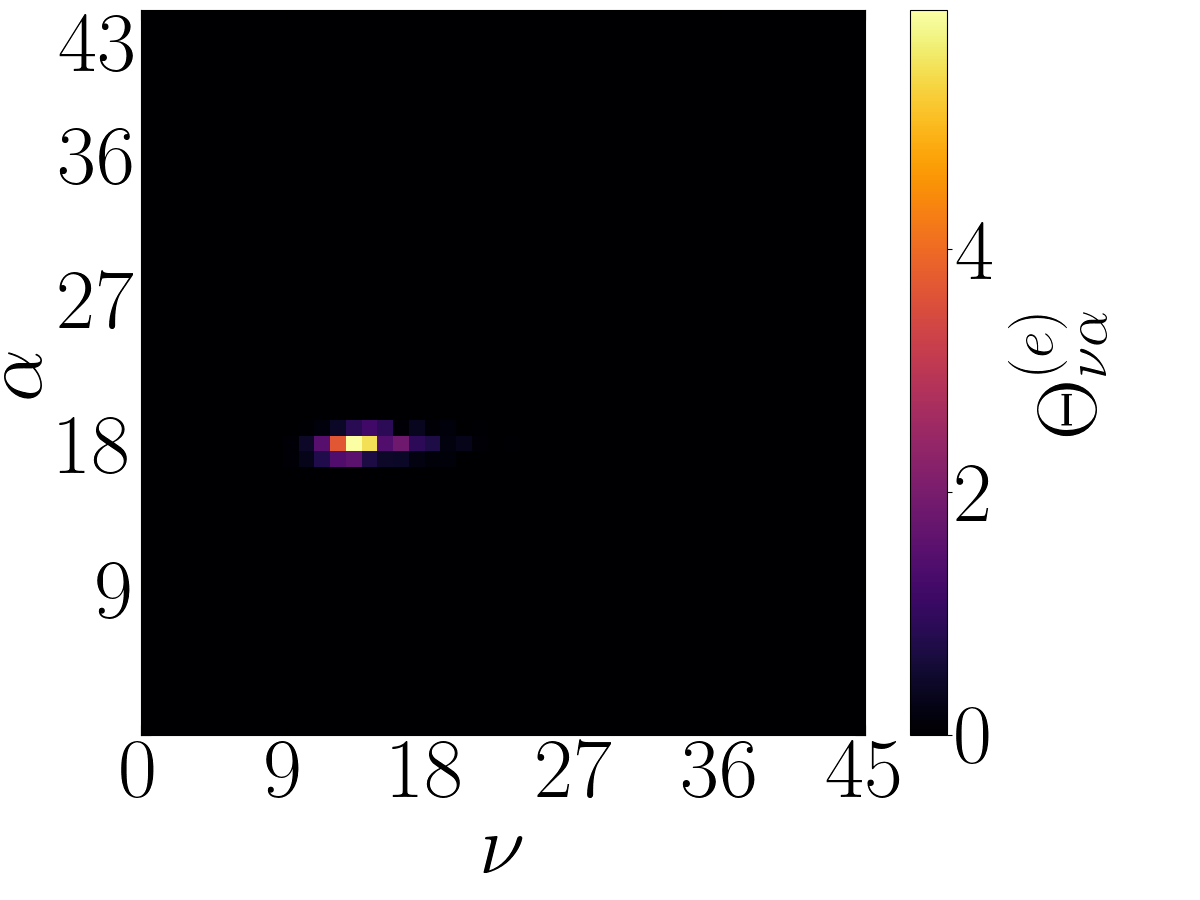} & 
    \includegraphics[width=.27\textwidth]{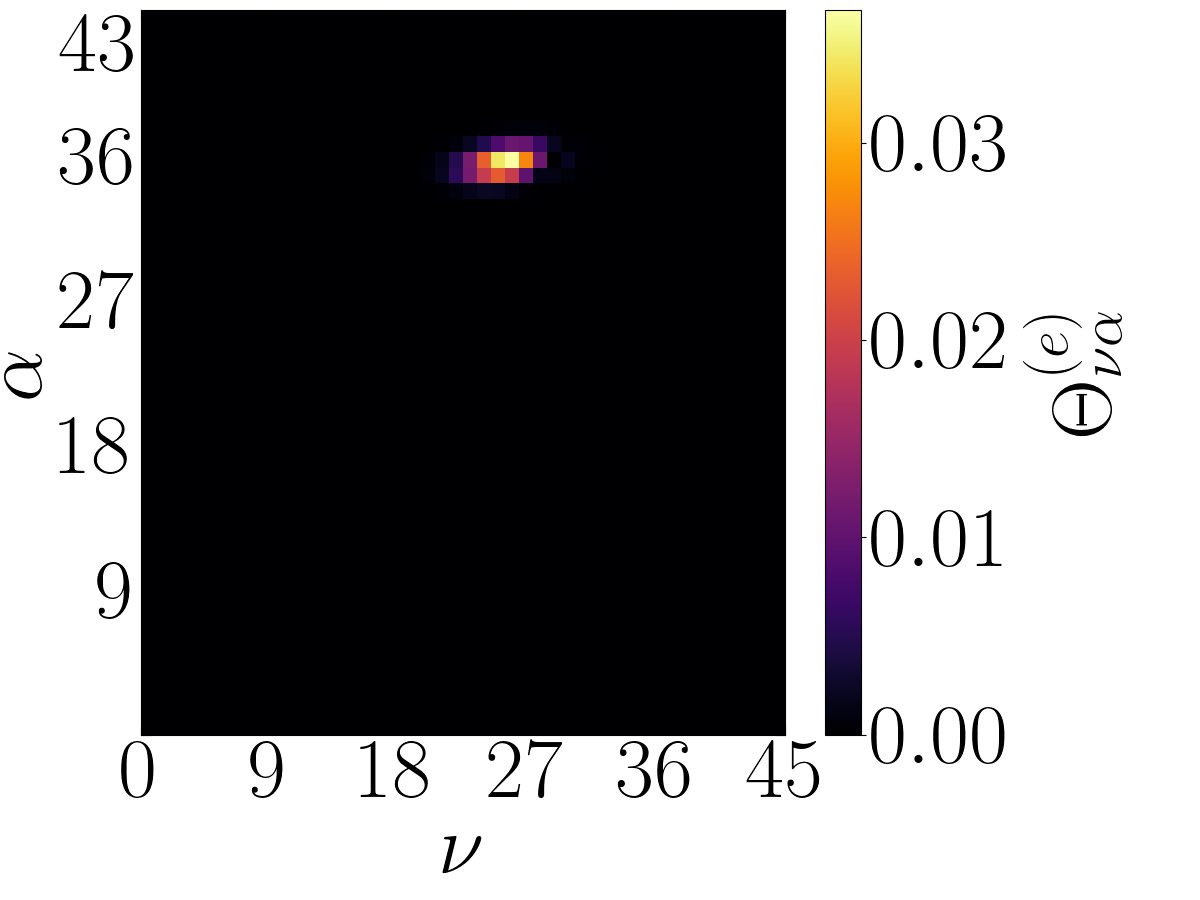}\\

  \end{tabular}
  \caption{Behaviour of (\ref{tran_matrix1}) when the central energy of the entangled photons is resonant with the excited vibrational level $\alpha = 7$, $2k_{0}\approx 3.8831$ eV, (first column), $\alpha = 18$ (second column), $2k_{0}\approx 4.0024$ eV, and $\alpha = 36$, $2k_{0}\approx 4.1614$ eV, (third column). The first row corresponds to the uncorrelated case, i.e, $\Theta _{\alpha \nu}^{(u)}$. The second and third row are the correlated case with $\sigma _{s} = 0.5\sigma$, and $\sigma _{s}=0.1\sigma$, respectively.}
  \label{fig_res_dif_sigmas}
\end{figure*}

To understand the influence of the vibrational structure on the behavior observed in Fig.~\ref{res_selec}, we study the behavior of the matrix elements of Eq. (\ref{tran_matrix1}) for different resonance scenarios.  In Fig.~\ref{fig_res_dif_sigmas} we show the case where the central energy of the entangled photons is resonant with the excited vibrational level $\alpha = 7$ (first column), $\alpha = 18$ (second column), and $\alpha = 36$ (third column). This corresponds to values of $2k_{0}\approx 3.8831$ eV, $2k_{0}\approx 4.0024$ eV, and $2k_{0}\approx 4.1614$ eV, respectively. The first row corresponds to the uncorrelated case, i.e, $\Theta _{\alpha \nu}^{(u)}$. The second and third row are the correlated case with $\sigma _{s} = 0.5\sigma$, and $\sigma _{s}=0.1\sigma$, respectively. We can observe that for $\alpha = 7$, the Gaussian functions are allowing the participation of more Franck-Condon factors product $F_{\nu} F_{\nu \alpha}$, than in the case of $\alpha = 18$, and $\alpha = 36$. For $\alpha = 36$ this participation is substantially reduced even in the uncorrelated case, where we can evidence that only a fraction of the upper branch of the Franck-Condon parabolas is contributing to the transition. Notice that for $\sigma _{s} = 0.1\sigma$ (high correlation degree), there is more dispersion for high energy vibrational levels, than observed for low levels. Thus, low energy vibrational levels are more reliable to perform tasks which require rapid enhancement of selectivity. 

This analysis shows that the transition matrix, Eq. (\ref{tran_matrix1}), could help to construct an optimal experimental set up to exploit vibrational selectivity due to entangled photons.

\section{Summary and Outlook}
\label{conclusions}

The discrepancies between the experimental results and theoretical simulations of the ETPA remain an active area of research. Most theoretical works use second-order PT within simplified models  and approximations (or fourth-order perturbation in the density matrix formalism)~\cite{PhysRevLett.78.1679, PhysRevB.69.165317, mollow1968two, RevModPhys.88.045008, Leon-Montiel_2013, Landes21, 10.1063/5.0049338}. Although these theoretical works have provided qualitative and phenomenological insights into ETPA processes, their numerical predictions often deviate significantly from experimental measurements.  A question then arises regarding the validity of the approximations used in these studies. In this paper, we have examined this question by relaxing typical approximations done when deploying  second-order order PT and  investigate the dynamics of vibronic populations of diatomic system excited by ultrabroadband frequency-entangled photons --a model system that has previously been studied via exact numerical  solutions of the quantum dynamics of the system \cite{PhysRevA.97.063859}.  We show that our efficient analytical framework reproduces the results obtained via numerical solution of the complete Schrödinger equation, while providing additional physical insight. 

Specifically, using second-order PT that keeps: i) the temporal integrals in a finite time ($-t_{0}\leq t < \infty$); ii) the functions to be integrated on the frequency domain without any asymptotic approximations regarding resonance or far from resonance; and iii) the rapid convergent series representation of the transition amplitudes, we obtained the same excited state dynamics as that predicted by solving a system of more than $25\times 10^{6}$ coupled differential equations \cite{oka2011}. These steps allowed us to arrive at analytical expressions that capture the light-matter correlations that affect the TPA processes. 

The structure of our analytical expressions in Eq. (\ref{tranunco1}), and Eq. (\ref{tranen1}) provides an understanding of the nature of the TPA processes depending on the correlations in the light source. First, Eq. (\ref{tranunco1}) shows that CTPA can be seen as a multiplicative single-photon processes of transitions from ground states to intermediate states and from intermediate to excited states. In contrast, ETPA transition (\ref{tranen1}) becomes a weighted average of two differentiated one-photon transitions, modulated by a Gaussian envelope that captures the resonance or off-resonance conditions and the degree of two-photon correlation. These results allow us to understand the key physical differences in vibronic excitation with uncorrelated and quantum-correlated photons. Furthermore, we have extended the analysis presented in Ref.~\cite{PhysRevA.97.063859} investigating the population dynamics and vibronic selectivity for intermediate photon correlations degrees. We showed that a high correlation degree is not always essential for achieving a substantial enhancement of the targeted vibrational state. 

Our analytical expression for the population of the targeted vibrational state reveals a factor that explains (and potentially quantifies) the quantum enhancement of the population selectivity. Combined with the product of the Franck-Condon factors $F_{\nu}F_{\nu \alpha}$, which we refer to as the transition matrix (defined in Eq. \ref{tran_matrix1}), this factor provides insight into the maximum achievable population with each correlation degree. Additionally, this matrix reveals how the correlation degree of the entangled photons shape the participation of the Franck-Condon factors in the transitions. Therefore, the structure defined in Eq. (\ref{tran_matrix1}) may assist in designing optimal experimental setups to leverage the vibrational selectivity through entangled photons.

We also explored different resonance scenarios to understand the selectivity properties. Taking advantage of the low computational effort required by our approach, we studied simulations of the population for different targeted levels as a function of the number of entangled modes in the biphoton state, and the molecular structure. We show that it is possible to obtain high selectivity with a low degree of correlation, if the targeted vibrational level has associated high values of the Franck-Condon factors transitions. In our particular case, we can expect high selectivity from a reduced entanglement degree if the targeted level is between levels $\alpha = 7$ and $\alpha = 10$. We observed that for low photon correlation degrees, the functional form of the selectivity is related with the Franck-Condon factors $F_{\mu}$ structure.

Our results are consistent with those in Ref.~\cite{varnavski2023colors}. Accounting for a non-simplified molecular electronic structure has yielded insights into experimentally observed phenomena, such as a red-shift of peaks between CTPA and ETPA. The analytical framework we derived supports assumptions that suggest that the quantum correlations of molecular states interact with the quantum correlations of entangled photons. These interactions constitute a foundational basis for understanding this phenomenon. Regarding this aspect, we obtained a structure that describes the involvement of molecular levels correlations (encoded in the Franck-Condon factors) and their contributions in ETPA dynamics.

We have examined ultrabroad-band frequency entangled photons which avoid usual labeling of the entangled photons generated by SPDC as idler and signal photons. An extension of this work is to implement other kinds of JSA's containing information on pump time, entanglement time and entanglement area, which are parameters that can be accessed and modified in experimental setups. 

It would also be interesting to extend the framework developed here to investigate the relevance of the vibrational structures for ETPA of widely investigated molecular systems such as  ZnTPP and rhodamine B (RhB), but also for the quantum-light activation of relevant photosensitive proteins  (e.g., light-oxygen-voltage (LOV) domains~\cite{zayner2014factors}) that are engineered as fluorescent labels or as  optogenetic tools~\cite{lindner2022optogenetics, losi2018blue}.  Investigating the activation of such  photosensors  with quantum states of light may provide insight into the possible advantages of ETPA for targeted control in biology. 

In our work, the system considered has a vibrational structure that is modeled by  a Morse potential with the vibrational transitions described within the Franck–Condon approximation. However, a variety of molecules of chemical and biological significance have a vibrational structure that includes  Herzberg–Teller couplings. A possible interesting extension of out work is  to investigate the effect of Herzberg–Teller coupling in  ETPA~\cite{doi:10.1021/acs.jpcb.2c00846, doi:10.1021/acs.jpclett.2c01963, doi:10.1021/acs.jpca.0c07896}.  Finally, our framework can be adapted to investigate the advantages of plasmonic structures  as a promising avenue to improve signal acquisition in ETPA measurements ~\cite{oka2015highly, oka2017generation, oka2013generation, rusak2019enhancement, IzadshenasJahromi:24, izadshenas2024molecular}.  These topics are the subject of our future studies.

 \section*{Acknowledgments}
We would like to thank Chawntell Kulkarni, Sougato Bose, Andrew Fisher, Robert Thew, Karolina Słowik, and Anita Dabrowska for helpful discussions. We gratefully acknowledge funding from the Engineering and Physical Sciences Research Council (EPSRC) (Grants No. EP/R513143/1 and No. EP/T517793/1) and the Gordon and Betty Moore Foundation (Grant 8820). 

\section*{DATA AVAILABILITY}

All datasets and the code developed to generate them are available from \href{https://github.com/christiancv11/Perturbation_Theory_ETPA}{Github: Perturbation Theory scope for predicting vibronic selectivity by entangled two photon absorption}.

\appendix
\section{Discretized Schrodinger equations}\label{app:disc_schro_eq}

In this section, we derive the discretized form of the Schr\"odinger equations and matrix definitions for subsequent computational implementation. To solve the equations (\ref{sch1}), (\ref{sch2}), and (\ref{sch3}), we discretize the photon fields by converting from $\int dk$ to $\sum _{k} \delta k$, and $\hat{a}(k)$ to $\hat{a}_{k}$. The discretized equations have the following form
\begin{eqnarray}
\frac{d}{dt} \psi ^{(2p)}_{kk'}(t)=&-&i(k+k')\psi ^{(2p)}_{kk'}(t)
\nonumber \\
&-&i\sum _{\nu}\sqrt{\frac{\gamma F_{\nu}}{2\pi}}\left[ \psi ^{(1pm)}_{k\nu}(t) + \psi ^{(1pm)}_{k'\nu}(t) \right]
\nonumber \\
\end{eqnarray}

\begin{eqnarray}
\frac{d}{dt} \psi ^{(1pm)}_{k\nu}(t) = &-&i(k+\omega _{m_{\nu}})\psi ^{(1pm)}_{k\nu}(t) 
\nonumber \\
&-&i \sqrt{\frac{2\gamma F_{\nu}}{\pi}}\sum _{k'}\delta k\,\psi ^{(2p)}_{kk'}(t) 
\nonumber \\
&-&i\sum _{\nu '}\sqrt{\frac{\gamma F_{\nu \nu '}}{\pi}}\psi ^{(e)}_{\nu '}(t)
\end{eqnarray}

\begin{equation}
\frac{d}{dt}\psi ^{(e)}_{\nu '}(t) = -i\omega _{e_{\nu}} \psi ^{(e)}_{\nu '}(t)-i \sum _{\nu \nu'}\sqrt{\frac{\gamma F_{\nu \nu '}}{\pi}} \sum _{k}\delta k\, \psi ^{(1pm)}_{k\nu}(t)
\end{equation}

Defining the following matrices which are containing the functions for each state

\begin{equation}
\psi ^{(2p)} = 
\begin{pmatrix}
\psi ^{(2p)} _{k_{0}k_{0}} & \psi ^{(2p)} _{k_{0}k_{1}} & \cdots & \psi ^{(2p)} _{k_{0}k_{M}} \\
\psi ^{(2p)} _{k_{1}k_{0}} & \psi ^{(2p)} _{k_{1}k_{1}} & \cdots & \psi ^{(2p)} _{k_{1}k_{M}} \\
\vdots & \vdots & \cdots & \vdots \\
\psi ^{(2p)} _{k_{M}k_{0}} & \psi ^{(2p)} _{k_{M}k_{1}} & \cdots & \psi ^{(2p)} _{k_{M}k_{M}} \\
\end{pmatrix}
\end{equation}

\begin{equation}
\psi ^{(1pm)} = 
\begin{pmatrix}
\psi ^{(1pm)} _{k_{0},0} & \psi ^{(1pm)} _{k_{0},1} & \cdots & \psi ^{(1pm)} _{k_{0},N} \\
\psi ^{(1pm)} _{k_{1},0} & \psi ^{(1pm)} _{k_{1},1} & \cdots & \psi ^{(1pm)} _{k_{1},N} \\
\vdots & \vdots & \cdots & \vdots \\
\psi ^{(1pm)} _{k_{M},0} & \psi ^{(1pm)} _{k_{M},1} & \cdots & \psi ^{(1pm)} _{k_{M}, N} \\
\end{pmatrix}
\end{equation}

\begin{equation}
\psi ^{e} = 
\begin{pmatrix}
\psi ^{e} _{0} \\
\psi ^{e} _{1} \\
\vdots \\
\psi ^{e} _{N} \\
\end{pmatrix}
\end{equation}

\begin{equation}
\gamma = 
\begin{pmatrix}
\gamma _{0} \\
\gamma _{1} \\
\vdots \\
\gamma _{N} \\
\end{pmatrix}
\end{equation}

\begin{equation}
\gamma ^{(gm)} = 
\begin{pmatrix}
\gamma _{0} & \gamma _{1} & \cdots & \gamma _{N} \\
\gamma _{0} & \gamma _{1} & \cdots & \gamma _{N} \\
\vdots & \vdots & \cdots & \vdots \\
\gamma _{0} & \gamma _{1} & \cdots & \gamma _{N} \\
\end{pmatrix}
\end{equation}

\begin{equation}
\gamma ^{(me)} = 
\begin{pmatrix}
\gamma _{00} & \gamma _{01} & \cdots & \gamma _{0N} \\
\gamma _{10} & \gamma _{11} & \cdots & \gamma _{1N} \\
\vdots & \vdots & \cdots & \vdots \\
\gamma _{N0} & \gamma _{N1} & \cdots & \gamma _{NN} \\
\end{pmatrix}
\end{equation}

the discretized form of the set of differential equations is given by

\begin{eqnarray}
\frac{d}{dt}\psi ^{(2p)}_{k_{\alpha}k_{\beta}}=&-&i(k_{\alpha}+k_{\beta})\psi ^{(2p)}_{k_{\alpha}k_{\beta}}
\nonumber \\
&-& i (\psi ^{(1pm)}\gamma)_{\alpha} - i (\psi ^{(1pm)}\gamma)_{\beta}
\label{schdeqs1}
\end{eqnarray}

\begin{eqnarray}
 \frac{d}{dt}\psi ^{(1pm)}_{k_{\alpha},j}=&-&i(k_{\alpha}+\omega_{m_{j}}) \psi ^{(1pm)}_{k_{\alpha},j} 
\nonumber \\ 
&-& i (\psi ^{(2p)}\gamma ^{(gm)})_{\alpha j} - i(\gamma ^{(me)}\psi ^{e})_{j} 
\nonumber \\
\label{schdeqs12}
\end{eqnarray}

\begin{equation}
 \frac{d}{dt}\psi ^{e}_{j} = -i \omega _{e _{j}}\psi ^{e}_{j}- i \sum _{n=0}^{N}(\psi ^{(1pm)}\gamma ^{(me)})_{nj} .
\label{schdeqs13}
\end{equation}

The matrix notation has allowed us to significantly improve the computational speed of solving the full set of differential equations. Specifically, it allows performing the numerical calculations with a number of modes that depends of each $\sigma _{s}$ value, without losing any information. For instance, we have used 2001 modes per photon for the uncorrelated case and for $\sigma _{s} = \sigma$ case; 3001 modes per photon for $\sigma _{s}=0.5\sigma$; 5001 modes per photon for $\sigma _{s}=0.25\sigma$; 6001 modes per photon for $\sigma _{s}=0.1\sigma$; and 7001 modes per photon for $\sigma _{s}=0.05\sigma$. This matrix structure also facilitates the use of optimized NumPy functions, enabling faster computations and making it easier to implement various pre-installed solver tools. 

\begin{figure}[h]
	\includegraphics[width=0.47\textwidth]{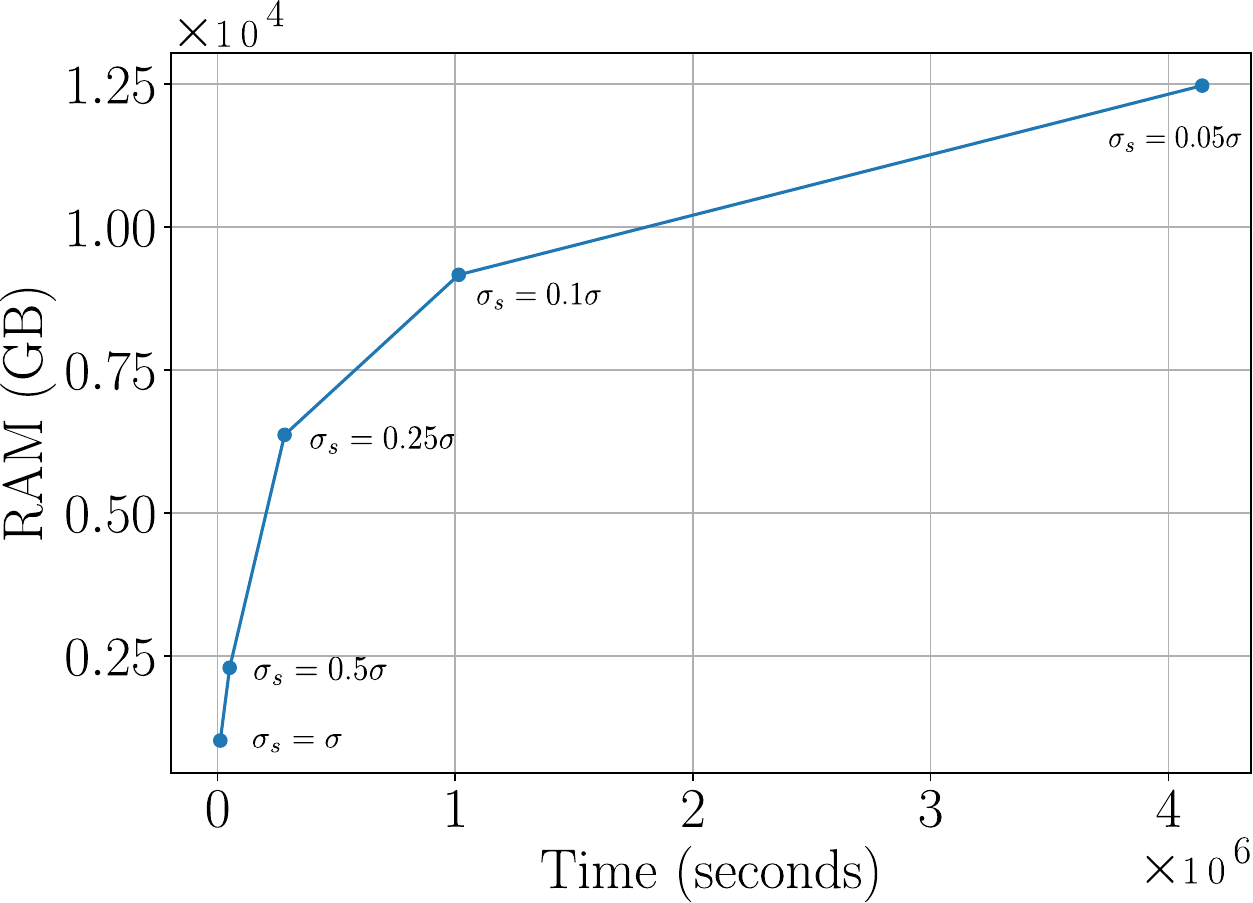}\\
	\caption{Computational consumption of the numerical method described by equations (\ref{schdeqs1}), (\ref{schdeqs12}), and (\ref{schdeqs13}). Each point corresponds to each studied situation.}
	\label{comp_num}
 \end{figure}

\begin{figure}[h]
	\includegraphics[width=0.47\textwidth]{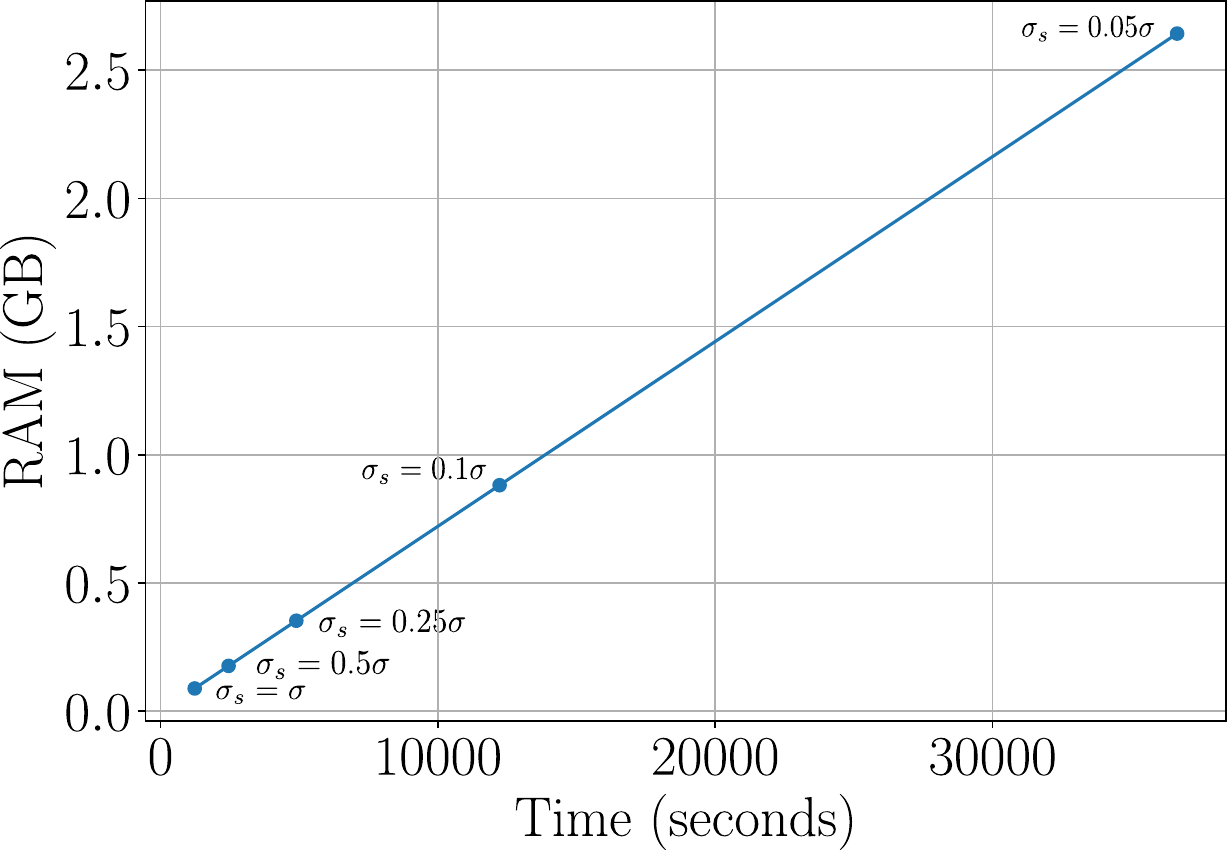}\\
	\caption{Computational consumption of the analytical method. Each point corresponds to each studied situation.}
	\label{comp_an}
 \end{figure}

The computational consumption of the numerical and analytical methods are shown in Fig. \ref{comp_num} and Fig. \ref{comp_an}, respectively. We can observe that our analytical approach reduces significantly the RAM and time consumption (approximately three orders of magnitude in RAM and two orders of magnitude in time). Since our method avoids the discretization of the photon field, the consumption with the analytical approach scales linearly with the level of entanglement. This confirms that the use of PT not only provides new physical insight about ETPA, but also reduces significantly the computational effort providing a path to explore more complex systems.

\newpage
\section{Population dynamics of intermediate states $\langle m _{\nu}\rangle$}

As a matter of completeness, we include the population dynamics of the first 23 intermediate states $\langle m _{\nu}\rangle$ for uncorrelated photons, Fig.~\ref{pop_med_class}, and for different degrees of photon correlation, Fig.~\ref{fig_inter_1b}. Here we observe the natural dynamics of a TSE process whereby the intermediate states are excited and the resultant populations do not depend on the degree of quantum correlation of the photons, as previously shown by H. Oka \cite{oka2011control}. For different large degrees of photon correlation, we can also see that absorption is slower and resonant, while for low correlation degrees, population oscillations, commonly found in near-resonant absorption, are observed.

\begin{widetext}
\begin{center}
\begin{figure}[H]
\begin{center}
	\includegraphics[width=0.3\textwidth]{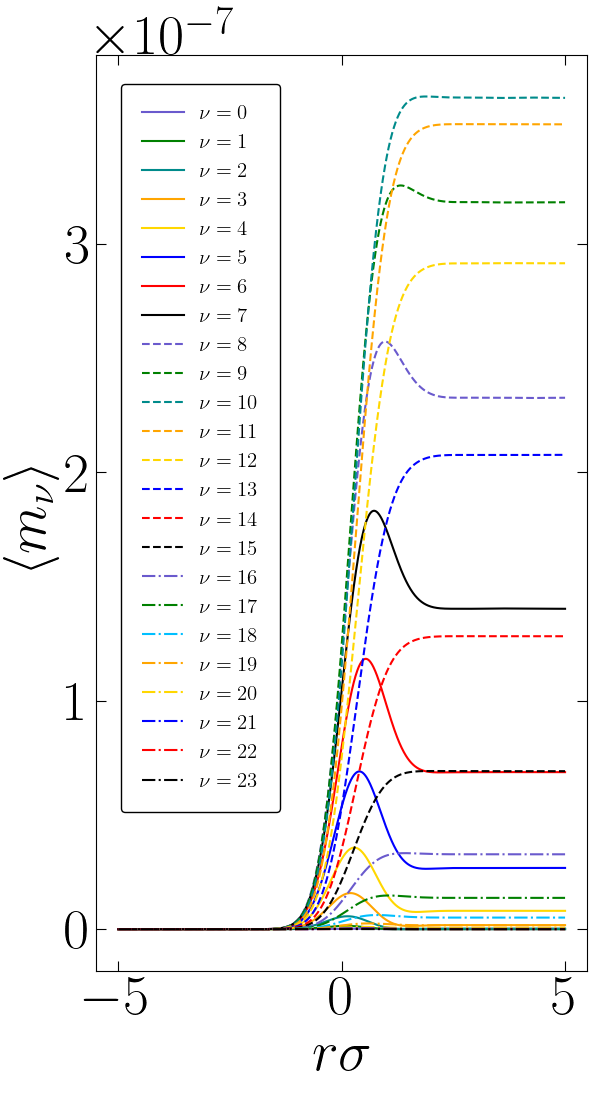}\\
	\caption{Population dynamics of intermediate states $\langle m _{\nu}\rangle$ as a function of $r\sigma$ for uncorrelated photons characterized by JSA given in (\ref{psi2unco}). }
	\label{pop_med_class}
\end{center}

 \end{figure}    
\end{center}

\begin{figure}[H]
    \centering
      \begin{tabular}{@{}ccc@{}}
    \begin{adjustbox}{valign=c}
    \includegraphics[width=.27\textwidth]{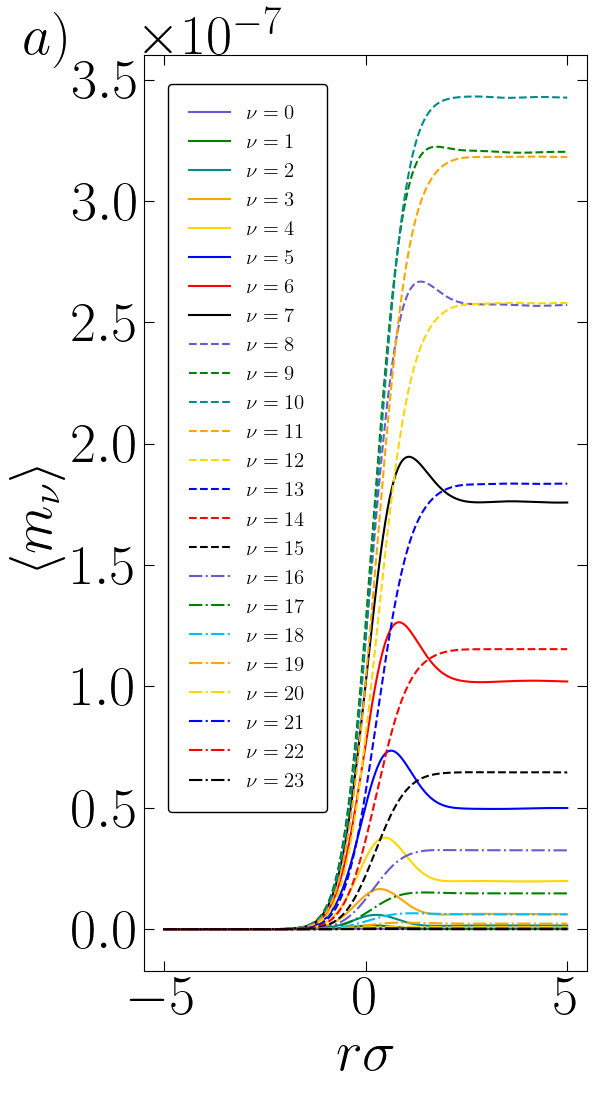}
    \end{adjustbox}&
    \begin{adjustbox}{valign=c}
    \includegraphics[width=.27\textwidth]{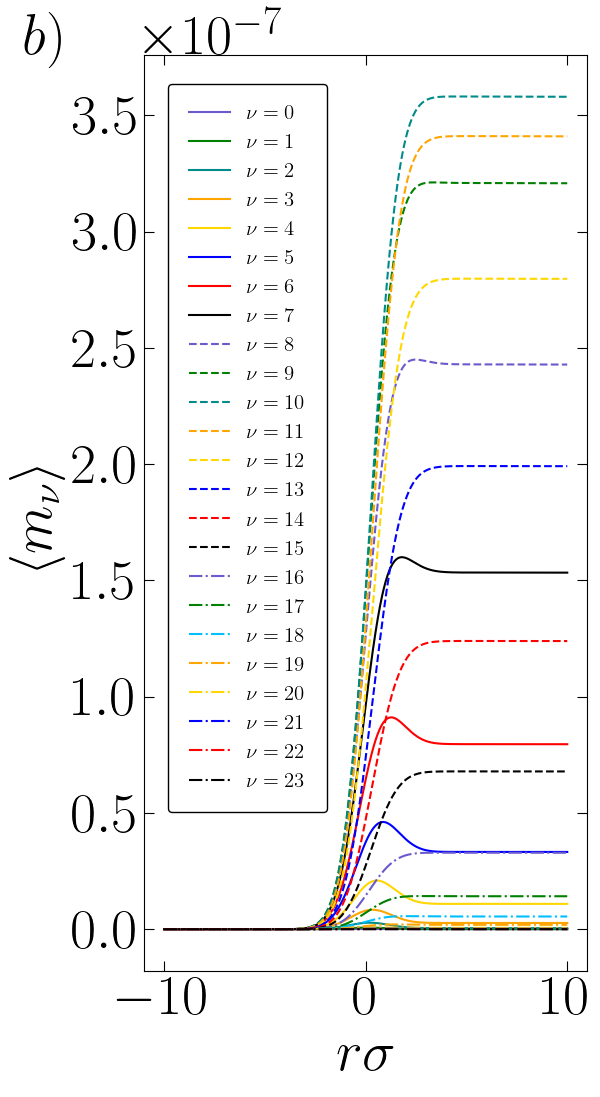}
    \end{adjustbox}& 
    \begin{adjustbox}{valign=c}
    \includegraphics[width=.27\textwidth]{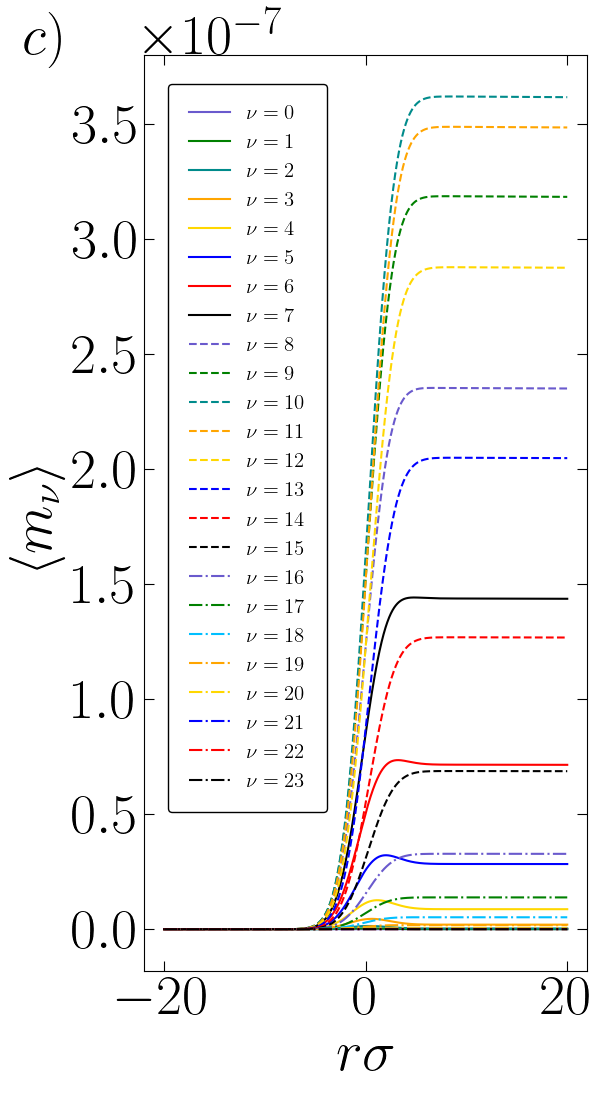}
    \end{adjustbox}\\      

  \end{tabular}
 
  \caption{Population dynamics of intermediate states $\langle m _{\nu}\rangle$ as a function of $r\sigma$ for correlated photons characterized by the JSA given in  (\ref{psientangled112}) for: $a)$ $\sigma _{s} = \sigma$; $b)$ $\sigma _{s} = 0.5\sigma$; and $c)$ $\sigma _{s} = 0.25\sigma$. For largely correlated photons, the absorption is slower and resonant, while for low degree of correlation, population oscillations, commonly found in near-resonant absorption of uncorrelated photons, are observed.}
  \label{fig_inter_1b}

\end{figure}   

\newpage

\section{Explicit form of $I_{\nu \alpha}^{(u)}(t, t_{0})$, $I_{\nu \alpha}^{(en,1)}(t, t_{0})$ and $I_{\nu \alpha}^{(en,2)}(t, t_{0})$}
\label{explicitfapp}

In this appendix, we present the explicit forms of the functions inside the representation series of the transition probabilities. Let us start with Eq. (\ref{uncoexp1}) for the case of uncorrelated photons. Function $f^{(u,1)}_{n,l}(t,t_{0})$ is defined as follows: 
\begin{equation}
f^{(u,1)}_{n,l}(t,t_{0}) = \frac{(2t\sigma ^{2}-i(k_{0} - \omega _{m_{\nu}}))^{2l}}{(2it\sigma ^{2}+k_{0}-\Omega _{\alpha \nu})^{-2n}} \Delta \text{F}_{1} ,
\end{equation}
being
\begin{eqnarray}
\Delta \text{F}_{1}  = \text{F}_{1} \left( 2; -2l, -2n; 3; \frac{2it\sigma ^{2}}{-2it\sigma ^{2}-k_{0} + \omega _{m_{\nu}}} , \frac{2it\sigma ^{2}}{2it\sigma ^{2}+k_{0}-\Omega _{\alpha \nu}} \right)
\nonumber \\
-\text{F}_{1} \left( 2; -2l, -2n; 3; \frac{2t_{0}\sigma ^{2}}{2t\sigma ^{2}-i(k_{0} - \omega _{m_{\nu}})} , \frac{2it_{0}\sigma ^{2}}{2it\sigma ^{2}+k_{0}-\Omega _{\alpha \nu}} \right) ,
\nonumber \\
\end{eqnarray}
where $\text{F}_{1}(a;b_{1},b_{2};c;x,y)$ is the Appell hypergeometric function of two variables.

On the other hand, the function $f^{(u,2)}_{n,l}(t,t_{0})$ is defined as follows:

\begin{eqnarray}
f^{(u,2)}_{n,l}(t,t_{0}) = &\,& \frac{(4t\sigma ^{2}-i(k_{0} - \omega _{m_{\nu}}))^{2l+1}}{(4it\sigma ^{2}-k_{0}+\Omega _{\alpha \nu})^{-2n-1}}\, _{2}\text{F}_{1}\left( 1, 2(n+l+1);2(n+1);\frac{-4it\sigma ^{2}-k_{0}+\Omega _{\alpha \nu}}{\omega _{e_{\alpha}} -2 \omega _{m_{\nu}}} \right)   
\nonumber \\
&-& \frac{(2(t-t_{0})\sigma ^{2}-i(k_{0} - \omega _{m_{\nu}}))^{2l+1}}{(2i(t-t_{0})\sigma ^{2}+k_{0}-\Omega _{\alpha \nu})^{-2n-1}}\, _{2}\text{F}_{1}\left( 1, 2(n+l+1);2(n+1);\frac{-2i(t-t_{0})\sigma ^{2}-k_{0}+\Omega _{\alpha \nu}}{\omega _{e_{\alpha}} -2 \omega _{m_{\nu}}} \right) ,
\end{eqnarray}
where $_{2}F_{1}(a, b; c; z)$ is the hypergeometric function.

For the entangled photons case, the term $I_{\nu \alpha}^{(en,1)}(t, t_{0})$ in (\ref{entexp1}) is defined as
\begin{eqnarray}
I_{\nu \alpha}^{(en,1)}(t, t_{0}) = \frac{1}{2\pi}\sum _{n,l=0}^{\infty}\frac{4^{-n-l}\sigma ^{-2l-3} \sigma _{s}^{-2n+1}}{n!l!(2l+1)(2l+3)} \left[ \frac{f^{(e,1)}_{n,l}(t) + f^{(e,2)}_{n,l}(t,t_{0})}{4it\sigma ^{2}\sigma _{s} ^{2} - \sigma ^{2}(2k _{0} - \omega _{e_{\alpha}}) + \sigma _{s}^{2}(k_{0} - \omega _{m_{\nu}})  }  \right] ,
\end{eqnarray}
being
\begin{eqnarray}
f^{(e,1)}_{n,l}(t) = \frac{(-1)^{l}(k_{0} - \omega _{m_{\nu}})^{2l+3}}{(4t\sigma _{s} ^{2} + i(2k _{0} - \omega _{e_{\alpha}}) )^{-2n-1}} \, _{2}\text{F}_{1}\left( 1, 2(n+l+2);2(l+2);\frac{\sigma _{s}^{2}(k_{0} - \omega _{m_{\nu}})}{4it\sigma ^{2}\sigma _{s} ^{2} - \sigma ^{2}(2k _{0} - \omega _{e_{\alpha}}) + \sigma _{s}^{2}(k_{0} - \omega _{m_{\nu}}) } \right) ,
\end{eqnarray}
and
\begin{eqnarray}
f^{(e,2)}_{n,l}(t,t_{0}) = \frac{i (2\sigma ^{2}(t+t_{0})-i(k_{0} - \omega _{m_{\nu}}))^{2l+3}}{(2\sigma _{s} ^{2}(t-t_{0}) + i(2k _{0} - \omega _{e_{\alpha}}) )^{-2n-1}} \, _{2}\text{F}_{1}\left( 1, 2(n+l+2);2(l+2);\frac{i\sigma _{s}^{2}(2\sigma ^{2}(t+t_{0})-i(k_{0} - \omega _{m_{\nu}}))}{4it\sigma ^{2}\sigma _{s} ^{2} - \sigma ^{2}(2k _{0} - \omega _{e_{\alpha}}) + \sigma _{s}^{2}(k_{0} - \omega _{m_{\nu}}) } \right)   .
\end{eqnarray}

Now, the term $I_{\nu \alpha}^{(en,2)}(t, t_{0})$ in (\ref{entexp2}) is defined as

\begin{eqnarray}
I_{\nu \alpha}^{(en,2)}(t, t_{0}) = \frac{4}{\pi \sqrt{\sigma ^{2} + \sigma _{s}^{2}}} \left\lbrace \frac{1}{\sigma ^{2}} \sum _{n,l=0}^{\infty} c_{nl}f^{(e,3)}_{n,l}(t, t_{0}) + \frac{i}{\sigma _{s} ^{2}} \sum _{n,l'=0}^{\infty} c_{nl'}f^{(e,4)}_{n,l'}(t, t_{0})\right\rbrace ,
\end{eqnarray}
being
\begin{equation}
c_{nl} = \frac{2^{-2(n+l)-3}\Gamma (2n+1)(\sigma \sigma _{s})^{-2n+1}}{n!l!(2l+1)(\sigma ^{2} + \sigma _{s} ^{2})^{n+l+2}}    
\end{equation}
\begin{eqnarray}
f^{(e,3)}_{n,l}(t, t_{0}) =&\, & \frac{(2i\sigma ^{2}(t+t_{0})+4it\sigma _{s}^{2}+k_{0}-\omega _{m_{\nu}})^{2l+3}}{(-2i\sigma^2 \sigma_s^2 (t-t_{0})- \sigma^2 (2k _{0} - \omega _{e_{\alpha}}) - \sigma _{s}^{2} (k_{0}-\Omega _{\alpha \nu}))^{-2n-1}}  
\nonumber \\
&\times & \, _{2}\text{F}_{1}^{\text{R}}\left( 1, 2(n+l+2);2(n+1);\frac{2i\sigma^2 \sigma_s^2 (t-t_{0})+ \sigma^2 (2k _{0} - \omega _{e_{\alpha}}) + \sigma _{s}^{2} (k_{0}-\Omega _{\alpha \nu})}{(\sigma ^{2} + \sigma _{s} ^{2}) (4it\sigma_s^2+2k_{0}+\omega _{e_{\alpha}})} \right)   
\nonumber \\
&+& \frac{(4it\sigma _{s}^{2}+k_{0}-\omega _{m_{\nu}})^{2l+3}}{(4it\sigma^2 \sigma_s^2 (t-t_{0})+ \sigma^2 (2k _{0} - \omega _{e_{\alpha}}) + \sigma _{s}^{2} (k_{0}-\Omega _{\alpha \nu}))^{-2n-1}}  
\nonumber \\
&\times & \, _{2}\text{F}_{1}^{\text{R}}\left( 1, 2(n+l+2);2(n+1);\frac{4it\sigma_s^2 + \sigma^2 (2k _{0} - \omega _{e_{\alpha}}) + \sigma _{s}^{2} (k_{0}-\Omega _{\alpha \nu})}{(\sigma ^{2} + \sigma _{s} ^{2}) (4it\sigma_s^2+2k_{0}+\omega _{e_{\alpha}})} \right) ,  \end{eqnarray}
and
\begin{eqnarray}
f^{(e,4)}_{n,l'}(t, t_{0}) =&\, & \frac{(-1)^{1-l'}(-2\sigma _{s} ^{2}(t-t_{0})+i(k_{0}-\omega _{m_{\nu}}))^{2l+3}}{(-2i\sigma^2 \sigma_s^2 (t-t_{0})- \sigma^2 (2k _{0} - \omega _{e_{\alpha}}) - \sigma _{s}^{2} (k_{0}-\Omega _{\alpha \nu}))^{-2n-1}}  
\nonumber \\
&\times & \, _{2}\text{F}_{1}^{\text{R}}\left( 1, 2(n+l'+2);2(n+1);\frac{2i\sigma^2 \sigma_s^2 (t-t_{0})+ \sigma^2 (2k _{0} - \omega _{e_{\alpha}}) + \sigma _{s}^{2} (k_{0}-\Omega _{\alpha \nu})}{(\sigma ^{2} + \sigma _{s} ^{2}) (k_{0}-\Omega _{\alpha \nu})} \right)   
\nonumber \\
&+& \frac{i(4it\sigma _{s}^{2}+k_{0}-\omega _{m_{\nu}})^{2l'+3}}{(4it\sigma^2 \sigma_s^2 (t-t_{0})+ \sigma^2 (2k _{0} - \omega _{e_{\alpha}}) + \sigma _{s}^{2} (k_{0}-\Omega _{\alpha \nu}))^{-2n-1}}  
\nonumber \\
&\times & \, _{2}\text{F}_{1}^{\text{R}}\left( 1, 2(n+l'+2);2(n+1);\frac{4it\sigma_s^2 + \sigma^2 (2k _{0} - \omega _{e_{\alpha}}) + \sigma _{s}^{2} (k_{0}-\Omega _{\alpha \nu})}{(\sigma ^{2} + \sigma _{s} ^{2}) (k_{0}-\Omega _{\alpha \nu})} \right) .\end{eqnarray}

Where $ _{2}\text{F}_{1}^{\text{R}}(a, b; c; z) $ is the regularized hypergeometric function given by $_{2}\text{F}_{1}^{\text{R}}(a, b; c; z)=\, _{2}F_{1}(a, b; c; z)/\Gamma (c)$, being $\Gamma (z)$ the Gamma function. These series representations, composed of error functions and hypergeometric functions, show the intricate relationships between the photons quantum correlations and the vibrational structure.
\end{widetext}

\bibliography{ref_pt}

\end{document}